\documentclass[ reprint, amsmath,amssymb, aps, pra,]{revtex4-2}

\usepackage[utf8]{inputenc}
\usepackage{graphicx}
\usepackage{dcolumn}
\usepackage{bm}
\usepackage{colortbl}  
\usepackage{diagbox}   
\usepackage[table]{xcolor}
\usepackage{amsthm}
\usepackage{amsmath}
\usepackage{amssymb}
\usepackage{mathrsfs}
\usepackage{comment}
\usepackage{color}
\usepackage[breaklinks=true,colorlinks,allcolors=blue]{hyperref}
\usepackage{braket}
\usepackage{stackengine}
\usepackage{scalerel}
\usepackage{dsfont}
\usepackage{float}
\usepackage{longtable}
\usepackage{tikz}
\usepackage{tabularx}

\usepackage{multirow}
\usepackage{array}

\newcommand*\circled[1]{\tikz[baseline=(char.base)]{
            \node[shape=circle,draw,inner sep=2pt] (char) {#1};}}

\begin{document}

\preprint{APS/123-QED} 

\title{Thermodynamics of coherent energy exchanges between lasers and two-level systems}

\author{Ariane Soret$^{\dagger,*}$, Massimiliano Esposito$^\dagger$}

\affiliation{$^\dagger$Complex Systems and Statistical Mechanics, Department of Physics and Materials Science, University of Luxembourg, L-1511 Luxembourg, Luxembourg}

\email{ariane.soret@gmail.com} 

\date{\today}

\begin{abstract}
We study the quantum thermodynamics of a coherent macroscopic electromagnetic field (laser) coupled to a two-level system (qubit) near resonance, from weak to strong driving regimes. This combined system is, in turn, weakly coupled to a thermal radiation field, and can be described by an autonomous quantum master equation. We show that the laser acts as an autonomous work source and that, in the macroscopic limit, the work produced is independent of the phase of the laser. Using the dressed qubit approach, we show that the variation of energy in the laser is not the work transferred to the dressed qubit, which is instead obtained from the ``dressed laser'' –  a coherent superposition of the laser and the qubit. Using a two-point measurement technique with counting fields, we obtain the full counting statistics for the work of the laser and dressed laser, and show that they satisfy Crooks fluctuation theorems. We then use these theorems as criteria to investigate the thermodynamic consistency of quantum master equations, first in the autonomous setup for the combined system, then in the non-autonomous setup for the quantum system where the coherent field is eliminated and effectively described by a time dependent external field. Treating the laser as an external field is known to yield expressions for the work which are in contradiction with quantum thermodynamics predictions in the strong driving regime. We show that these inconsistencies stem from a confusion between the laser and dressed laser, and show how to correct them. We also derive a new master equation, thermodynamically consistent across all driving regimes, from which the Bloch and Floquet master equations can be obtained using additional approximations (of which we also examine the consistency). 
\end{abstract}

\maketitle

\section{Introduction}

Quantum optics is the study of the interaction of matter (atoms and molecules) with quantized radiation fields \cite{milburn2008}. In many cases, the radiation fields can be treated as baths, and their degrees of freedom can be traced out from the equations of motion, leading to a quantum optical master equation describing the dynamics of the system of atoms and molecules. This procedure is well suited, for example, to describe the decay of a two-level system in vacuum, or the resonance fluorescence when an external coherent field is also present \cite{breuer2002theory}. In the context of quantum computing and technologies, many implementations rely on the ability to coherently monitor a two-level system, or qubit, using a laser, while competing with spontaneous emission processes which act as a damping mechanism. The dynamics of the qubit is then described by the optical Bloch equation \cite{grynberg1998,redfield1955,elouard2020thermodynamics}, or, in the strong driving regime, by the Floquet master equation \cite{alicki2013,langemeyer2014,mori2023}, derived from the quantum Floquet theory \cite{shirley1965}. 

The optical Bloch equation was primarily used in spectroscopy \cite{allen1975}, which motivated early works on its thermodynamic consistency \cite{skinner1995}.
The rapid and recent development of quantum technologies has revived interest in the thermodynamic consistency of quantum optical master equations \cite{alicki2013, langemeyer2014, cuetara2015rapid, hofer2017markovian, elouard2020thermodynamics,prasad2024closing}, with the Bloch and Floquet master equations being increasingly used to study the thermodynamics of driven quantum systems \cite{kosloff2015,uzdin2016,poem2019,donvil2018,schaller2018, zhang2022}. 
More precisely, the consistency of the Bloch master equation has been studied at the average level \cite{elouard2020thermodynamics,skinner1995}, while the full counting statistics has been done for the steady state of the Floquet master equation \cite{cuetara2015rapid}. In these approaches, the driving field is described by an external time dependent field, which interacts with the qubit through a time dependent Hamiltonian $\hat{V}(t)$. 
Interestingly, it was found \cite{elouard2020thermodynamics} that, in the strong drive regime, the rate of work performed by the driving field on the qubit is \textit{not} equal to the expression predicted by quantum thermodynamics, namely $\text{Tr}[\hat{\rho}(t)d_t \hat{V}(t)]$ \cite{spohn1978irreversible, alicki1979quantum}. Instead, the work depends on the rates of dissipation to the bath. The qualitative interpretation of this feature is that the work results from a non conservative force, arising from neglecting the fluctuation of the number of photons \cite{elouard2020thermodynamics}. We show that this explanation is not sufficient, and that the complete answer is that the Floquet master equation describes the energy transfers in the dressed qubit basis (the eigenbasis of the joint qubit-coherent field system), in which the work source is not the original coherent field. 
Moreover, the consistency at the fluctuating level, i.e., whether the master equations preserve fluctuation theorems, has never been addressed.

In this paper, we study thermodynamics at the average and fluctuating level of a qubit driven by a coherent monochromatic radiation (further on called the laser) and dissipating to a thermal cavity (the bath). 
Importantly, everything is derived starting from a microscopic and autonomous description of the combined qubit-laser-bath system where the laser is modeled as a monochromatic mode in a macroscopic (large number of photons) coherent state. 
In section~\ref{sec:laser}, we provide a general description of lasers as work sources, and show that the work transferred by a single laser source to a generic quantum system is independent of the phase of the laser in the macroscopic limit (large number of photons). In section~\ref{sec:uni}, we introduce the qubit-laser-bath model, starting from the full unitary level. We study the model's dynamics, first in the autonomous description using the dressed qubit approach \cite{grynberg1998}, then in the equivalent non-autonomous description for the qubit after a frame rotation. Using the two-point measurement method with counting fields \cite{esposito2009nonequilibrium}, we then derive and compare the laws of thermodynamics and work fluctuation theorems in both descriptions. 
In section~\ref{sec:cons-qme}, we identify the thermodynamic consistency conditions implied by the previous results at the level of quantum master equations in both the autonomous and non-autonomous pictures. In section~\ref{sec:cons-obe-fme}, we use the formalism of quantum maps to derive a new master equation, called generalized Bloch equation, valid at all driving strengths and which is thermodynamically consistent. We then examine the Bloch and Floquet quantum master equations, which are obtained from the generalized Bloch equation using additional approximations. We find that the Floquet equation is fully consistent (the fluctuation theorems are preserved and the first and second law hold at the average and fluctuating levels) and that the work performed on the dressed qubit is indeed not expressed in the canonical form $\text{Tr}[\hat{\rho}(t)d_t \hat{V}(t)]$, but has become a non-conservative force of which we explain the origin.
The Bloch equation instead satisfies the fluctuation theorems, but the first law of thermodynamics is only satisfied at the average level. In section~\ref{sec:red}, we comment on an alternative derivation of the Bloch master equation, commonly used in the literature, which relies on the Redfield equation, see for instance \cite{breuer2002theory, elouard2020thermodynamics}. We show that the Bloch master equation, dressed with counting fields, obtained from the Redfield equation, is different from the one obtained using the quantum maps, although both approaches yield the same Bloch master equation once the counting fields are set to zero. The Redfield approach breaks the fluctuation theorems, which implies that the thermodynamics at the fluctuating level should be examined using master equations derived from the quantum maps. In section~\ref{sec:steady-state}, we compare the steady-state heat and work flows predicted by the Bloch and Floquet master equations in their common regime of validity. A summary of results is given in section~\ref{sec:summary}, while conclusions are drawn in section~\ref{sec:concl}. Throughout the paper, we set $\hbar=1$ and $k_B=1$.

\section{Work from a laser source}\label{sec:laser}

In this section, we discuss the properties of a laser coupled to a generic quantum system. The total system-laser is a closed system, described by a density matrix $\hat{\rho}(t)$ evolving according to a unitary operator $\hat{U}$, $\hat{\rho}(t)=\hat{U}\hat{\rho}(0)\hat{U}^\dagger$. We further assume that initially 
\begin{equation}
    \hat{\rho}(0)=\hat{\rho}_S(0)\otimes\hat{\rho}_L(0) \, .
    \label{eq:ini-general}
\end{equation}
In later sections, the system will be a qubit.  

\subsection{Lasers: autonomous work sources}

Quantum mechanically, a monochromatic radiation field of frequency $\omega_L$, is described by the Hamiltonian 
\begin{equation}
 \hat{H}_L=\omega_L(\hat{a}^\dagger\hat{a}+1/2) \, ,
 \label{eq:def-hl}
\end{equation}
where $\hat{a}^\dagger, \hat{a}$ are bosonic creation and annihilation operators. A laser can in turn be modelled as a field in a pure coherent state, described by the density matrix
\begin{equation}
\begin{array}{ll}
\hat{\rho}^{\text{coh}}_L&=|\alpha\rangle\langle\alpha| \;, \ \ \text{with}\\
\\
 |\alpha\rangle&\equiv e^{-|\alpha|^2/2}\sum\limits_{N\geq 0}\frac{\alpha^{N}}{\sqrt{N!}}|N\rangle \, ,
\end{array}
\label{eq:coh-state}
\end{equation}
where $\alpha=|\alpha|e^{i\phi}$ and $|\alpha|$ and $\phi$ are respectively the amplitude and phase. 
The corresponding average number of photons is $\langle N\rangle = |\alpha|^2$, and the standard deviation is $\sigma(N) \equiv \sqrt{\langle N^2\rangle-\langle N\rangle^2}=|\alpha|$. 

From a thermodynamics viewpoint, a laser is an autonomous work source, because its change in von Neumann entropy, $S \equiv \text{Tr}[\hat{\rho}_L\log\hat{\rho}_L]$, while interacting with the system, is negligible compared to its corresponding change in energy $E_L \equiv \text{Tr}[\hat{H}_L\hat{\rho}_L]$ \cite{strasberg2017repeated},
\begin{equation}
    \frac{\Delta S_L}{\Delta E_L}\rightarrow 0 \;.
    \label{eq:def-work-source}
\end{equation}
This remains true when averaging over the phase $\phi$ of the laser, after which the laser source is described by a phase averaged state, also called a Poisson state, 
\begin{align}
\begin{aligned}
  \hat{\rho}_L^{\text{poi}}&\equiv  e^{-|\alpha|^2}\sum\limits_{N_\geq 0}\frac{|\alpha|^{2N}}
    {N!}|N\rangle \langle N|\\
  &  = \int\limits_0^{2\pi} d\phi ||\alpha|e^{i\phi}\rangle\langle|\alpha|e^{i\phi}| \, ,
\end{aligned}
\label{eq:poisson}
\end{align}
which yields, as for a coherent state, $\langle N\rangle = |\alpha|^2$ and  $\sigma(N)=|\alpha|$. It will furthermore be useful to think of a Poisson state as a thermal state at infinite temperature. Indeed, it is known that,  in the large $|\alpha|^2$ limit, a Poisson distribution converges to a Gaussian distribution of average $|\alpha|^2$ and standard deviation $|\alpha|$. In turn, such a Gaussian state is equivalent to a Gibbs state at temperature $\beta_L^{-1}\equiv|\alpha|^2$. See appendix~\ref{app:work-source} for a proof of \eqref{eq:def-work-source}, for both, a coherent and Poisson state, and Fig.\ref{fig:plots} for a numerical check in the case where the system is a qubit. 

As a result, we may then identify minus the variation of energy in the laser as work, 
\begin{equation}
    W_L\equiv -\Delta E_L = -\text{Tr}[\hat{H}_L(\hat{\rho}_L(t)-\hat{\rho}_L(0))] \, .
    \label{eq:def-wl}
\end{equation}

\subsection{Irrelevance of the phase}\label{subsec:irrelevence}

The phase $\phi$ may be difficult to determine in practice. However, as we now show, in the macroscopic limit $|\alpha|\rightarrow +\infty$, the work \eqref{eq:def-wl} becomes independent of $\phi$, and the laser source can equivalently be described by a Poisson state \eqref{eq:poisson}.
%


\subsubsection{Average work}

The expectation value of $\hat{H}_L$ is given by 
\begin{align}
   & \text{Tr}[\hat{H}_L\hat{\rho}(t)]=  \sum\limits_{N,N',N''\geq 0}\langle N|\hat{H}_L|N\rangle \times\\
& \text{Tr}_S[\langle N|\hat{U}|N'\rangle\hat{\rho}_S(0)\langle N'|\hat{\rho}_L(0)|N''\rangle\langle N''|\hat{U}^\dagger|N\rangle]] \nonumber \, .
\end{align}
In the macroscopic limit $|\alpha|\gg 1$, since the distribution $e^{-|\alpha|^2/2}\frac{\alpha^{N}}{\sqrt{N!}}$ is peaked around $N= |\alpha|^2$, for both a coherent and a Poisson state, we may approximate
\begin{align}
   & \text{Tr}[\hat{H}_L\hat{\rho}(t)]\sim  \sum\limits_{N\geq 0}\sum\limits_{N'\in\Delta}\langle N|\hat{H}_L|N\rangle \times\\
& \text{Tr}_S[\langle N|\hat{U}|N'\rangle\hat{\rho}_S(0)\langle N'|\hat{\rho}_L(0)|N'\rangle\langle N'|\hat{U}^\dagger|N\rangle]] \, ,
\end{align}
where $\Delta\equiv [|\alpha|^2-|\alpha|,|\alpha|^2+|\alpha|]$. This means that for a macroscopic coherent and a Poisson state,  \eqref{eq:def-work-source} and \eqref{eq:def-wl} hold. We numerically confirm this expectation in Fig.~\ref{fig:plots}, where the system is a qubit. 

\subsubsection{Work fluctuations}\label{subsubsec:work-fluct}

The above discussion is for average quantities. We now show that, at the level of fluctuations, performing a two-point measurement \cite{esposito2009nonequilibrium} when the laser is initialized in a Poisson state is equivalent to performing a series of measurements on a system initialized in a coherent state and averaging over the initial phase. 

Let us first recall the two-point measurement approach with counting fields: given a (possibly time dependent) observable $\hat{A}$, of eigenvalues $\{a_m(\tau)\}$ at time $\tau$, the probability to observe a fluctuation $\Delta a$ when measuring $\hat{A}$ at times $0$ and $t$ is given by
\begin{equation}
    p(\Delta a) = \sum\limits_{a_l(0),a_m(t)} P[a_m(t),a_l(0)]\delta[\Delta a -(a_m(t)-a_l(0))] \, ,
\end{equation}
where $P[a_m(t),a_l(0)]$ is the joint probability to measure $a_l(0)$ at time $0$ and $a_m(t)$ at time $t$. The statistics of $p(\Delta a)$ is conveniently described using the moment generating function, defined as the Fourier transform of $p(\Delta a)$,
\begin{equation}
    \mathcal{G}(\lambda,t) \equiv \int\limits_{-\infty}^{+\infty}  e^{i\lambda\Delta a}p(\Delta a) \, d\Delta a \, ,
    \label{eq:def-mgf}
\end{equation}
where $\lambda\in\mathbb{R}$ is called a \textit{counting field}. The time dependence of $\mathcal{G}(\lambda,t)$ lies in $\Delta a$, which corresponds to a fluctuation observed between the times $0$ and $t$.
The moment generating function $\mathcal{G}(\lambda,t)$ is also equal to the trace of the ``tilted'' density matrix $\hat{\rho}^\lambda(t)$, obtained from the evolution operator $\hat{U}(t,0)=\mathcal{T}_{\leftarrow}[e^{-i\int_0^t ds\hat{H}(s)}]$ (with $\mathcal{T}_{\leftarrow}$ denoting time-ordering), dressed with the counting field $\lambda$:
\begin{equation}
    \begin{array}{ll}
    \mathcal{G}(\lambda,t) &= \text{Tr}[\hat{\rho}^\lambda(t)] \\
    \\
\hat{\rho}_\lambda(t) &\equiv \hat{U}_{\lambda}(t,0)\hat{\bar{\rho}}(0)\hat{U}^{\dagger}_{-\lambda}(t,0)\\
\\
\hat{U}_\lambda(t,0) &\equiv e^{i\hat{A}(t)\lambda/2}\hat{U}(t,0)e^{-i\hat{A}(0)\lambda/2}
  \, ,
    \end{array}
    \label{eq:rho-lambda}
\end{equation}
where $\hat{\bar{\rho}}(0)$ is the diagonal part of $\hat{{\rho}}(0)$ in the eigenbasis of $\hat{A}(0)$ chosen for the measurement. From \eqref{eq:rho-lambda}, and assuming the initial condition \eqref{eq:ini-general}, it is clear that the generating function obtained when the laser is initialized in a Poisson state, $\mathcal{G}^{\text{poi}}(\lambda,t)$, is related to the generating function, $\mathcal{G}^{\text{poi}}(\lambda,t)$, obtained with a coherent state, by
\begin{equation}
    \mathcal{G}^{\text{poi}}(\lambda,t) = \int\limits_0^{2\pi}d\phi \, \mathcal{G}^{\text{coh}}(\lambda,t) \, .
    \label{eq:equiv-mgf}
\end{equation}
The work $W_L$ defined in \eqref{eq:def-wl} is then obtained by measuring $-\hat{H}_L$. Since the Poisson state \eqref{eq:poisson} commutes with $\hat{H}_L$, the initial projective measurement does not modify the density matrix, $\overline{\hat{\rho}}(0)=\hat{\rho}(0)$, hence does not modify the dynamics, and the work fluctuations can be rigorously computed. This is not the case for a coherent state. However, from \eqref{eq:equiv-mgf}, we see that we may obtain the statistics of the work during a series of measurements performed on coherent states, by performing a single measurement on a Poisson state. Therefore, for all practical purposes, when dealing with measurements of thermodynamic observables, we will always assume that the laser is initialized in a Poisson state. 

To summarize, we showed that, from a thermodynamics viewpoint, a laser can equivalently be described as a pure coherent state or as a Poisson state. The advantage of the Poisson state description is that it allows a rigorous investigation of the work statistics using the two-point measurement scheme with counting fields. However, in many applications, a laser is described as an external, time-dependent field. 
We will refer to the time-dependent field description as the non-autonomous case, and we will investigate the thermodynamics in this case as well. 

As a final remark, we point out that knowing the initial phase only becomes important if other lasers with different phases were used later on, since they would induce a dephasing. In the case of multiple coherent sources, one could resort instead to the recently developed photon resolved Floquet theory \cite{engelhardt2023unified}, which is consistent with full counting statistics methods \cite{platero2023photonresolved}. The results presented in this work could in principle be extended to multiple light sources using the framework of \cite{platero2023photonresolved} for the counting statistics of the laser's photons. However, for the sake of clarity and without loss of generality, we focus on the case of a single laser. 

We now proceed to analysing the thermodynamics of a qubit driven by a laser.

\section{Unitary description}\label{sec:uni}

\subsection{Model}\label{subsec:model}

We denote by $X$ the system consisting of qubit $A$ and laser $L$. 
The total qubit-laser-bath system evolves in the product Hilbert space 
\begin{equation}
\begin{array}{ll}
\mathcal{H} &= \mathcal{H}_X\otimes\mathcal{H}_B \, ,\\
\\
\mathcal{H}_X &= \mathcal{H}_A\otimes\mathcal{H}_L \, , 
\end{array}
\end{equation}
where $\mathcal{H}_A,\mathcal{H}_L,\mathcal{H}_B$ are respectively the Hilbert spaces of the qubit, laser and bath, and where $\mathcal{H}_X$ is the Hilbert space of the qubit-laser system.
The qubit is characterised by a ground state $|a\rangle$ and an excited state $|b\rangle$, separated by $\omega_A$. The total Hamiltonian is
\begin{equation}
\hat{H}=\hat{H}_X\otimes\mathds{I}_B+\hat{V}_{AB}+\mathds{I}_X\otimes \hat{H}_B \, ,
    \label{eq:ham}
\end{equation}
where 
\begin{equation}
    \hat{H}_X=\hat{H}_A\otimes \mathds{I}_L+\hat{V}_{AL}+\mathds{I}_A\otimes\hat{H}_L \, .
\end{equation}
The laser Hamiltonian $\hat{H}_L$ was defined in \eqref{eq:def-hl}, while the qubit and bath Hamiltonians are respectively $\hat{H}_A=\frac{\omega_A}{2}\hat{\sigma}_z$ and $\hat{H}_B=\sum_k \omega_k (\hat{b}_k^\dagger \hat{b}_k+\frac{1}{2})$, where $\hat{b}_k^\dagger, \hat{b}_k$ are bosonic creation and annihilation operators. The  qubit-laser and  qubit-bath interaction Hamiltonians are respectively
\begin{equation}
    \begin{array}{ll}
         \hat{V}_{AL}&=\frac{g_0}{2}(\hat{\sigma}_+ +\hat{\sigma}_-)\otimes(\hat{a}+\hat{a}^\dagger)\otimes\mathds{I}_B,  \\
         \\
    \hat{V}_{AB}&=(\hat{\sigma}_++\hat{\sigma}_-)\otimes\mathds{I}_L\otimes(\hat{B}+\hat{B}^\dagger),
    \end{array}
    \label{eq:ops}
\end{equation}
with
\begin{equation}
\begin{array}{ll}
 \hat{\sigma}_+&=|b\rangle\langle a| \, ,\\
 \\
 \hat{\sigma}_-&=|a\rangle\langle b| \, ,\\
 \\
 \hat{\sigma}_z&=(|b\rangle\langle b|-|a\rangle\langle a|) \, ,
\end{array}
\end{equation}
and $\hat{B}=\sum_k \frac{g_k}{2}\hat{b}_k$, where $g_0,g_k\in\mathbb{C}$ are coupling amplitudes.
To alleviate the notation, we will drop the tensor products with identity operators when there is no ambiguity. The total qubit-laser-bath system is described by a density matrix $\hat{\rho}(t)$, which follows a unitary dynamics
\begin{equation}
    \frac{d\hat{\rho}(t)}{dt}=-i[\hat{H}(t),\hat{\rho}(t)] \, ,
    \label{eq:liouville}
\end{equation}
the solution of which is
\begin{equation}
    \hat{\rho}(t) = \hat{U}(t,0)\hat{\rho}(0)\hat{U}^\dagger(t,0) \, ,
\end{equation}
where $\hat{U}(t,0)\equiv e^{-i\hat{H}t}$ is the propagator. The combined qubit-laser system is described by a density matrix $\hat{\rho}_X\equiv \text{Tr}_B[\hat{\rho}]$, where $\text{Tr}_B$ denotes the partial trace over the space $\mathcal{H}_B$. The state of the qubit $A$ is in turn given by the density matrix $\hat{\rho}_A\equiv \text{Tr}_L[\hat{\rho}_X]$, and the state of the laser $L$ is described by the density matrix $\hat{\rho}_L\equiv \text{Tr}_A[\hat{\rho}_X]$. The density matrix is initially factorized,
\begin{equation}
    \hat{\rho}(0)=\hat{\rho}_A(0)\otimes\hat{\rho}^{\text{coh}}_L(0)\otimes\hat{\rho}_B \, ,
    \label{eq:rho-ini-alb}
\end{equation}
where the bath is in a Gibbs state, 
\begin{equation}
    \hat{\rho}_B=e^{-\beta_B \hat{H}_B}/Z_B \, ,
    \label{eq:rho-b}
\end{equation}
with $Z_B=\text{Tr}[ e^{-\beta_B \hat{H}_B}]$ and $\beta_B^{-1}$ the temperature.

\begin{figure}
    \centering
    \includegraphics[scale=0.5]{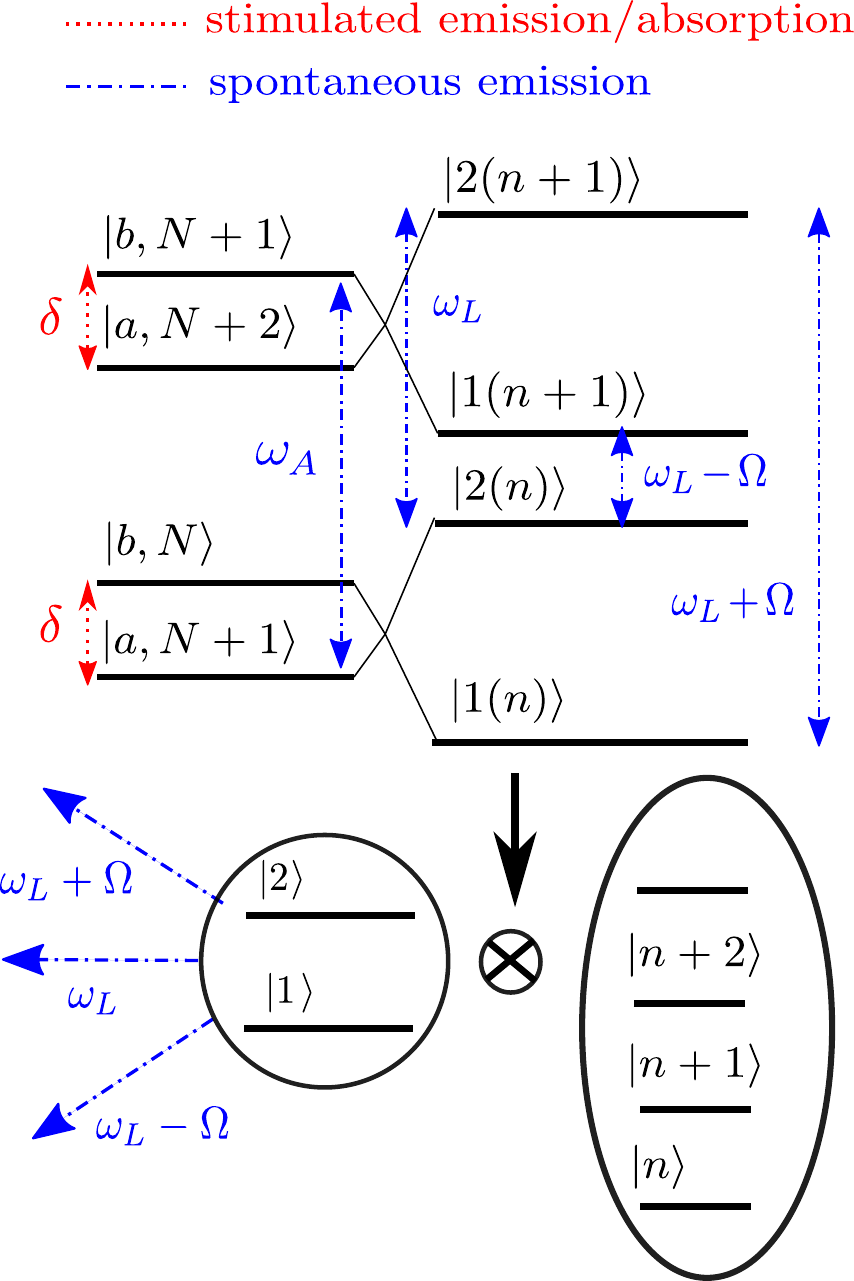}
    \caption{Schematic representation of the mapping to the dressed qubit space. Top left: the three processes at play represented in the product basis $\{|b,N\rangle,|a,N+1\rangle\}$ of the product Hilbert space $\mathcal{H}_A\otimes\mathcal{H}_L$: spontaneous emission and stimulated emission/absorption. Top right: change of basis to the eigenbasis of $\hat{H}_A+\hat{V}_{AL}+\hat{H}_L$. Bottom: mapping to $\mathcal{H}_{DA}\otimes\mathcal{H}_{DL}$. }
    \label{fig:triplet}
\end{figure}

There are two processes involved in the evolution of the qubit: the spontaneous absorption/emission and the stimulated absorption/emission. In the product basis $\{|b,N\rangle,|a,N\rangle\}$ of the Hilbert space $\mathcal{H}_A\otimes\mathcal{H}_L$, where $\{|N\rangle\}_{N\in\mathbb{N}}$ is the Fock basis for the photons of the mode $\omega_L$, these processes are represented as follows (see Fig.\ref{fig:triplet}): During a spontaneous emission (resp. absorption), the qubit emits (resp. absorbs) a photon from the bath, while the number of photons in the laser remains constant. Hence, a spontaneous emission (resp. absorption) induces a transition $|b,N\rangle\rightarrow|a,N\rangle$ (resp. $|a,N\rangle\rightarrow|b,N\rangle$) between states separated by an energy gap $\omega_A$. On the other hand, during a stimulated emission (resp. absorption), the qubit exchanges a photon with the laser, which corresponds to the transitions $|b,N\rangle\rightarrow|a,N+1\rangle$ (resp. $|a,N+1\rangle\rightarrow|b,N\rangle$) between states separated by an energy gap $|\delta|$, where 
\begin{equation}
    \delta\equiv \omega_A-\omega_L
\end{equation}
is the detuning between the qubit and laser frequencies. 

\begin{figure}
    \centering
    \includegraphics[scale=0.4]{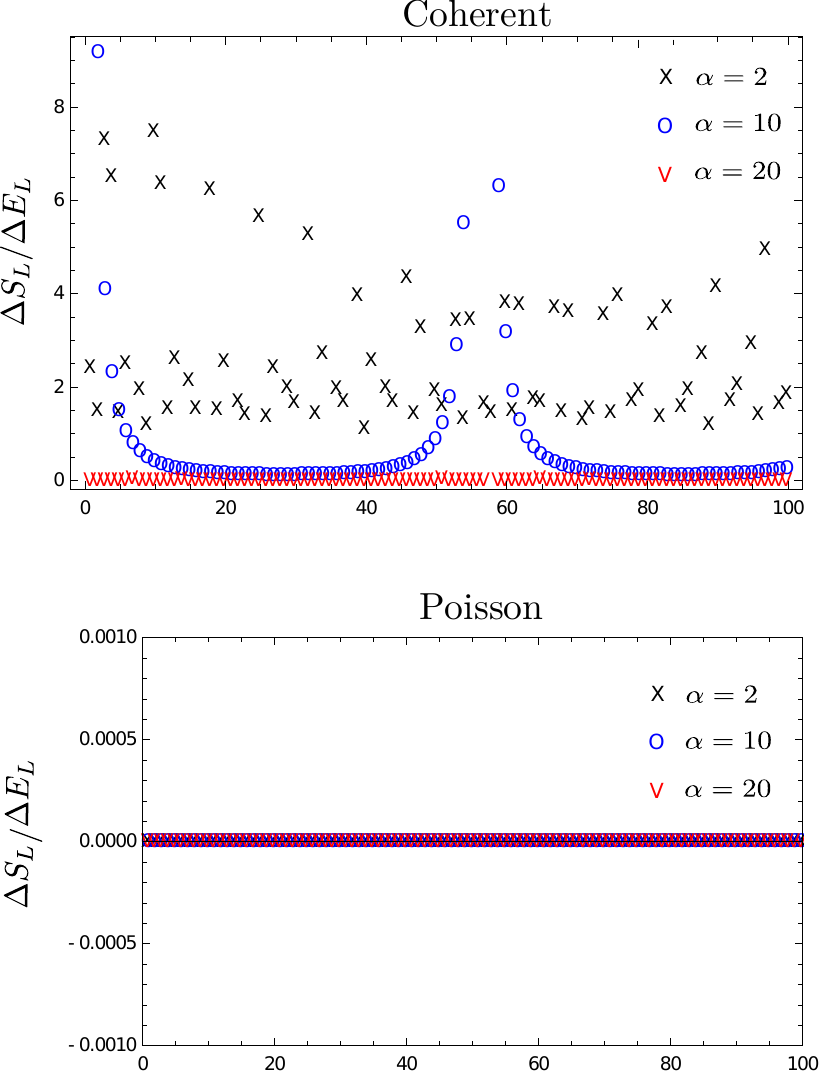}
    \caption{Ratio of the variation of the von Neumann entropy $\Delta S_L$ and of the energy $\Delta E_L$ of a laser interacting with a two-level system. The coupling Hamiltonian is identical to $\hat{V}_{AL}$, defined in \eqref{eq:ops}, without the $\mathds{I}_B$ component. Top: laser in a coherent state. Bottom: laser in a Poisson state.
    }
    \label{fig:plots}
\end{figure}

\subsection{Dressed qubit approach}\label{subsec:per-spec}

In this section, we introduce the two assumptions and approximations underlying near-resonant coherent driving, which we will make throughout this paper. We then present the dressed qubit approach \cite{grynberg1998}, and show that this approach leads to a change of basis, which allows us to write the Hilbert space $\mathcal{H}_X$ as a tensor product of two new Hilbert spaces.

\subsubsection{Assumptions}

We make the following two assumptions, later referred to as assumptions 1 and 2:
\begin{enumerate}
    \item The laser is nearly resonant with the qubit: \\ $\omega_L\gg |\omega_L-\omega_A| $.
    \item The photon statistics of the laser satisfies: \\ $\langle N\rangle\gg \sigma(N)\gg 1$,
\end{enumerate}
where $\sigma(N)= \sqrt{\langle N^2\rangle-\langle N\rangle^2}$ is the standard deviation.
Assumption 1 allows us to perform the rotating wave approximation, i.e. neglect the off-resonant terms $\hat{\sigma}_+\hat{a}^\dagger$ and $\hat{\sigma}_-\hat{a}$ in $\hat{V}_{AL}$ in \eqref{eq:ops} \cite{breuer2002theory}. Consequently, $\hat{H}_X$ becomes block diagonal in the product basis $\{|a,N+1\rangle,|b,N\rangle\}$, $\hat{H}_X =\sum_{N\in\mathbb{N}} \hat{H}_X^{(N)} 
$,
where $\hat{H}_X^{(N)}$ only acts on the subspace $\mathcal{E}(N)$ spanned by $\{|b,N\rangle,|a,N+1\rangle\}$.
%
%
Assumption 2 is satisfied in the macroscopic limit $|\alpha|\gg 1$, and implies that the relative variations of $\sqrt{N}$ in the range $\sigma(N)$ around $\langle N\rangle$ are small, which allows us to neglect the fluctuations of $\sqrt{N}$ in the subspace $\mathcal{E}(N)$, and to replace 
\begin{equation}
    g_0\sqrt{N+1}\approx g_0\sqrt{\langle N\rangle}\equiv g \, .
    \label{eq:g}
\end{equation}
Consequently, 
\begin{align}
&\hat{H}_X^{(N)}=\left[(N+1+\frac{1}{2})\omega_L-\frac{\omega_A}{2}\right]|a,N+1\rangle\langle a,N+1| \nonumber\\
&\hspace{1.35cm}+ \left[(N+\frac{1}{2})\omega_L+\frac{\omega_A}{2}\right]|b,N\rangle\langle b,N| \label{eq:hxn-2}\\
&\hspace{1.35cm}+\frac{g}{2}(|a,N+1\rangle\langle b,N|+|b,N\rangle\langle a,N+1|)  \, . \nonumber
\end{align}

\subsubsection{The dressed qubit and dressed laser Hilbert spaces}\label{subsubsec:mapping}

Under the assumptions 1 and 2, the restrictions $\hat{H}^{(N)}_X$ of $\hat{H}_X$ on $\mathcal{E}(N)$ defined in \eqref{eq:hxn-2} can be diagonalized by a unitary transformation which is identical in every $\mathcal{E}(N)$,
\begin{equation}
\begin{array}{l}
\hat{H}_X^{(N)} \\
\\
=(N+1+\frac{\Omega}{2})|2(n)\rangle\langle 2(n)| \\
\\
+ (N+1-\frac{\Omega}{2})|1(n)\rangle\langle 1(n)|\, ,
\end{array}
\label{eq:hx-diag}
\end{equation}
where
\begin{equation}
\begin{array}{ll}
|2(n)\rangle &\equiv\sqrt{\frac{\Omega+\delta}{2\Omega}}|b,N\rangle+ \sqrt{\frac{\Omega-\delta}{2\Omega}}|a,N+1\rangle\\
\\
|1(n)\rangle &\equiv -\sqrt{\frac{\Omega-\delta}{2\Omega}}|b,N\rangle +\sqrt{\frac{\Omega+\delta}{2\Omega}}|a,N+1\rangle
\end{array}
\label{eq:eig-bas}
\end{equation}
and where
\begin{equation}
    \Omega=\sqrt{\delta^2+g^2}
    \label{eq:omegaN}
\end{equation}
is the Rabi frequency \cite{grynberg1998}.

Using the eigenbasis $\{|1(n)\rangle,|2(n)\rangle\}_{n\in\mathbb{N}}$, we see that the total Hilbert space $\mathcal{H}_X=\mathcal{H}_A\otimes\mathcal{H}_L$ is equivalent to a tensor product of two new Hilbert spaces, defined by the change of basis (see Fig.\ref{fig:triplet})
\begin{equation}
    \begin{array}{ll}
    \mathcal{H}_{A}\otimes\mathcal{H}_{L}\rightarrow \mathcal{H}_{DA}\otimes\mathcal{H}_{DL}\\
\\
 |j(n)\rangle \mapsto |j\rangle\otimes|n\rangle \, .
    \end{array}
    \label{eq:phida}
\end{equation}
In this new basis, the Hamiltonian \eqref{eq:hx-diag} becomes
\begin{align}
&\hat{H}_X=\hat{H}_{DA}\otimes\mathds{I}_{DL}+\mathds{I}_{DA}\otimes\hat{H}_{DL} \, ,
 \label{eq:hx-split-da}\\
& \hat{H}_{DA}=\frac{\Omega}{2}\left(|2\rangle\langle 2|-|1\rangle\langle 1|\right) \, ,\nonumber \\
& \hat{H}_{DL}=\sum\limits_{n\geq 0} \omega_L(n+1)|n\rangle\langle n| \, .\nonumber 
\end{align}
We use different notations $N, n$ in order to distinguish between the Fock basis $\{|N\rangle\}$ of $\hat{H}_L$ and the Fock basis $\{|n\rangle\}$ of $\hat{H}_{DL}$. 
The subscript $DA$ stands for ``dressed qubit'' \cite{grynberg1998}, denoting the qubit ``dressed'' with the photons from the driving field. 
By symmetry we introduce $DL$ for ``dressed laser'' -- the state of the laser slightly modified by the interaction with the qubit. 
The identity \eqref{eq:hx-split-da} describes the physical phenomenon at play: the laser and the qubit form a new quantum state, consisting of the dressed qubit and another monochromatic macroscopic field. Moreover, the change of basis \eqref{eq:eig-bas} and the mapping \eqref{eq:phida} allow to re-write the initial condition \eqref{eq:rho-ini-alb}, up to corrections of the order $1/|\alpha|$, as 
\begin{equation}
 \hat{\rho}(0) = \hat{\rho}_{DA}(0)\otimes\hat{\rho}^{\text{coh}}_{DL}(0)\otimes\hat{\rho}_B \, ,   
 \label{eq:rho-ini-dadlb}
\end{equation}
with $\hat{\rho}^{\text{coh}}_{DL}(0)$ in a coherent state in the basis $\{|n\rangle\}$.

Notice that, by construction, the Hamiltonian $\hat{H}_{DL}$ is equal to
\begin{equation}
    \hat{H}_{DL}=\mathds{I}_A\otimes\hat{H}_L +\frac{\omega_L}{2}\hat{\sigma}_z\otimes\mathds{I}_L \, .
    \label{eq:hdl}
\end{equation}
%

\subsection{Non-autonomous approach}

An alternative approach to study the problem of \ref{subsec:model} is to trace out the degrees of freedom of the laser and account for its effect using an external, time-periodic field. 

\medskip

Consider the density matrix $\hat{\tilde{\rho}}(t)$ obtained by performing the following unitary transformation on $\hat{\rho}(t)$, called the Mollow transformation \cite{mollow1975},
\begin{align}
    \hat{\tilde{\rho}}(t)&\equiv \hat{D}^\dagger[\alpha(t)]\hat{\rho}(t)\hat{D}[\alpha(t)] \nonumber \\
    &=\hat{\tilde{U}}(t,0)\hat{\tilde{\rho}}(0)\hat{\tilde{U}}^\dagger(t,0)\, ,
    \label{eq:mollow}
\end{align}
with $\alpha(t)\equiv \alpha e^{-i\omega_Lt}$, $\hat{\tilde{U}}(t,0)\equiv \hat{D}^\dagger[\alpha(t)]\hat{U}(t,0)\hat{D}[\alpha(0)]$, and where we introduced the displacement operator \cite{grynberg1998} $\hat{D}[\alpha(t)] \equiv e^{\alpha(t) \hat{a}^\dagger - \alpha(t)^*  \hat{a}}$, a unitary operator acting on the creation and annihilation operators as $\hat{D}[\alpha(t)]^\dagger\hat{a}\hat{D}[\alpha(t)] = \hat{a}+\alpha(t)$, $\hat{D}[\alpha(t)]\hat{a} \hat{D}^\dagger[\alpha(t)] = \hat{a}-\alpha(t)$, 
and creating a coherent state from the vacuum,  $\hat{D}[\alpha(t)]|0\rangle= |\alpha(t)\rangle$. The density matrix 
$\hat{\tilde{\rho}}(t)$ is the solution of \cite{grynberg1998}
\begin{align}
 \frac{d\tilde{\hat{\rho}}(t)}{dt}&=-i[\tilde{\hat{H}}(t),\tilde{\hat{\rho}}(t)] \, , \label{eq:liouville-time}\\
 \hat{\tilde{\rho}}(0) &= \hat{\rho}_A(0)\otimes\hat{\rho}_B' \, , \label{eq:ini-rho-a}
\end{align}
where $\hat{\rho}_B'\equiv|0\rangle\langle 0|\otimes\hat{\rho}_B$ and 
\begin{equation}
    \hat{\tilde{H}}(t)= \hat{H}_A(t)+\hat{V}_{AB}'+\hat{H}_B' \, ,
    \label{eq:htilde}
\end{equation}
with
\begin{equation}
    \hat{H}_A(t) = \hat{H}_A+\hat{V}(t)
\end{equation}
and
\begin{equation}
    \hat{V}(t)=\frac{1}{2}(g\hat{\sigma}_+ e^{-i\omega_L t}+g^*\hat{\sigma}_- e^{i\omega_L t}) \, ,
    \label{eq:hat}
\end{equation}
where $g= g_0\alpha$ is the same as defined in \eqref{eq:g}, and where we regrouped $\hat{V}_{AB}'=\hat{V}_{AL}+\hat{V}_{AB}$ and $\hat{H}_{B}'=\hat{H}_{L}+\hat{H}_{B}$ \footnote{To obtain the expression for $\hat{\tilde{H}}(t)$, we used 
$\hat{D}^\dagger(\alpha(t))\hat{H}_L\hat{D}(\alpha(t)) =\hat{H}_L+\omega_L(\alpha(t)^*\hat{a}+\alpha(t)\hat{a}^\dagger +|\alpha|^2) $ and $
-i\partial_t (\hat{D}^\dagger(\alpha(t)))\hat{D}(\alpha(t))= \omega_L(\alpha(t)^*\hat{a}+\alpha(t)\hat{a}^\dagger +|\alpha|^2) $.
}. 

\subsubsection{Floquet basis}

Since the Hamiltonian \eqref{eq:htilde} is $2\pi/\omega_L$ periodic, it is convenient to describe the evolution of $\hat{\tilde{\rho}}(t)$ using Floquet states. The Floquet states, $\{|u_n(t)\rangle\}$, are by definition $2\pi/\omega_L$-periodic and solutions of the eigenvalue problem
\begin{equation}
    (\hat{H}_A+\hat{V}(t)-i\partial_t)|u_n(t)\rangle=\epsilon_n |u_n(t)\rangle \, .
    \label{eq:eig-val-prob}
\end{equation}
The $\{|u_n(t)\rangle\}$ form an orthonormal basis of $\mathcal{H}_A$. Quite conveniently, in the present case, the $\{|u_n(t)\rangle\}$ are simply related to the states $\{|j\rangle\}$ by (see details in appendix~\ref{app:floquet-basis})
\begin{equation}
    e^{i\omega_L\hat{\sigma}_z t/2}|u_j(t)\rangle=|j\rangle
    \label{eq:floquet-states}
\end{equation}
for $j=1,2$. 

\subsubsection{Equivalence between dressed qubit and rotating frame}\label{subsubsec:floquet-rot}

The operation \eqref{eq:floquet-states}, which defines a change of basis from the Floquet basis to the dressed qubit basis, is equivalent to going to the rotating frame, where the qubit-bath system is described by the density matrix
\begin{equation}
    \hat{\tilde{\rho}}^{\text{rot}}(t)\equiv \left[e^{i\omega_L\hat{\sigma}_z t/2}\otimes\mathds{I}_{B}\right]\hat{\tilde{\rho}}(t)\left[e^{-i\omega_L\hat{\sigma}_z t/2} \otimes\mathds{I}_{B}\right]\, ,
    \label{eq:rho-tilde-rot}
\end{equation}
and follows the dynamics
\begin{align}
    \frac{d\hat{\tilde{\rho}}^{\text{rot}}(t)}{dt}&=-i[\hat{H}_{A}^{\text{rot}}+\hat{V}_{AB}'(t)+\hat{H}_B',\hat{\tilde{\rho}}^{\text{rot}}(t)] \, ,\label{eq:rot-frame}
\end{align}
where 
\begin{align}
 \hat{V}_{AB}'(t)&\equiv e^{i\omega_L\hat{\sigma}_z t/2}\hat{V}_{AB}'e^{-i\omega_L\hat{\sigma}_z t/2} \, 
 \label{eq:vabprimet}\\
 &= e^{i\omega_L t }\hat{\sigma}_+ \hat{B} + e^{-i\omega_L t }\hat{\sigma}_- \hat{B}^\dagger \, .\nonumber
\end{align}
As expected from the relation \eqref{eq:floquet-states}, we check that 
\begin{align}
\hat{H}_{A}^{\text{rot}}&\equiv \left[e^{i\omega_L\hat{\sigma}_z t/2}(\hat{H}_A+\hat{V}(t))e^{-i\omega_L\hat{\sigma}_z t/2}-\frac{\omega_L}{2}\hat{\sigma}_z\right]\, \nonumber
\\
&=\hat{H}_{DA} \, .
\end{align}
Moreover, we can show (see proof in appendix~\ref{app:floquet-basis}) that the evolution of the dressed qubit and bath in the autonomous description, obtained by tracing out the dressed laser's degrees of freedom in \eqref{eq:liouville}, is equivalent to the evolution of the qubit and bath in the rotating frame in the non-autonomous description, in other words
\begin{equation}
    \hat{\tilde{\rho}}^{\text{rot}}=\text{Tr}_{DL}[\hat{\rho}] \, .
    \label{eq:equiv-rot-da}
\end{equation}
We will use this equivalence extensively in the rest of this work.


\subsection{Thermodynamics at the average level}\label{subsec:thermo-uni-av}

We now derive and compare the first and second laws of thermodynamics in the autonomous and non-autonomous descriptions. 
Starting with the autonomous description, we show that the laser acts as a work source for the qubit, while the proper work source in the dressed qubit approach is the dressed laser. We then discuss the non-autonomous description, showing the equivalence with the autonomous case: the laws of thermodynamics for the qubit are equivalent in the autonomous and non-autonomous descriptions, and the laws of thermodynamics in the dressed qubit picture (in the autonomous description) are equivalent to the laws of thermodynamics in the rotating frame (in the non-autonomous picture). The results are summarized in the Fig.~\ref{fig:summary}.

\subsubsection{Autonomous description}\label{subsubsec:auto-uni}

\begin{figure}
    \includegraphics[scale=0.4]{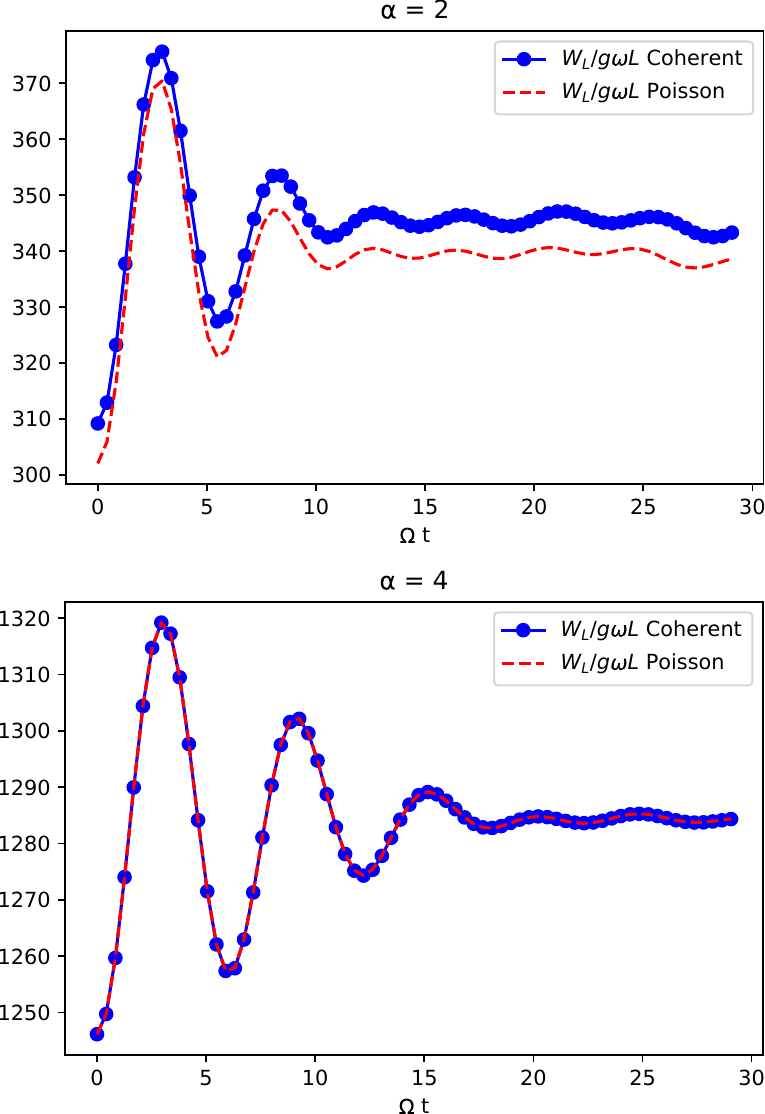}
    \caption{Work $W_L$ transferred from the laser to the qubit, with the time in units of the Rabi frequency $\Omega^{-1}$. The joint qubit-laser system is coupled to a heat bath. The Hamiltonian is given in \eqref{eq:ham}. The parameters are $\beta=5/D$ where $D=20$ is the spectral width of the bath, $g_0=g_k=0.1$ for all $k$, and the bath is modelized with $N_B=50$ modes. The blue dotted line is the work obtained when the laser is in a coherent state, while the red dashed line corresponds to a Poisson state. Top: $\alpha =2$. Bottom: $\alpha=4$. The number of photons is chosen as $2(|\alpha|^2+|\alpha|)$.    }
    \label{fig:wl-poi-coh}
\end{figure}

\begin{figure}
    \includegraphics[scale=0.4]{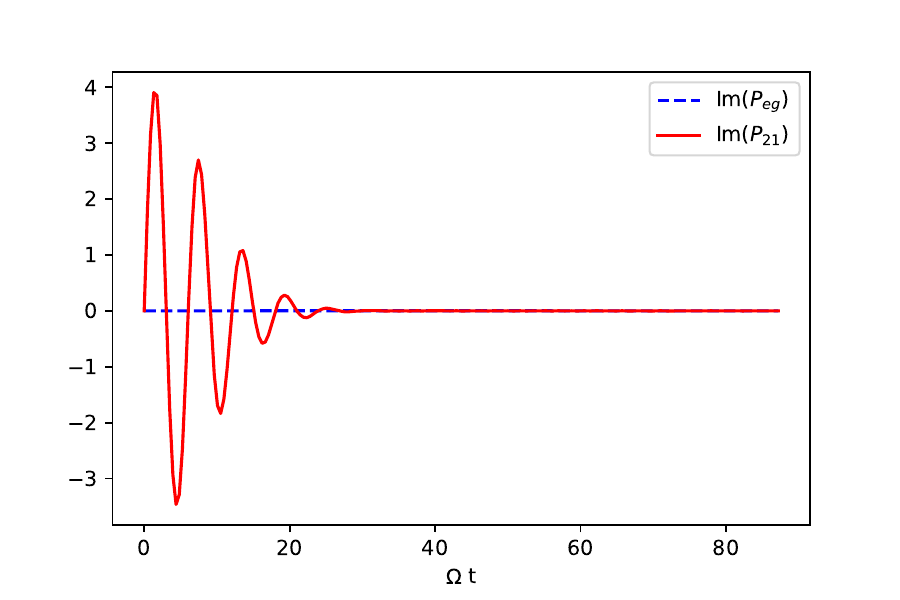}
    \caption{ Dashed line: coherences in the dressed qubit basis; full line: coherences in the qubit basis. Parameters: $N_B=50$ modes in the bath, $2|\alpha|^2+2|\alpha|$ photons in the laser, $\beta=5/D$ where $D=20$ is the spectral width of the bath, $g_0=g_k=0.1$ for all $k$, and where the time is in units of the Rabi frequency.}
    \label{fig:im-rho}
\end{figure}

In the autonomous description, the total qubit-laser-bath system is closed, hence its total energy is conserved, $
\text{Tr}[(\hat{\rho}(t)-\hat{\rho}(0))\hat{H}] = 0$ for all $t\geq 0$.
Since the bath is assumed by be initially at thermal equilibrium, 
the variation of energy in the bath is identified as minus the heat \cite{esposito2010entropy}, 
\begin{equation}
    Q\equiv -\text{Tr}[(\hat{\rho}(t)-\hat{\rho}(0))\hat{H}_B] \, .
    \label{eq:q}
\end{equation}
On the other hand, as discussed in section~\ref{sec:laser}, we identify minus the variation of energy in the laser as work,
\begin{equation}
    W_L\equiv -\text{Tr}[(\hat{\rho}(t)-\hat{\rho}(0))\hat{H}_L] \, .
\end{equation}
%
The conservation of energy leads to define the energy of the qubit as 
\begin{align}
     E_A(t)&\equiv \text{Tr}[\hat{\rho}(t)(\hat{H}_A+\hat{V}_{AL}+\hat{V}_{AB})] \, ,
     \label{eq:ea}
\end{align}
which leads to the first law for the qubit,
\begin{equation}
    \Delta E_A = Q+W_L \, .
    \label{eq:1st-law-a}
\end{equation}
We point out that the rate of the work is proportional to the coherences in the dressed qubit basis: from \eqref{eq:liouville}, we obtain the rate $\dot W_L=-\text{Tr}[d_t\hat{\rho}(t)\hat{H}_L]$,
\begin{align}
&    \dot W_L = i \omega_L g_0 \text{Tr}[(\hat{a}^\dagger\hat{\sigma}_--\hat{a}\hat{\sigma}_+)\hat{\rho}(t)] \label{eq:wl-rate} \\
&=i \omega_L g_0\sum_{n\geq 0}(\langle a,n+1|\hat{\rho}(t)|b,n\rangle - \langle b,n|\hat{\rho}(t)|a,n+1\rangle ) \nonumber \\
    &\approx i\omega_L g\sum_{n\geq 0}(\langle 2,n|\hat{\rho}(t)|1,n\rangle - \langle 1,n|\hat{\rho}(t)|2,n\rangle )\nonumber \\
    &= -\omega_L g \, \text{Im}(\langle 2|\hat{\rho}_{DA}(t)|1\rangle) \, ,
    \label{eq:wl-a}
\end{align}
where we used the assumption 2 for the third line. We highlight that the assumption 2 (macroscopic limit) ensures that these coherences survive the phase averaging. To see this, consider the case where the interaction with the bath is neglected, hence where the evolution of $\hat{\rho}_X(t)$ depends only on $\hat{H}_X$. The generalization to the case where the bath is taken into account is straightforward by repeating the reasoning at the level of quantum maps, which will be introduced in section~\ref{sec:cons-qme} [specifically in \eqref{eq:map-x}]. When the bath is neglected, we have $\hat{\rho}_X(t) = e^{-it\hat{H}_X}\hat{\rho}_X(0)e^{it\hat{H}_X}$. Under the assumption 2, $\hat{H}_X$ can be written as a sum of terms \eqref{eq:hx-diag}, which, using \eqref{eq:rho-ini-alb}, yields
\begin{equation}
    \langle 2,n|\hat{\rho}(t)|1,n'\rangle \propto e^{-|\alpha|^2}\frac{|\alpha|^{n+n'}e^{i\phi(n-n')}}{\sqrt{n!n'!}}e^{i(n'-n+\Omega)t} \, ,
    \label{eq:im-p21}
\end{equation}
hence when setting $N=N'$ the dependence in the phase $\phi$ disappears. Numerically, we find that this feature happens as early as $\alpha=4$, see Fig.~\ref{fig:wl-poi-coh}. This illustrates the general statement of section~\ref{subsec:irrelevence}, that the phase of the laser is irrelevant for the work. However, we point out that the coherent term $\langle b|\hat{\rho}_A|a\rangle\equiv\sum_N \langle b,N|\hat{\rho}|a,N\rangle$ does not survive phase averaging, since 
\begin{equation}
    \langle b,N|\hat{\rho}|a,N\rangle \propto e^{i\phi} \,  ,
    \label{eq:im-peg}
\end{equation}
as illustrated in Fig.\ref{fig:im-rho}.

\medskip

Let us now turn to the dressed qubit picture. We define instead 
\begin{equation}
    E_{DA}(t)\equiv \text{Tr}[\hat{\rho}(t)(\hat{H}_{DA}+\hat{V}_{AB})] \, ,
    \label{eq:ea-def}
\end{equation}
and the conservation of energy leads to the first law
\begin{equation}
    \Delta E_{DA} = Q+W_{DL} \, ,
    \label{eq:1st-law-a-da}
\end{equation}
where, 
\begin{equation}
    W_{DL}\equiv -\text{Tr}[(\hat{\rho}(t)-\hat{\rho}(0))\hat{H}_{DL}] \, ,
    \label{eq:wdl}
\end{equation}
is identified as work.

Using \eqref{eq:hdl}, we can split
\begin{equation}
    W_{DL} = W_L-Tr\left[\frac{\omega_L}{2}\hat{\sigma}_z(\hat{\rho}(t)-\hat{\rho}(0))\right] \, ,
    \label{eq:wdl-split}
\end{equation}
which leads to the following identity connecting the internal energies of the qubit and dressed qubit,
\begin{equation}
    \Delta E_{A} = \Delta E_{DA} + Tr\left[\frac{\omega_L}{2}\hat{\sigma}_z(\hat{\rho}(t)-\hat{\rho}(0))\right] \, .
\end{equation}

We now examine the second law of thermodynamics. Since the density matrix is initially factorized \eqref{eq:rho-ini-alb}, the von Neumann entropy $S_X\equiv -\text{Tr}[\hat{\rho}_X\log\hat{\rho}_X]$ can be split into two terms \cite{esposito2010entropy}, $\Delta S_X =\beta_B Q + D(\hat{\rho}||\hat{\rho}_X(t)\otimes\hat{\rho}_B)$, 
%
%
where $D(\hat{\rho}_1||\hat{\rho}_2)\equiv \text{Tr}[\hat{\rho}_1\log\hat{\rho}_1]-\text{Tr}[\hat{\rho}_1\log\hat{\rho}_2]$ denotes the relative entropy between two density matrices. Since a relative entropy is always positive, the identity leads to a second law of thermodynamics for $X$,
\begin{equation}
  \Sigma_X\equiv  \Delta S_X - \beta_B Q\geq 0\, ,
  \label{eq:sigma-x}
\end{equation}
where $\Sigma_X$ is the entropy production for $X$.
Using again \eqref{eq:rho-ini-alb}, the von Neumann entropy variation $\Delta S_X$ can then be written as
\begin{equation}
    \Delta S_X =\Delta S_{A}+\Delta S_{L}-D(\hat{\rho}_X(t)||\hat{\rho}_{A}(t)\otimes\hat{\rho}_{L}(t)) \, ,
    \label{eq:split-sx}
\end{equation}
where $S_{A}, S_{L}$ are respectively the von Neumann entropies associated to the subsystems $A$ and $L$.
As long as the laser stays close to a coherent state or a Poisson state, $\Delta S_L$ is negligible (see appendix~\ref{app:work-source} and Fig.\ref{fig:plots}), and we obtain the second law for the qubit,
\begin{equation}
  \Sigma_A\equiv  \Delta S_{A}  - \beta_B Q \geq D(\hat{\rho}_X(t)||\hat{\rho}_{A}(t)\otimes\hat{\rho}_{L}(t))\geq 0 \, ,
\end{equation}
where $\Sigma_A$ is the entropy production of the qubit. 
Likewise, for the dressed qubit, using \eqref{eq:rho-ini-dadlb}, we find
\begin{align}
  \Sigma_{DA}\equiv   \Delta S_{DA}-\beta_B Q & \geq D(\hat{\rho}(t)||\hat{\rho}_{DA}(t)\otimes\hat{\rho}_{DL}(t)) \geq 0 \, .
    \label{eq:sigma-da}
\end{align}
%

\subsubsection{Non-autonomous description}\label{subsubsec:na-dyn}

In the non-autonomous description, the qubit-bath system is isolated, with energy changes due solely to the time-dependence of the Hamiltonian, identified as work,
\begin{equation}
\begin{array}{ll}
\dot W&\equiv d_t \text{Tr}[\hat{\tilde{\rho}}(t)\hat{\tilde{H}}(t)] = \text{Tr}[\hat{\tilde{\rho}}(t)d_t \hat{V}(t)] \, . 
\label{eq:wl}
\end{array}
\end{equation}

Interestingly, this definition becomes equivalent to the definition of work in the autonomous description \eqref{eq:def-wl} in the macroscopic limit $|\alpha|\gg 1$. Indeed, applying the transformation \eqref{eq:mollow} in \eqref{eq:wl-rate}, we find
\begin{align}
    \dot W_L=\dot W + i\omega_L g_0 \text{Tr}[(\hat{\sigma}_-\hat{a}^\dagger-\hat{\sigma}_+\hat{a})\hat{\tilde{\rho}}]\label{eq:wl-vt} \, .
\end{align}
Since the coupling amplitude of $\hat{V}(t)$ is $g=g_0|\alpha|$, the first term on the r.h.s. is dominant compared to the second one when $|\alpha|\gg 1$.

Defining now
\begin{equation}
    \tilde{E}_A(t)\equiv \text{Tr}[(\hat{H}_A+\hat{V}(t)+\hat{V}_{AB'})\hat{\tilde{\rho}}(t)] \, ,
    \label{eq:ea-tilde}
\end{equation}
the conservation of energy leads to
\begin{equation}
    d_t \tilde{E}_A(t)=\dot W-\text{Tr}[\hat{H}_{B'}d_t\hat{\tilde{\rho}}(t)] \, .
\end{equation}
Note that, using \eqref{eq:mollow} which connects $\hat{\tilde{\rho}}(t)$ to $\hat{\rho}(t)$, we obtain the identity $
 \tilde{E}_A(t)=E_A(t) $.
A standard calculation also shows that 
\begin{equation}
    -\text{Tr}[\hat{H}_{B'}d_t\hat{\tilde{\rho}}(t)]=\dot Q +\dot W_L-\dot W \, ,
\end{equation}
which simplifies to 
\begin{equation}
    -\text{Tr}[\hat{H}_{B'}d_t\hat{\tilde{\rho}}(t)]=\dot Q 
    \label{eq:def-q-hb'}
\end{equation}
in the macroscopic limit, 
hence the first law for the qubit in the non-autonomous description is consistent with that of the qubit in the autonomous description, 
\begin{equation}
    d_t\tilde{E}_A(t)= \dot Q + \dot W_L \, .
    \label{eq:1st-law-na}
\end{equation}

Similarly, we find that the first law in the rotating frame coincides with that of the dressed qubit, namely, 
\begin{equation}
    d_t \tilde{E}^{\text{rot}}_A(t) = \dot Q+\dot W_{DL}\, ,
    \label{eq:1st-law-rot}
\end{equation}
where (see appendix~\ref{app:proof60} for details)
\begin{equation}
\tilde{E}^{\text{rot}}_A(t)\equiv \text{Tr}[(\hat{H}_{A}^{\text{rot}}+\hat{V}_{AB}'(t))\hat{\tilde{\rho}}^{\text{rot}}(t)] \, .
\end{equation}
This shows that the equivalence between the dressed qubit, in the autonomous picture, and the qubit in the rotating frame, in the non-autonomous picture, also holds for the thermodynamics at the average level.

%
%
As a final remark, we point out that, on one hand
\begin{equation}
\text{Tr}[\hat{\tilde{\rho}}(t)d_t \hat{V}(t)] = -\omega_L g \text{Im}(\langle b|\hat{\tilde{\rho}}^{\text{rot}}|a\rangle) \, ,
\end{equation}
while on the other hand, a standard calculation shows that 
\begin{equation}
    \text{Im}(\langle b|\hat{\tilde{\rho}}^{\text{rot}}|a\rangle) = \text{Im}(\langle 2|\hat{\rho}|1\rangle) \, .
    \label{eq:equiv-im}
\end{equation}
This is consistent with the fact that the work in the autonomous picture \eqref{eq:wl-a} is equal to the one in the non-autonomous picture \eqref{eq:wl}.

We now turn to the second law. Using the initial condition in \eqref{eq:liouville-time}, and following the same reasoning as in the autonomous description, we obtain
\begin{equation}
    \Delta \tilde{S}_A-\beta_B Q\geq D(\hat{\tilde{\rho}}_X||\hat{\tilde{\rho}}_A(t)\otimes\hat{\tilde{\rho}}_L(t))\geq 0 \, 
    \label{eq:2nd-law-exact-na}
\end{equation}
and
\begin{align}
    \Delta \tilde{S}_A^{\text{rot}}-\beta_B Q&\geq D(\hat{\tilde{\rho}}_X||\hat{\tilde{\rho}}^{\text{rot}}_A(t)\otimes\hat{\tilde{\rho}}_L(t)) \geq 0 \,  .
     \label{eq:2nd-law-exact-na-rot}
\end{align}

\subsection{Thermodynamics at the fluctuating level}\label{subsec:thermo-uni-fluct}

Here, we focus on the thermodynamics at the fluctuating level. We will resort to the two-point measurement technique with counting fields, introduced in section~\ref{subsubsec:work-fluct}. This technique is efficient to relate the fluctuations of observables corresponding to operators which commute (e.g.  $\hat{H}_A$ and $\hat{H}_L$) when the interactions between these operators are weak, but fails in the presence of strong interactions, which is the case in the macroscopic limit $|\alpha|\gg 1$ for example. In this section, we show that the dressed qubit picture, on the one hand, and the Mollow transformation, on the other hand, allow to overcome the difficulty raised by the  coupling term $\hat{V}_{AL}$, yielding two different work fluctuation theorems, respectively for $W_{DL}$ and $W_L$.


\subsubsection{Joint generating functions}

One can measure several observables simultaneously provided that they commute. 
In the case of simultaneous measurements, we resort to a joint moment generating function
\begin{equation}
\begin{array}{ll}
 \mathcal{G}(t,\boldsymbol{\lambda})&\equiv \text{Tr}[\hat{\rho}_{\boldsymbol{\lambda}}(t)] \\
\\
\hat{\rho}_{\boldsymbol{\lambda}}(t)
&\equiv\hat{U}_{\boldsymbol{\lambda}}(t,0)\bar{\hat{\rho}}(0) \hat{U}^\dagger_{-\boldsymbol{\lambda}}(t,0)\\
\\
\hat{U}_{\boldsymbol{\lambda}}(t,0)&\equiv
e^{i\boldsymbol{\lambda}\cdot\boldsymbol{\hat{H}}/2} \hat{U}(t,0) e^{-i\boldsymbol{\lambda}\cdot\boldsymbol{\hat{H}}/2} \, ,    
\end{array}\label{eq:mgf}
\end{equation}
where $\boldsymbol{\hat{H}},\boldsymbol{\lambda}$ respectively denote a vector of Hamiltonians and a vector of counting fields. By convention, we use the same subscripts for the counting fields and the corresponding Hamiltonians; for instance, when measuring 
$\boldsymbol{\hat{H}}=(\hat{H}_{A}(t),\hat{H}_{B'})$ we use $\boldsymbol{\lambda}=(\lambda_{A},\lambda_{B'})$, while the counting fields for $\boldsymbol{\hat{H}}=(\hat{H}_{DA},\hat{H}_{DL},\hat{H}_B)$ are denoted $\boldsymbol{\lambda}=(\lambda_{DA},\lambda_{DL},\lambda_B)$.

Let us now connect the outcomes of the measurements described by \eqref{eq:mgf} with the thermodynamic quantities introduced in section~\ref{subsec:thermo-uni-av}. From \eqref{eq:q} and \eqref{eq:def-q-hb'}, it is straightforward to see that the heat $Q$ leaked from the bath is equivalently obtained by measuring $-\hat{H}_B$ or $-\hat{H}_{B'}$, while from \eqref{eq:def-wl} and
\eqref{eq:wdl} we see that the work terms $W_{L}$ and $W_{DL}$ are obtained by measuring respectively $-\hat{H}_{L}$ and $-\hat{H}_{DL}$. Finally, we may measure the energy of the qubit (resp. dressed qubit), defined in \eqref{eq:ea-tilde} (resp. \eqref{eq:ea-def}), using a two-point measurement of $\hat{H}_A(t)$ (resp. $\hat{H}_{DA}$) if we require that the coupling $\hat{V}_{AB'}$ (resp. $\hat{V}_{AB}$) is switched on only after the initial measurement and switched off before the final one.

\begin{figure*}
    \centering
    \includegraphics[width=\textwidth]{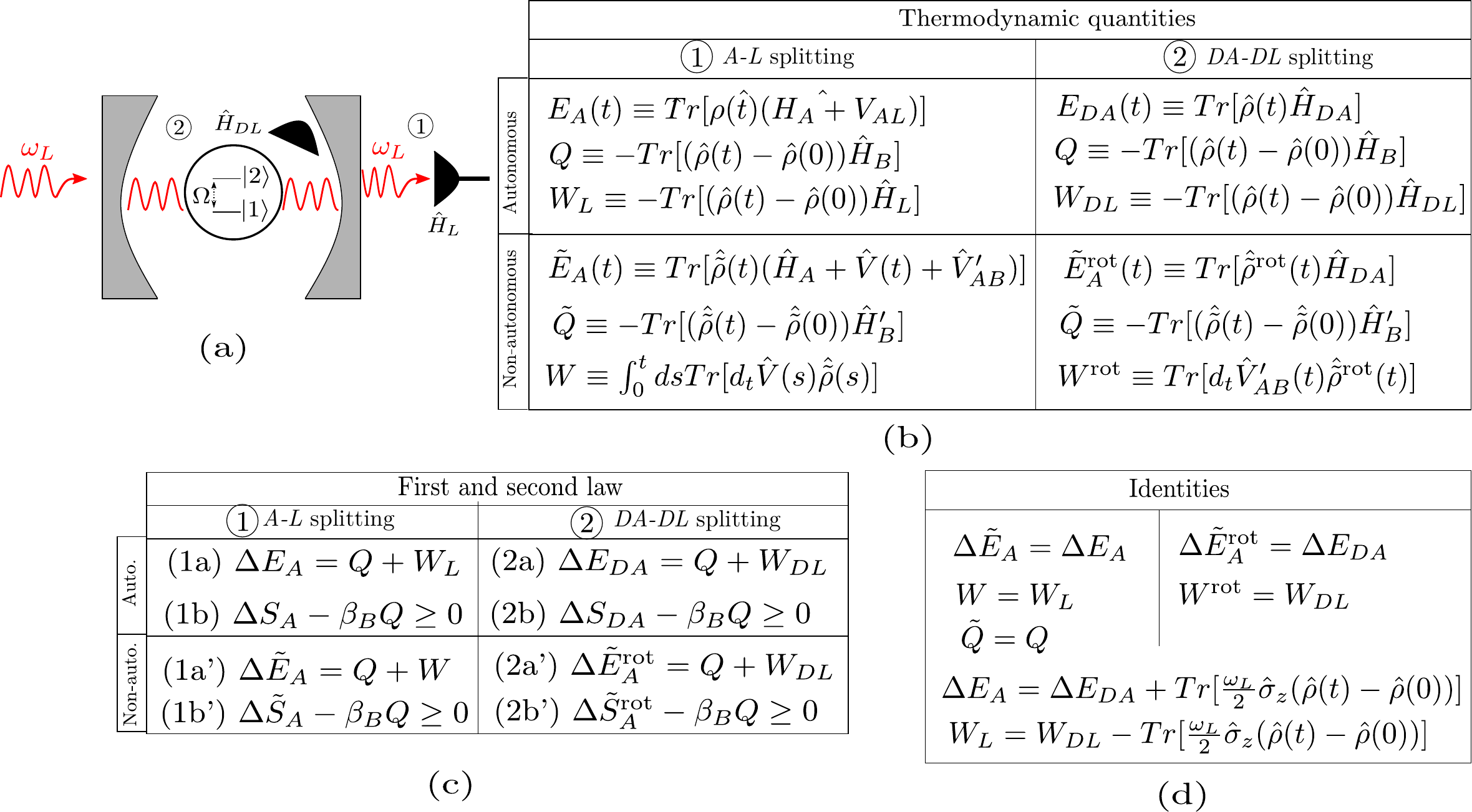}
    \caption{Summary of the laws of thermodynamics and identities at the unitary level. (a) In the situation 1, $\hat{H}_L$ is first measured before the interaction in turned on and the last measurement is performed after the interaction is switched off. In this case, it is more convenient to use the non-dressed approach. In the situation 2, the interaction is always on, but we measure instead $\hat{H}_{DL}$, which implies that the dressed qubit approach is more convenient. (b) Thermodynamic observables in both the dressed and non-dressed approaches. (c) Laws of thermodynamics. (d) Identities between the thermodynamic observables of both approaches.}
    \label{fig:summary}
\end{figure*}

\subsubsection{Work and entropy fluctuation theorems -- autonomous picture}\label{subsubsec:ft}

Fluctuation theorems are symmetries relating the energy or entropy fluctuations generated during a given forward process and its time reversed counterpart. In the context of two-point measurement schemes with counting fields, such theorems can be expressed as symmetries between the moment generating functions of the forward and backward dynamics.

In this section, we derive a work fluctuation theorem and an entropy fluctuation theorem in the autonomous
description.

We first need to introduce the moment generating function of the reversed dynamics, which is defined as \cite{esposito2009nonequilibrium,soret2023-3}
\begin{align}
\begin{aligned}
\mathcal{G}^{R}(\boldsymbol{\lambda},t)&\equiv \text{Tr}[\hat{\rho}^R_{\boldsymbol{\lambda}}(t)] \\
&=Tr\left[\hat{U}_{\boldsymbol{\lambda}}^{\dagger}(t,0)\bar{\hat{\rho}}^R(0) \hat{U}_{\boldsymbol{-\lambda}}(t,0)\right], 
\end{aligned}\label{eq:mgfrev}
\end{align}
where $\bar{\hat{\rho}}^R(0)$ is the diagonal part of the initial density matrix of the reverse dynamics $\hat{\rho}^R(0)$ in the common eigenbasis of $\hat{H}_{DA},\hat{H}_{DL},\hat{H}_B$ chosen for the measurement.   
Now, we assume that the initial density matrix of the reverse dynamics is factorized, as for the forward dynamics, 
\begin{align}
\begin{aligned}
   & \hat{\rho}^R(0)=\hat{\rho}^R_{DA}(0)\otimes\hat{\rho}^R_{DL}(0)\otimes\hat{\rho}_B  \, .
\end{aligned}\label{eq:fact-ini-mat}
\end{align}
Notice that, since $\hat{H}$ is time-independent, \eqref{eq:fact-ini-mat} implies that the time-reversed density matrix is simply given by $\hat{\rho}^R(t)=\hat{\rho}(-t)$. 
Let's furthermore assume that
\begin{equation}
    \begin{array}{ll}
&\hat{\rho}_{DA}(0)=\hat{\rho}^R_{DA}(0)=e^{-\beta_{DA}\hat{H}_{DA}}/Z_{DA} \, .
    \end{array}
    \label{eq:init-cond}
\end{equation}
As explained in section~\ref{subsubsec:work-fluct}, the two-point measurement technique with counting fields allows to compute rigorously the fluctuations of the work performed by a Poisson state, which correspond to the fluctuations of the work performed by a coherent state during a series of experiments where the initial phase of the coherent state is randomly chosen. We therefore assume that $\hat{\rho}_{DL}(0)$ and $\hat{\rho}^R_{DL}(0) $ are Poisson states \eqref{eq:poisson}. As explained in section~\ref{subsubsec:work-fluct}, Poisson states can be written as Gibbs states in the macroscopic limit $|\alpha|\gg 1$, specifically (see appendix~\ref{app:ft})
\begin{equation}
    \hat{\rho}_{DL}(0) =  \hat{\rho}_{DL}^R(0) = e^{-\beta_{DL}\hat{H}'_{DL}}/Z_{DL} \, ,
    \label{eq:cond-rhod}
\end{equation}
with $\beta_{DL}\equiv 1/|\alpha|^2$ and where $\hat{H}_{DL}'=(\hat{a}^\dagger\hat{a}-\langle N\rangle)^2/2$. Under these assumptions, the moment generating function \eqref{eq:mgf} satisfies the following symmetry (see appendix~\ref{app:ft}), 
\begin{equation}
     \mathcal{G}(\boldsymbol{\lambda},t)=\mathcal{G}^R(-\boldsymbol{\lambda}+i\boldsymbol{\nu},t) \, ,
     \label{eq:ft-2}
\end{equation}
with $\boldsymbol{\nu}=(\beta_{DA},\beta_{DL},\beta_B)$. Given that, under the assumption 2, fluctuations of the order $1/|\alpha|$ can be neglected, we further on replace $\beta_{DL} = 0$. 

On its own, the symmetry \eqref{eq:ft-2} is formal, but combined with the notion of energy conservation, it yields a work fluctuation theorem. Energy conservation is conveniently expressed using the generating function \eqref{eq:mgf}: in the absence of external driving, the total energy of the system should be conserved. Setting $\lambda_{DA}=\lambda_{DL}=\lambda_B\equiv\lambda$, this condition is satisfied at the average level iff $\partial_\lambda \mathcal{G}(\boldsymbol{\lambda},t)|_{\lambda=0}=0$, which yields the first law \eqref{eq:1st-law-a-da}. Imposing instead energy conservation at the fluctuating level takes the form of a strict energy conservation condition \cite{soret2022}, 
\begin{equation}
 \hat{\rho}_{\boldsymbol{\lambda}}(t)=\hat{\rho}_{\boldsymbol{\lambda}+\chi\boldsymbol{1}}(t) \, ,
 \label{eq:strict-en-cons}
\end{equation}
where $\boldsymbol{1}=(1,1,1)$. Together with the symmetry \eqref{eq:ft-2}, the strict energy conservation condition \eqref{eq:strict-en-cons} implies a work fluctuation theorem: setting $\beta_{DA}=\beta_{B}\equiv\beta$, we find
\begin{equation}
\begin{array}{ll}
\mathcal{G}(0,\lambda_{DL},0,t)&=\mathcal{G}^R(i\beta,-\lambda_{DL},i\beta,0)\\
&=\mathcal{G}^R(0,-\lambda_{DL}-i\beta,0,t) \, .
\end{array}
\end{equation}
Applying a reverse Fourier transform allows to rephrase the above equality in terms of the probabilities $p(W_{DL})$ the [respectively $p^R(W_{DL})$] to observe a variation $W_{DL}$ in the forward (respectively time-reversed) dynamics [see \eqref{eq:def-mgf}], which yields 
\begin{equation}
    \frac{p(W_{DL})}{p^R(-W_{DL})} = e^{\beta W_{DL}} \, .
    \label{eq:crooks-a}
\end{equation}
The symmetry \eqref{eq:ft-2} and fluctuation theorem \eqref{eq:crooks-a} constitute a important results of this work, and will serve as a criteria of consistency for quantum optical quantum master equations. 

A similar relation as \eqref{eq:ft-2} can be obtained for the entropy production, $\Sigma$, obtained from measuring the variations of the operator 
\begin{equation}
    \hat{\Sigma}_{DA}(t)\equiv-\log\hat{\rho}_{DA}(t)-\log\hat{\rho}_{DL}(t)+\beta_B \hat{H}_B \, .
    \label{eq:op-entropy}
\end{equation}
Assuming the initial conditions \eqref{eq:rho-ini-dadlb} and \eqref{eq:fact-ini-mat}, we obtain an entropy fluctuation theorem
\begin{equation}
   \mathcal{G}_\Sigma(\lambda_\Sigma,t)=\mathcal{G}_\Sigma^R(-\lambda_{\Sigma}+i,t) \, ,
     \label{eq:ft-sigma}
\end{equation}
where $\lambda_\Sigma$ is the counting field associated to $\hat{\Sigma}$.
In particular, $\mathcal{G}_\Sigma(i,t)=1$, which leads to the integral fluctuation theorem $\langle e^{-\Sigma}\rangle = 1$ and by convexity to $\langle\Sigma\rangle \geq 0$, which is the second law \eqref{eq:sigma-da}.

\subsubsection{Work and entropy fluctuation theorems - non-autonomous picture}\label{subsubsec:ft-w-na}

We now turn to the non-autonomous description, starting with the work fluctuation theorem. 

Performing a two point-measurement of the Hamiltonian $\hat{H}_L$, then applying the Mollow transformation \eqref{eq:mollow}, yields the tilted density matrix
\begin{align}
\hat{\tilde{\rho}}_{\lambda_L} &\equiv D^\dagger[\alpha(t)]\hat{\rho}_{\lambda_L}D[\alpha(t)] \label{eq:mgf-mollow}\\
&= \mathcal{T}[e^{-i\int_0^t ds \hat{H}_{\lambda_L}(s)}]\hat{\rho}(0) \mathcal{T}[e^{i\int_0^t ds \hat{H}_{-\lambda_L}(s)}] \, , \nonumber
\end{align}
where $\hat{\rho}_{\lambda_L}(t)$ is obtained by choosing $\boldsymbol{\lambda}=(0,\lambda_L,0)$ with $\boldsymbol{H}=(0,\hat{H}_L,0)$ in \eqref{eq:mgf}, 
where $\mathcal{T}$ denotes time ordering and where $\hat{H}_{\lambda_L}(t) = \hat{H}_A+\hat{H}_{B'}+\hat{V}_{AB}+\hat{V}_{AL}^{\lambda_L}+\hat{V}_{\lambda_L}(t)$ with
\begin{align}
\hat{V}_{AL}^{\lambda_L} &= \frac{g_0}{2}(\hat{\sigma}_+\hat{a}e^{-i\omega_L\lambda_L}+\hat{\sigma}_-\hat{a}^\dagger e^{i\omega_L\lambda_L}) \label{eq:v-1}\, , \\
\hat{V}_{\lambda_L}(t) &= \frac{g_0}{2}(\hat{\sigma}_+\alpha(t)e^{-i\omega_L\lambda_L}+\hat{\sigma}_-\alpha^*(t) e^{i\omega_L\lambda_L}) \label{eq:v-2} \\
&= \hat{V}(t+\lambda_L) \nonumber \, .
\end{align}
The fluctuations of $\hat{H}_L$, measured by $\lambda_L$, are now carried both by $\hat{V}_{AL}^{\lambda_L}$ and the time dependent term $\hat{V}_{\lambda_L}(t)$. Performing now additional projective measurements with counting fields on $\hat{H}_A+\hat{V}(t)$ and $\hat{H}_B$ yields the tilted density matrix
\begin{equation}
    \hat{\tilde{\rho}}_{\boldsymbol{\lambda}}\equiv \hat{\tilde{U}}_{\boldsymbol{\lambda}} \hat{\tilde{\rho}}(0)\hat{\tilde{U}}_{\boldsymbol{-\lambda}}^\dagger \, ,
\end{equation}
with $\boldsymbol{\lambda}=(\lambda_A,\lambda_L,\lambda_B)$ and
\begin{align}
    \hat{\tilde{U}}_{\boldsymbol{\lambda}} = e^{i\frac{\lambda_B}{2}\hat{H}_B}e^{i\frac{\lambda_A}{2}(\hat{H}_A+\hat{V}(t))} & \mathcal{T}[e^{-i\int_0^t ds \hat{H}_{\lambda_L}(s)}]\\
    &\times e^{-i\frac{\lambda_B}{2}\hat{H}_B}e^{-i\frac{\lambda_A}{2}(\hat{H}_A+\hat{V}(0))} \, .\nonumber
\end{align}
The tilted density matrix for the reversed dynamics is in turn given by
\begin{equation}
\hat{\tilde{\rho}}_{\boldsymbol{\lambda}}^R(t)\equiv\hat{\tilde{U}}^\dagger_{\boldsymbol{\lambda}} \hat{\tilde{\rho}}(0)\hat{\tilde{U}}_{\boldsymbol{-\lambda}}  \, .
\label{eq:rev-auto}
\end{equation}
Noticing that the Mollow transformation does not change the trace, we may apply the same reasoning as in the autonomous description. Let us introduce the partition function $Z_A(t)\equiv \text{Tr}[e^{-\beta_A(\hat{H}_A+\hat{V}(t))}]$. Since $e^{i\hat{\omega_L\sigma}_zt/2}\hat{V}(t)e^{-i\omega_L\hat{\sigma}_zt/2}=\hat{V}(0)$, the partition function is in fact time-independent, $Z_A(t)=Z_A(0)\equiv Z_A$. We now assume that the initial density matrix of the reverse process is factorized as in \eqref{eq:ini-rho-a}, and further that
\begin{align}
    \hat{\tilde{\rho}}_A(0)&=\frac{e^{-\beta_A(\hat{H}_A+\hat{V}(0))}}{Z_A}\label{eq:ini-cond-na}\\
    \hat{\tilde{\rho}}_A^R(0) &= \frac{e^{-\beta_A(\hat{H}_A+\hat{V}(t))}}{Z_A} \nonumber\, .
\end{align}
We point out that, since, in the rotating frame, the evolution of the qubit is equivalent to that of the dressed qubit (as shown in section~\ref{subsubsec:floquet-rot}), which is governed by a time-independent Hamiltonian, we deduce, as in the autonomous case, the relation $\hat{\tilde{\rho}}^R(t) = \hat{\tilde{\rho}}(-t)$. For an alternative proof using the Floquet states, see appendix~\ref{app:ft-na}. 
We then obtain, in the macroscopic limit $|\alpha|\gg 1$, the following symmetry for the generating function in the non-autonomous description $\mathcal{\tilde{G}}(\boldsymbol{\lambda},t)\equiv \text{Tr}[\hat{\tilde{\rho}}_{\boldsymbol{\lambda}}]$,
\begin{equation}
    \mathcal{\tilde{G}}(\boldsymbol{\lambda},t)=\mathcal{\tilde{G}}^R(-\boldsymbol{\lambda}+i\boldsymbol{\nu},t)  \, ,
\label{eq:ft-3}
\end{equation}
with $\boldsymbol{\nu}=(\beta_{A},0,\beta_B)$. Similarly as in the autonomous case, the energy conservation condition is here satisfied if $\partial_\lambda\mathcal{\tilde{G}}(\boldsymbol{\lambda},t)|_{\lambda=0}=0$ when $\boldsymbol{\lambda}=(\lambda,\lambda,\lambda)$. When satisfied, this condition yields the first law \eqref{eq:1st-law-na}. In turn, the strict energy conservation reads 
\begin{equation}
    \hat{\tilde{\rho}}_{\boldsymbol{\lambda}+\chi\boldsymbol{1}}(t)= \hat{\tilde{\rho}}_{\boldsymbol{\lambda}}(t) \, .
    \label{eq:en-cons-na}
\end{equation}
%

The combination of \eqref{eq:en-cons-na} and \eqref{eq:ft-3} then yields the following work fluctuation theorem for the laser, similar to the Crooks relation \cite{crooks1999entropy,tasaki2000jarzynski}: setting $\beta_A=\beta_B=\beta$, we find
\begin{equation}
    \frac{p(W_L)}{p^{R}(-W_L)} = e^{\beta W_L} \, .
    \label{eq:crooks-na}
\end{equation}
Notice that \eqref{eq:crooks-na} implies the Jarzynski equality \cite{jarzynski1997nonequilibrium}.  
%


We now turn to the entropy fluctuation theorem. We use the same reasoning as in the previous section: since the entropy fluctuations of the laser can be neglected in the autonomous picture (see appendix~\ref{app:work-source}), measuring 
\begin{equation}
    \hat{\Sigma}_A(t)\equiv \log \hat{\tilde{\rho}}_A(t)+\log \hat{\tilde{\rho}}_L(t)+\beta_B\hat{H}_B
\end{equation}
amounts to measuring the entropy production $\Delta \tilde{S}_A-\beta_B Q$. Then, assuming that the density matrices of the forward and time-reversed dynamics are initially factorized with    $\hat{\tilde{\rho}}^R(0)=\hat{\tilde{\rho}}_A(t)\otimes\hat{\rho}_{B'}$, we obtain the entropy fluctuation theorem
\begin{equation}   \tilde{\mathcal{G}}_\Sigma(\lambda_{\Sigma_{A}},t)=\tilde{\mathcal{G}}_\Sigma^R(-\lambda_{\Sigma_{A}}+i,t) \, .
     \label{eq:ft-sigma-na}
\end{equation}

We mention that recent works \cite{engelhardt2023unified, platero2023photonresolved} have examined the full counting statistics of the work performed by a laser using the counting field method, but no symmetry such as \eqref{eq:ft-3} had so far been derived. We also highlight that, in those approaches, the statistics of the laser is described solely using the term \eqref{eq:v-2}. Our approach shows that the full statistics are in fact given by both the terms \eqref{eq:v-1} and \eqref{eq:v-2}. In the macroscopic limit $|\alpha|\gg 1$, the term \eqref{eq:v-2} dominates, and both approaches should become equivalent. It would be interesting to study the low number of photons limit, where \eqref{eq:v-1} and \eqref{eq:v-2} become comparable.


\section{Thermodynamic consistency of quantum master equations}\label{sec:cons-qme}

We now turn to the effective description in terms of quantum master equations and examine their thermodynamic consistency, i.e. under which conditions the laws of thermodynamics derived in section~\ref{subsec:thermo-uni-av}, the symmetries  \eqref{eq:ft-2} and \eqref{eq:ft-3}, and the fluctuation theorem \eqref{eq:crooks-a} and \eqref{eq:crooks-na} hold. We derive the quantum master equations using the theory of quantum maps.
%
%
%
%
%
To keep track of the energy transfers fluctuations during the derivations, we start from the tilted unitary dynamics defined in \eqref{eq:mgf}. 

\subsection{For the qubit-laser system $X$}\label{subsubsec:tilted-lx}

The coupling to the thermal bath is assumed to be weak. Since the density matrix is initially a tensor product of the matrices  of $X$ and $B$ \eqref{eq:rho-ini-alb}, the evolution of $\hat{\rho}^{\boldsymbol{\lambda}}_X(t)\equiv \text{Tr}_B[\hat{\rho}^{\boldsymbol{\lambda}}(t)]$ is described by a quantum map, 
\begin{equation}
\begin{array}{ll}
\hat{\rho}_X^{\boldsymbol{\lambda}}(t)=\sum\limits_{\mu,\nu}\hat{W}^{\boldsymbol{\lambda}}_{\mu,\nu}(t,0)\hat{\rho}_X(0) \hat{W}_{\mu,\nu}^{-\boldsymbol{\lambda}\dagger}(t,0)\\
\\
 \equiv \hat{M}^{\boldsymbol{\lambda}}(t,0)\hat{\rho}_X(0) \, ,
\end{array}
\label{eq:map-x}
\end{equation}
where $\hat{W}^{\boldsymbol{\lambda}}_{\mu,\nu}(t,0)$ are Kraus operators \cite{breuer2002theory}, 
\begin{equation}
    \hat{W}^{\boldsymbol{\lambda}}_{\mu,\nu}(t,0)=\sqrt{\eta_\nu}\langle \mu |\hat{U}_{\boldsymbol{\lambda}}(t,0)|\nu\rangle \, ,
    \label{eq:kraus}
\end{equation}
where $\hat{U}_{\boldsymbol{\lambda}}(t,0)$ was defined in \eqref{eq:mgf} and with $\{|\nu\rangle\}$ the eigenstates of $\hat{H}_B$ of eigenvalues $\nu$ and $ \eta_\nu=e^{-\beta_B\omega_\nu}/Z_B$.

We then make the Markov approximation, or semigroup hypothesis in the context of quantum maps \cite{breuer2002theory}: $\hat{M}^{\boldsymbol{\lambda}}(t,0)=\hat{M}^{\boldsymbol{\lambda}}(t,s)\hat{M}^{\boldsymbol{\lambda}}(s,0)$ for all $0\leq s \leq t$. This leads to a time local equation of motion of the form
\begin{align}
    \frac{d\hat{\rho}_X^{\boldsymbol{\lambda}}(t)}{dt} &= \lim_{\delta\to\delta_0}\frac{1}{\delta}(\hat{M}^{\boldsymbol{\lambda}}(t+\delta,t)-\mathds{I})\hat{\rho}_X(t)\nonumber\\
    &\equiv\mathcal{L}_{\boldsymbol{\lambda}}^X(\hat{\rho}_X^{\boldsymbol{\lambda}}(t)) \, ,
    \label{eq:l}
\end{align}
where the coarse graining time $\delta_0$ is chosen larger than the relaxation time of the bath and smaller than the relaxation time of $X$. We will discuss precisely those time scales in section~\ref{sec:cons-obe-fme}. 

The thermodynamic consistency condition for the master equation \eqref{eq:l}, where only the thermal bath has been traced out, has been identified in our previous work \cite{soret2022}. It reads 
\begin{equation}
\mathcal{L}^{X,R}_{0,0,-\lambda_B}[...] = \mathcal{L}^{X\dagger}_{0,0,-\lambda_B +i \beta_B}[...],
\label{eq:detailed}
\end{equation}
where we introduced the adjoint $\mathcal{O}^{\dagger}$ of a superoperator $\mathcal{O}$ as the one satisfying $\text{Tr}[(\mathcal{O}(X))^\dagger Y]=\text{Tr}[X^\dagger \mathcal{O}^\dagger(Y)] $ for all operators $X,Y$. 

\subsection{For the dressed qubit: Autonomous description}\label{subsubsec:cons-dq}

We now proceed to trace out the degrees of freedom of the dressed laser, and examine the energy exchanges in the dressed qubit picture. We therefore set $\boldsymbol{\lambda}=(\lambda_{DA},\lambda_{DL},\lambda_B)$. In the dressed qubit picture, the dressed qubit and dressed laser interact indirectly through the bath. The consistency condition for the master equations for $\hat{\rho}_{DA}$ can then be derived following the same logic as in \cite{soret2022}. Using the initial condition \eqref{eq:rho-ini-dadlb} with \eqref{eq:cond-rhod}, tracing out the degrees of freedom of $DL$ in \eqref{eq:map-x} leads to 
\begin{equation}  
\hat{\rho}_{DA}^{\boldsymbol{\lambda}}(t) =  \sum_{\boldsymbol{\kappa,\kappa'}} \hat{W}_{\boldsymbol{\kappa,\kappa'}}^{\boldsymbol{\lambda}} (t,0) \hat{\rho}_{DA}(0)
\hat{W}_{\boldsymbol{\kappa,\kappa'}}^{\boldsymbol{-\lambda}\, \dagger}(t,0) \, ,  
\label{eq:genopen} 
\end{equation}
where the sum runs over the pairs $\boldsymbol{\kappa}=(\mu,n), \boldsymbol{\kappa'}=(\nu,n')$ and where
\begin{equation}
    \hat{W}_{\boldsymbol{\kappa,\kappa'}}^{\boldsymbol{\lambda}} = \sqrt{\eta_{\nu}\xi_{n}} \langle n, \mu | \hat{U}_{\boldsymbol{\lambda}}(t,0)  |n',\nu \rangle \, ,
\label{eq:kraus-2}
\end{equation}
with $\xi_{n}=\langle n|\hat{\rho}_{DL}(0)|n\rangle$. Notice that the Kraus operators \eqref{eq:kraus-2} satisfy the property 
\begin{equation}
 \hat{W}_{\boldsymbol{\kappa,\kappa'}}^{\boldsymbol{\lambda}}(t,0)=   e^{\frac{\lambda_{DA}}{2}\hat{H}_{DA}} \hat{W}_{\boldsymbol{\kappa,\kappa'}}^{0,\lambda_{DL},\lambda_B}(t,0)e^{\frac{-\lambda_{DA}}{2}\hat{H}_{DA}} \, ,
 \label{eq:kraus-da}
\end{equation}
which implies that
\begin{equation}
\begin{array}{l}
    \hat{\rho}_{DA}^{\boldsymbol{\lambda}}(t) =e^{i\frac{\lambda_{DA}}{2}\hat{H}_{DA}} \\
    \times e^{t \mathcal{L}_{0,\lambda_{DL},\lambda_B} }[e^{-i\frac{\lambda_{DA}}{2} \hat{H}_{DA}} \hat{\rho}_{DA} (0)  e^{-i\frac{\lambda_{DA}}{2} \hat{H}_{DA}}] \\
    \times e^{i\frac{\lambda_{DA}}{2} \hat{H}_{DA}} \, ,
\end{array}
\label{eq:gen-tot}
\end{equation} 
where $\mathcal{L}_{\boldsymbol{\lambda}}$ is the superoperator dressed with counting fields describing the evolution of $\hat{\rho}_{DA}^{\boldsymbol{\lambda}}$.
The symmetry \eqref{eq:ft-2} is then satisfied if
\begin{equation}
    \mathcal{L}^{R}_{0,-\lambda_{DL},-\lambda_B+i\beta_B}[...] = \mathcal{L}^{\dagger}_{0,-\lambda_{DL},-\lambda_B}[...] \, ,
    \label{eq:detailed-2}
\end{equation}
as can be seen by computing separately the moment generating functions $\mathcal{G}(\boldsymbol{\lambda},t)$ and $\mathcal{G}^R(-\boldsymbol{\lambda}+i\boldsymbol{\nu},t)$ and using \eqref{eq:gen-tot} combined with the property of the adjoint $
\text{Tr}[(e^{t\mathcal{L}}[\hat{X}])^\dagger \hat{Y}] = \text{Tr}[\hat{X}^\dagger e^{t\mathcal{L}^\dagger}[\hat{Y}]]$. Note that the generalization to many uncoupled heat baths is straightforward by linearity.
The condition \eqref{eq:detailed-2} is another important result of this work: it is a simple criteria of thermodynamic consistency of quantum master equations for quantum systems coupled to heat baths and a coherent light source. 

Using the same reasoning as in section~\ref{subsubsec:ft}, we deduce that \eqref{eq:detailed-2} ensures that the entropy fluctuation theorem \eqref{eq:ft-sigma} holds at the level of master equations, hence that the second law is satisfied on average and at the level of the rates,
\begin{equation}
    d_t S_{DA}-\beta_B\dot Q\geq 0 \, .
    \label{eq:sigma-da-dl}
\end{equation}

The strict energy conservation condition \eqref{eq:strict-en-cons} takes the form 
\begin{equation}
    \mathcal{L}_{\boldsymbol{\lambda}}[...] = \mathcal{L}_{\boldsymbol{\lambda}+\chi\boldsymbol{1}}[...] \, ,
    \label{eq:strict-qme}
\end{equation}
and guarantees that the first law is satisfied at fluctuating level. It also implies energy conservation on average, i.e., that $\partial_\lambda Tr(\mathcal{L}_{\boldsymbol{\lambda}})|_{\lambda=0}=0$ when all the counting fields are set equal to $\lambda$, which in turn implies that the first law is satisfied on average at the level of the rates,
\begin{equation}
\begin{array}{ll}
d_t E_{DA} &= \dot Q +\dot W_{DL} \, .
\end{array}
\label{eq:1st-law-da-rate}
\end{equation}
A quantum master equation is said to be fully thermodynamically consistent iff it satisfies both \eqref{eq:detailed-2} and \eqref{eq:strict-qme}. 
Using the same argument as in \cite{soret2022}, we find that satisfying these two conditions requires to use the secular approximation in the dressed qubit basis. More precisely, we examine under which condition does \eqref{eq:genopen} becomes $\boldsymbol{\lambda}$ independent. Expressing the Kraus operators in the joint eigenbasis of $\hat{H}_{DA},\hat{H}_{DL}$ and $\hat{H}_B$, it appears that the only way to achieve this condition is to perform the secular approximation. We do not provide a detailed proof here since the reasoning and calculations are almost identical as in \cite{soret2022}, where thorough details are provided in the appendix. We will explicitly perform the secular approximation in the section~\ref{sec:cons-obe-fme}, when deriving the autonomous Floquet equation.

\begin{figure*}
    \centering
    \includegraphics[scale=0.25]{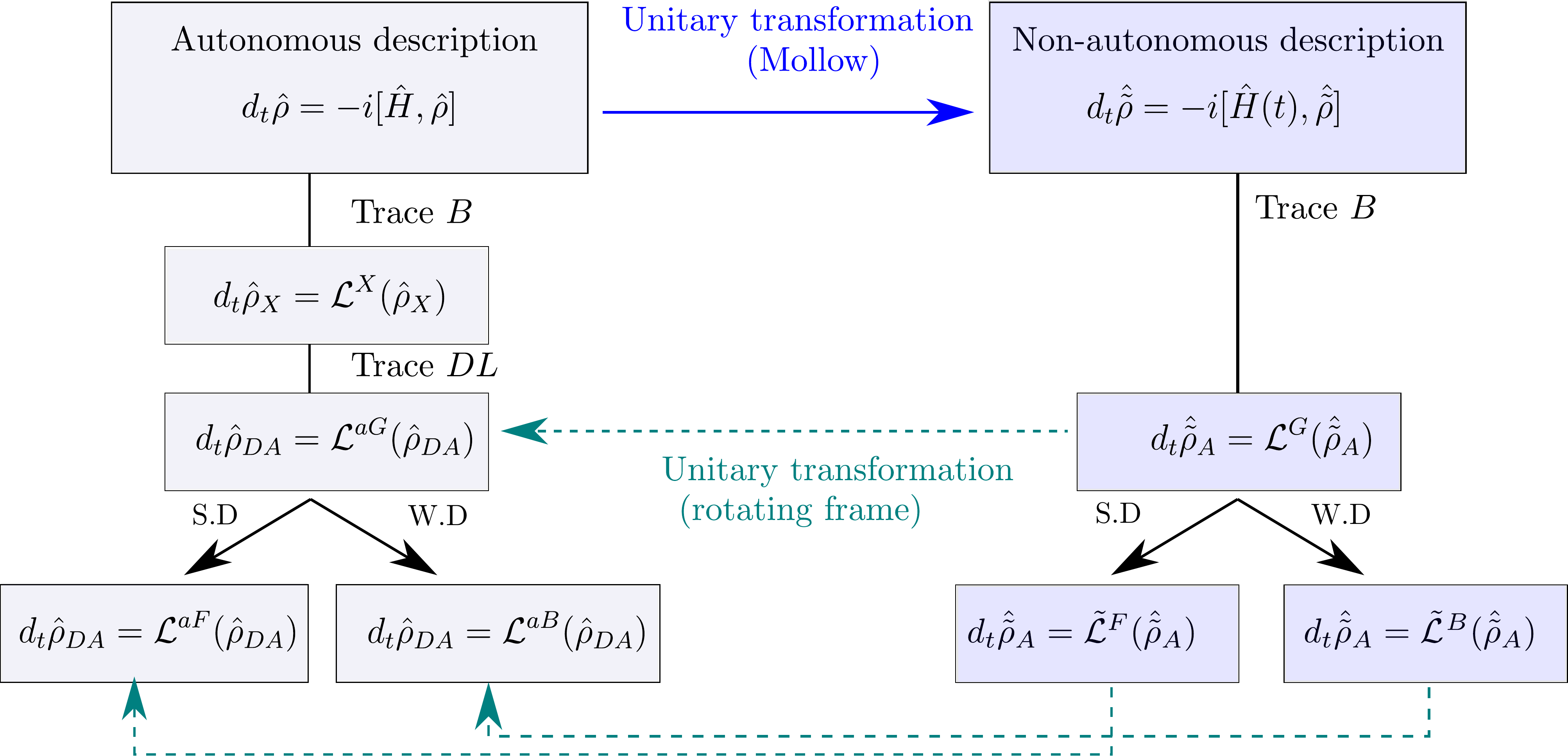}
    \caption{Schematic representation of the approximations performed in order to derive the generalized Bloch, Bloch and Floquet master equations, and summary of the unitary transformations connecting the autonomous and non-autonomous pictures, both at the unitary level and at the level of the master equations. S.D. stands for strong driving, and the weak and intermediate driving regimes are grouped under W.D. }
    \label{fig:recap}
\end{figure*}

\subsection{For the qubit: Non-autonomous description}

We now turn to the non-autonomous picture. In order to derive a master equation which preserves the symmetry \eqref{eq:ft-3}, we begin by noticing that, using the identity $e^{i(\hat{H}_A+\hat{V}(t))}=e^{-i\omega_L\hat{\sigma}_z t/2}e^{i(\hat{H}_A+\hat{V})}e^{i\omega_L\hat{\sigma}_z t/2}$ with $\hat{V}\equiv \frac{g}{2}(\hat{\sigma}_+ +\hat{\sigma}_-) $ and the cyclicity of the trace, the generating function in \eqref{eq:ft-3} can be re-written as
\begin{equation}
    \tilde{\mathcal{G}}(\boldsymbol{\lambda},t) = \text{Tr}[ \hat{U}^{\text{rot}}_{\boldsymbol{\lambda}}\hat{\tilde{\rho}}(0)\hat{U}^{\text{rot}\dagger}_{\boldsymbol{-\lambda}}]
\end{equation}
where 
\begin{align}
     \hat{U}^{\text{rot}}_{\boldsymbol{\lambda}} = &e^{i\frac{\lambda_B}{2}\hat{H}_B}e^{i\frac{\lambda_A}{2}(\hat{H}_A+\hat{V})} e^{i\omega_L\hat{\sigma}_z t/2} \label{eq:u-rot-lambda} \\
    &\times \mathcal{T}[e^{-i\int_0^t ds \hat{H}_{\lambda_L}(s)}] e^{-i\frac{\lambda_B}{2}\hat{H}_B}e^{-i\frac{\lambda_A}{2}(\hat{H}_A+\hat{V})} \, .\nonumber
\end{align}
When the counting fields are set to zero, $\hat{U}^{\text{rot}}_{\boldsymbol{\lambda}}$ becomes the propagator of the dynamics in the rotating frame \eqref{eq:rot-frame}. 
We may now apply the same reasoning as in the autonomous case in section~\ref{subsubsec:cons-dq} to compute a master equation for $\hat{\tilde{\rho}}_A^{\text{rot}}$. Since the counting fields $\lambda_A$ are now associated to a time-independent Hamiltonian in \eqref{eq:u-rot-lambda}, the Kraus operators for $\hat{\tilde{\rho}}_A^{\text{rot}}$ will have the same form as \eqref{eq:kraus-da}; applying the same reasoning as in section~\ref{subsubsec:cons-dq}, we obtain the following condition, which guarantees the symmetry \eqref{eq:ft-3},
\begin{equation}
    \tilde{\mathcal{L}}^{\text{rot}R}_{0,\lambda_{L},\lambda_B+i\beta_B}[...] = \tilde{\mathcal{L}}^{\text{rot}\dagger}_{0,\lambda_{L},\lambda_B}[...] \, .
    \label{eq:sym-qme-na}
\end{equation}
The strict energy conservation condition writes 
\begin{equation}
    \tilde{\mathcal{L}}_{\boldsymbol{\lambda}+\chi\boldsymbol{1}}[...] = \tilde{\mathcal{L}}_{\boldsymbol{\lambda}}[...] \, ,
    \label{eq:strict-a}
\end{equation} 
and is equivalent to \eqref{eq:strict-qme}, since we showed in section~\ref{subsubsec:ft-w-na} that the energy conservation conditions are equivalent in the autonomous and non-autonomous pictures. When satisfied, it implies that the first law holds at the level of the rates,
\begin{equation}
d_t E_A = \dot{W}_L+\dot Q\, .
\label{eq:1st-law-a-rate}
\end{equation}

\section{Generalized Bloch, optical Bloch and Floquet master equations}\label{sec:cons-obe-fme}

In this section, we derive three master equations which can be used in practice to study the coherent driving of a qubit, and examine whether they satisfy the general conditions of consistency identified in the previous section. Each master equation is derived both in the autonomous and non-autonomous pictures. 
A schematic representation of the approximations made and of the correspondences between the autonomous and non-autonomous picture is given in Fig.~\ref{fig:recap}.

We first derive a standard Markovian master equation in the joint qubit-laser space, in the autonomous picture. 
We then trace out the dressed laser to derive master equations in the dressed qubit basis. 
We begin with a new master equation, called the generalized Bloch equation, valid at all qubit-laser coupling strengths. 
Then, we identify three relevant qubit-laser coupling regimes (or driving regimes): strong, intermediate and weak. The strong driving regime leads to the Floquet master equation, while the intermediate and weak driving regimes gives rise to the Bloch master equation.  
We then proceed to show how master equations in the qubit space can be derived using the correspondence between the dressed qubit (in the autonomous picture) and the evolution in the rotating frame (in the non-autonomous picture), showed in section~\ref{subsubsec:floquet-rot}.  

\subsection{qubit-laser}\label{subsec:wc-auto}

For convenience, we use the interaction picture \cite{breuer2002theory}. 
In the interaction picture, the map \eqref{eq:map-x} becomes
\begin{equation}
\begin{array}{ll}
\hat{\rho}_X^{\boldsymbol{\lambda}I}(t)=\sum\limits_{\mu,\nu}\hat{W}^{\boldsymbol{\lambda}I}_{\mu,\nu}(t,0)\hat{\rho}_X(0) \hat{W}_{\mu,\nu}^{-\boldsymbol{\lambda}I\dagger}(t,0) \, ,
\end{array}
\label{eq:map-x-i}
\end{equation}
where we recall that $\boldsymbol{\lambda}=(\lambda_{DA},\lambda_{DL},\lambda_B)$ and
where the Kraus operators $\hat{W}^{\boldsymbol{\lambda}I}_{\mu,\nu}(t,0)$ are given by
\begin{equation}
    \hat{W}^{\boldsymbol{\lambda}I}_{\mu,\nu}(t,0)=\sqrt{\eta_\nu}\langle \mu |\mathcal{T}[e^{-i\int_0^t ds \hat{V}^{\boldsymbol{\lambda}}_{AB}(s)}]|\nu\rangle \, ,
    \label{eq:kraus-i}
\end{equation}
and where we remind that $\mathcal{T}$ denotes time ordering. 
To push the derivation further, we perform a perturbative expansion to second order in $\hat{V}_{AB}$.

The Hamiltonian $\hat{V}^{\boldsymbol{\lambda}}_{AB}(t)$ in \eqref{eq:kraus-i} is the Hamiltonian $\hat{V}^{\boldsymbol{\lambda}}_{AB}$ in the interaction picture, given by (see appendix~\ref{app:qme-x})
\begin{align}
&\hat{V}^{\boldsymbol{\lambda}}_{AB}(t)\equiv e^{i\hat{H}_0t}\hat{V}^{\boldsymbol{\lambda}}_{AB}e^{-i\hat{H}_0t}  \label{eq:vab}  \\
&=\left(\hat{S}^{\boldsymbol{\lambda}}_z(t)+\hat{S}^{\boldsymbol{\lambda}}_-(t)+\hat{S}^{\boldsymbol{\lambda}}_+(t)\right) \hat{B}^\dagger_{\lambda_B}(t) + h.c.\nonumber
\end{align}
with  $\hat{B}_{\lambda_B}(t)\equiv\sum_k g_k \hat{b}_k e^{-i\omega_k (t+\lambda_B/2)}$ and
\begin{align}
\hat{S}^{\boldsymbol{\lambda}}_z(t)&=e^{-i\omega_L t}e^{-i\omega_L\lambda_{DL}}\hat{S}_z\nonumber \\
\hat{S}^{\boldsymbol{\lambda}}_+(t)&=e^{-i(\omega_L -\Omega)t}e^{-i(\omega_L\lambda_{DL}-\Omega\lambda_{DA})}\hat{S}_+\nonumber\\
\hat{S}^{\boldsymbol{\lambda}}_-(t)&=e^{-i(\omega_L+\Omega)t}e^{-i(\omega_L\lambda_{DL}+\Omega\lambda_{DA})}\hat{S}_-\, ,
\end{align}
where
%
%
\begin{align}
\hat{S}_z&=\frac{g}{2\Omega}(|2\rangle\langle 2|-|1\rangle\langle 1|) \otimes\sum\limits_{n\geq 0} |n-1\rangle\langle n|\nonumber \\
&\equiv \hat{s}_z\otimes \sum\limits_{n\geq 0} |n-1\rangle\langle n|\nonumber \\
\hat{S}_+&\equiv-\frac{\Omega-\delta}{2\Omega}|2\rangle\langle1|\otimes\sum\limits_{n\geq 0} |n-1\rangle\langle n|\nonumber\\
&\equiv \hat{s}_+\otimes\sum\limits_{n\geq 0} |n-1\rangle\langle n|\nonumber\\
\hat{S}_-&\equiv\frac{\Omega+\delta}{2\Omega}|1\rangle\langle 2|\otimes\sum\limits_{n\geq 0} |n-1\rangle\langle n|\nonumber\\
&\equiv \hat{s}_- \otimes\sum\limits_{n\geq 0} |n-1\rangle\langle n|\, ,\label{eq:s-ops}
\end{align}
where we introduced the reduced operators 
\begin{align}
    \hat{s}_z&\equiv\frac{g}{2\Omega}(|2\rangle\langle 2|-|1\rangle\langle 1|) \equiv \frac{g}{2\Omega}\hat{\Sigma}_z\nonumber \\
\hat{s}_+&\equiv-\frac{\Omega-\delta}{2\Omega}|2\rangle\langle1|\equiv -\frac{\Omega-\delta}{2\Omega}\hat{\Sigma}_+\nonumber\\
\hat{s}_-&\equiv\frac{\Omega+\delta}{2\Omega}|1\rangle\langle 2|\equiv 
\frac{\Omega+\delta}{2\Omega}\hat{\Sigma}_-  ,\label{eq:s-ops-red}
\end{align}
and where $\hat{\Sigma}_z=|2\rangle\langle 2|-|1\rangle\langle 1|,\hat{\Sigma}_+=|2\rangle\langle 1|=\hat{\Sigma}_-^\dagger$. Later on, we will use the reduced operators dressed with counting fields,
\begin{align}
 \hat{s}_z^{\boldsymbol{\lambda}} & \equiv e^{-i\lambda_{DL}\omega_L}   \hat{s}_z \nonumber \\
 \hat{s}_+^{\boldsymbol{\lambda}} & \equiv e^{i\lambda_{DA}\Omega/2} e^{-i\lambda_{DL}\omega_L/2} e^{i\lambda_B(\omega_L-\Omega)/2}\hat{s}_+  \label{eq:slambda-1} \\
 \hat{s}_-^{\boldsymbol{\lambda}} & \equiv e^{-i\lambda_{DA}\Omega/2} e^{-i\lambda_{DL}\omega_L/2} e^{i\lambda_B(\omega_L+\Omega)/2}\hat{s}_- \, . \nonumber
\end{align}
It will further be useful to use the identity 
\begin{equation}
    \hat{S}_z+\hat{S}_++\hat{S}_-=\sum_{N_L} |a,N_L\rangle\langle b,N_L| \, .
    \label{eq:id-s}
\end{equation}
Let's now introduce the set $\{\hat{\sigma}_{mn}\}$ of jump operators between the eigenstates $\{|j,n\rangle\}$ of $\hat{H}_X$. The set $\{\hat{\sigma}_{mn}\}$ forms a basis of jump operators acting on $\mathcal{H}_X$. To alleviate the notations, we relabel the eigenstates
\begin{equation}
    |j,n\rangle\ \rightarrow |E_n\rangle \, ,
\end{equation}
so that, by definition, $\hat{\sigma}_{mn}=|E_n\rangle\langle E_m|$. We introduce $\omega_{mn}=E_m-E_n$, the corresponding Bohr frequencies. Note that different $\hat{\sigma}_{mn}$ may be associated to the same frequency $\omega_{mn}$. We now express the Kraus operators \eqref{eq:kraus-i} in the basis $\{\hat{\sigma}_{mn}\}$ and perform the Markov approximation, to obtain a time local equation of motion of the form of \eqref{eq:l}, 
\begin{align}
 &   \mathcal{L}_{\boldsymbol{\lambda}}(t)\hat{\rho}_S(t) \label{eq:aux-lfl}\\
 &=\lim_{\delta\to \delta_0}\frac{1}{\delta}\sum\limits_{mn,m'n'}d^{\boldsymbol{\lambda}}_{mn,m'n'}(t,\delta)\hat{\sigma}_{mn}\hat{\rho}^{\boldsymbol{\lambda}I}_X(t)\hat{\sigma}_{m'n'}^\dagger- \hat{\rho}^{\boldsymbol{\lambda}I}_X(t),
    \nonumber
\end{align}
where 
\begin{align}
&d^{\boldsymbol{\lambda}}_{mn,m'n'}(t,\delta)\equiv \label{eq:dij}\\
&\sum\limits_{\mu,\nu}\eta_\nu \text{Tr}_S[\hat{\sigma}_{mn}^\dagger\hat{W}^{\boldsymbol{\lambda}I}_{\mu,\nu}(t+\delta,t)]\text{Tr}_S[\hat{\sigma}_{m'n'}\hat{W}_{\mu,\nu} ^{-\boldsymbol{\lambda}I\dagger}(t+\delta,t)]\, . \nonumber
\nonumber
\end{align}
A perturbative expansion to second order in $\hat{V}_{AB}^{\boldsymbol{\lambda}}$, combined with $\hat{\rho}_B=\sum_\nu \eta_\nu |\eta_\nu\rangle\langle\eta_\nu|$, yields
\begin{widetext}
\begin{equation}
    \begin{array}{l}
\mathcal{L}_{\boldsymbol{\lambda}}(t)\hat{\rho}^I_X(t) = \lim_{\delta\to \delta_0}\frac{1}{\delta}\left[\sum\limits_{n,n'} \hat{\sigma}_{nn}\hat{\rho}_X^{\boldsymbol{\lambda}I}(t)\hat{\sigma}_{n'n'}^\dagger- \hat{\rho}_X^{\boldsymbol{\lambda}I}(t)\right .\\
    \\
    +\sum\limits_{mn,m'n'}Tr [\int_t^{t+\delta} ds\, \hat{\sigma}_{mn}^\dagger\hat{V}_{AB}^{\boldsymbol{\lambda}}(s) \hat{\rho}_B\int_t^{t+\delta} ds\, \hat{\sigma}_{m'n'}\hat{V}_{AB}^{-\boldsymbol{\lambda}\dagger}(s) ]\hat{\sigma}_{mn}\hat{\rho}_X^{\boldsymbol{\lambda}I}(t)\hat{\sigma}_{m'n'}^\dagger\\
    \\
    -\frac{1}{2}\sum\limits_{mn,m'n'}\text{Tr}_X[\hat{\sigma}_{m'n'}]Tr [\hat{\sigma}_{mn}^\dagger\int_t^{t+\delta} ds \hat{V}_{AB}^{\boldsymbol{\lambda}}(s) \int_t^{s} ds' \hat{V}_{AB}^{\boldsymbol{\lambda}}(s') \hat{\rho}_B]\hat{\sigma}_{mn}\hat{\rho}_X^{\boldsymbol{\lambda}I}(t)\hat{\sigma}_{m'n'}^\dagger\\
    \\
   \left. -\frac{1}{2}\sum\limits_{mn,m'n'}\text{Tr}_X[\hat{\sigma}_{mn}^\dagger]Tr [\hat{\sigma}_{m'n'}\int_t^{t+\delta} ds \hat{V}_{AB}^{-\boldsymbol{\lambda}\dagger}(s) \int_s^{t+\delta} ds' \hat{V}_{AB}^{-\boldsymbol{\lambda}\dagger}(s') \hat{\rho}_B]\hat{\sigma}_{mn}\hat{\rho}_X^{\boldsymbol{\lambda}I}(t)\hat{\sigma}_{m'n'}^\dagger \right].
    \end{array}
    \label{eq:gme-aux}
\end{equation}
\end{widetext}
Notice that the r.h.s. term of the first line cancels out since the $\hat{\sigma}_{nn}\hat{\rho}_X^{\boldsymbol{\lambda}I}(t)\hat{\sigma}_{n'n'}^\dagger=(\hat{\rho}^{\boldsymbol{\lambda}I}_X(t))_{nn'}|E_n\rangle\langle E_{n'}|$ and $\{|E_n\rangle\}_n$ is a basis of the system. Writing explicitly $\hat{V}^{\boldsymbol{\lambda}}_{AB}(t)$ in \eqref{eq:gme-aux}, we find that the trace over the bath yields terms of the following form for the coefficients of the master equation (see appendix~\ref{app:qme-x} for the full expression),
\begin{align}
& \frac{1}{\delta_0}   \int\limits_t^{t+\delta_0} ds\int\limits_t^{t+\delta_0} ds' \text{Tr}_B[ \hat{B}_{\lambda_B}^\dagger(s)\hat{B}_{-\lambda_B}(s')\hat{\rho}_B ]
e^{-i(\omega_{\alpha}s-\omega_{\alpha'}s')}\nonumber \\
=&\sum_k \text{ sinc}\left(\frac{\omega_k-\omega_\alpha}{2}\delta_0\right)\text{ sinc}\left(\frac{\omega_k-\omega_{\alpha'}}{2}\delta_0\right) \nonumber \\
&\times G_-(\omega_k)e^{i\lambda_B\omega_k}\delta_0e^{i(t+\delta_0/2)(\omega_{\alpha'}-\omega_\alpha)} \, , \\
& \frac{1}{\delta_0}   \int\limits_t^{t+\delta_0} ds\int\limits_t^{t+\delta_0} ds' \text{Tr}_B[ \hat{B}_{\lambda_B}(s)\hat{B}^\dagger_{-\lambda_B}(s')\hat{\rho}_B ]
e^{-i(\omega_{\alpha}s-\omega_{\alpha'}s')}\nonumber \\
=&\sum_k \text{ sinc}\left(\frac{\omega_k-\omega_\alpha}{2}\delta_0\right)\text{ sinc}\left(\frac{\omega_k-\omega_{\alpha'}}{2}\delta_0\right) \nonumber \\
&\times G_+(\omega_k)e^{-i\lambda_B\omega_k}\delta_0e^{i(t+\delta_0/2)(\omega_{\alpha'}-\omega_\alpha)} \, ,
\end{align}
where $G_\pm(\nu)$ is the real part of the half Fourier transform of the bath correlation functions,
\begin{align}
    \int\limits_0^{+\infty} d\tau \text{Tr}[\hat{B}(\tau)\hat{B}^\dagger(0)\hat{\rho}_B] e^{i\nu\tau} &\equiv G_+(\nu) + i I_+(\nu) \nonumber \\
    \int\limits_0^{+\infty} d\tau \text{Tr}[\hat{B}(\tau)\hat{B}^\dagger(0)\hat{\rho}_B] e^{i\nu\tau} &\equiv G_-(\nu) + i I_-(\nu) \label{eq:semi-fourier} \, .
\end{align}
The product of sinc functions may be approximated by
\begin{align}
 &  \delta_0 \text{ sinc}\left(\frac{\omega_k-\omega_\alpha}{2}\delta_0\right)\text{ sinc}\left(\frac{\omega_k-\omega_{\alpha'}}{2}\delta_0\right) \label{eq:approx-sinc}\\
 & \approx \delta_0\text{ sinc}\left(\frac{2\omega_k-\omega_\alpha-\omega_{\alpha'}}{2}\delta_0\right) \nonumber\\
    & \approx \left\{ 
    \begin{array}{ll}
    \delta[2\omega_k-(\omega_\alpha+\omega_{\alpha'})] \text{ if } |\omega_\alpha-\omega_{\alpha'}|<\delta_0^{-1}\\
    \\
    0 \text{ otherwise.}
    \end{array}
    \right.\nonumber
\end{align}
%


Going back to the Schr\"odinger picture, \eqref{eq:gme-aux} takes the form
\begin{equation}
    \frac{d\hat{\rho}^{\boldsymbol{\lambda}}_X(t)}{dt}=-i[\hat{H}_X+\hat{H}_{LS}]+\mathcal{D}_{\boldsymbol{\lambda}}(\hat{\rho}^{\boldsymbol{\lambda}}_X(t)) \, ,
    \label{eq:qme-x}
\end{equation}
where the dissipator dressed with counting fields $\mathcal{D}_{\boldsymbol{\lambda}}$ is expressed in terms of the operators \eqref{eq:s-ops}, and where $\hat{H}_{LS}$ is a Lamb shift contribution. The Lamb shift term can be written in terms of the jump operators \eqref{eq:s-ops}, with amplitudes given by the imaginary part $I_\pm(\nu)$ of the half Fourier transform of the bath correlation functions in \eqref{eq:semi-fourier}. 
We further on neglect the Lamb shift term $\hat{H}_{LS}$, given that it induces negligibly small corrections to the qubit's frequency \cite{farley1981,carmichael2002,elouard2020thermodynamics}. 

We now proceed to deriving master equations for the dressed qubit and the qubit, by tracing out respectively the degrees of freedom of the dressed laser and laser.

\subsection{Dressed qubit}

Since, in \eqref{eq:approx-sinc}, $\omega_\alpha\in\{\omega_L,\omega_L\pm\Omega\}$, we have, for any $\alpha\neq\alpha'$, $|\omega_\alpha-\omega_{\alpha'}|=\Omega$ or $|\omega_\alpha-\omega_{\alpha'}|=2\Omega$. This allows to identify three regimes, depending on the value of $\Omega$ (equivalently of the coupling $g$),
\begin{align}
    & 2\Omega<\delta_0^{-1} \mbox{ : weak driving} \label{eq:regimes-1} \\
    & \Omega < \delta_0^{-1} < 2\Omega \mbox{ : intermediate driving} \nonumber \\
    & \delta_0^{-1} < \Omega \mbox{ : strong driving} \, .\nonumber
\end{align}
However, these definitions are meaningless as long as we do not connect $\delta_0$ with the relevant time scales of the problem. In order to identify these time scales, we examine the real parts $G_\pm$ of the half Fourier transforms of the bath correlation functions, introduced in \eqref{eq:semi-fourier}; we may re-write them as
\begin{align}
    G_+(\nu)&\equiv\frac{1}{2}\int\limits_{-\infty}^{+\infty}d\tau \text{Tr}[\hat{B}(\tau)\hat{B}^\dagger(0)\hat{\rho}_B] e^{i\nu\tau}\\
     G_-(\nu)&\equiv\frac{1}{2}\int\limits_{-\infty}^{+\infty}d\tau \text{Tr}[\hat{B}^\dagger(\tau)\hat{B}(0)\hat{\rho}_B] e^{i\nu\tau} \, .
\end{align}
The functions $G_\pm(\nu)$ are related to the bath's zero-temperature spectral function, $\Gamma(\nu)\equiv\sum_k |g_k|^2\delta_D(\nu-\omega_k)$, where $\delta_D$ is the Dirac delta function, by \cite{elouard2020thermodynamics}
\begin{align}
    G_+(\nu) &= \Gamma(\nu)(n_B(\nu)+1) \\
    G_-(\nu) & = \Gamma(\nu)n_B(\nu) \, ,
\end{align}
where $n_B(\nu)\equiv (e^{\beta_B\nu}-1)^{-1}$. Note that $G_+(\nu)=e^{\beta_B\nu}G_-(\nu)$, which is the Kubo–Martin–Schwinger (KMS) condition \cite{kubo1957,breuer2002theory}.
Let us now define 
\begin{equation}
    \gamma_{max}\equiv \mbox{max}_{\alpha=z,+,-}\{ G_\pm(\omega_\alpha)\} \, ,
\end{equation}
where we recall that $\omega_\alpha$ are the frequencies appearing in the Fourier transform of $\hat{V}_{AB}(t)$. These frequencies, together with $\gamma_{max}$, are the relevant time scales to which $\delta_0$ should be compared in order to push further the derivation of the master equation \eqref{eq:qme-x}. A necessary condition on $\delta_0$ is that
\begin{equation}
    \delta_0^{-1} \gg \gamma_{max} \, .
\end{equation}
Moreover, we require that $\omega_L, \omega_A\gg \delta_0^{-1}$, which is a reasonable assumption in practice. 
Combined with \eqref{eq:approx-sinc}, these conditions allows to re-define the three driving regimes as  
\begin{align}
    & \omega_L, \omega_A \gg \delta_0^{-1} \gg \Omega, \gamma_{max} \mbox{ : weak driving} \label{eq:regimes-2} \\
    & \omega_L, \omega_A \gg \delta_0^{-1} > \Omega, \gamma_{max} \mbox{ : intermediate driving} \nonumber \\
    & \omega_L, \omega_A, \Omega \gg \delta_0^{-1} \gg  \gamma_{max} \mbox{ : strong driving} \, .\nonumber
\end{align}
We now proceed to deriving master equations. 
We begin by deriving a new master equation, called generalized Bloch master equation, valid at all coupling strengths. The Floquet and Bloch master equations are then obtained from the generalized Bloch equation by performing additional approximations, respectively in the strong and weak/intermediate driving regimes. 
A summary of the regimes of validity and approximations made for each master equation is given in the Table~\ref{table2}.

\begin{table*}
    \centering
\begin{tabular}{|c|c|c|c|c|}
\hline
\multicolumn{1}{|c|}{Driving} & \multicolumn{1}{c|}{$\begin{array}{l}
    \mbox{Weak}\\
    \Omega<\gamma_{max}
\end{array}$} & \multicolumn{1}{c|}{$\begin{array}{l}
    \mbox{Intermediate}\\
    \Omega\sim\gamma_{max}
\end{array}$} & \multicolumn{1}{c|}{$\begin{array}{l}
    \mbox{(Common regime of validity)}\\
  \omega_L,\omega_A \gg  \Omega \gg \gamma_{max}
\end{array}$} & \multicolumn{1}{c|}{$\begin{array}{l}
 \,\,\,\,\,\,\,\,\,\,\,\,\,\,\,   \mbox{Strong}\\
   \omega_L,\omega_A, \Omega\gg \gamma_{max}
\end{array}$} \\
\hline
$\begin{array}{c}
       \\
     \mbox{Time scales }\\
     \\
\end{array}$ 
& $\omega_L,\omega_A\gg\delta_0^{-1}\gg\Omega,\gamma_{max}$ &$\omega_L,\omega_A\gg\delta_0^{-1}>\Omega,\gamma_{max}$ & 
$\omega_L,\omega_A \gg \Omega\gg\delta_0^{-1}\gg\gamma_{max}$ & $\omega_L,\omega_A,\Omega\gg\delta_0^{-1}\gg\gamma_{max}$ \\
\hline

 & \multicolumn{4}{c|}{ $\begin{array}{c}
\mbox{\bf Generalized Bloch} \\
     \bullet \mbox{Approximation \eqref{eq:approx-gob}} \\
     \bullet \mbox{Consistency: } \\
     \mbox{Full consistency in strong coupling} \\
     \mbox{Weak/intermdediate: symmetries (\ref{eq:detailed-2}, \ref{eq:sym-qme-na}), and 1st and 2nd laws average \& rates}
\end{array}$ } \\
\cline{2-5}
\multirow{2}{*}{QME}  & \multicolumn{3}{c|}{$\begin{array}{c}
\mbox{\bf Bloch} \\
     \bullet \mbox{Approximation: $G_\pm(\nu)\rightarrow \overline{G}_\pm$} \\
     \bullet \mbox{Consistency: } \\
     \mbox{Symmetries (\ref{eq:detailed-2}, \ref{eq:sym-qme-na})} \\
     \mbox{1st and 2nd laws: average \& rates}
\end{array}$ } &  \\ 
\cline{2-5}
 & \multicolumn{2}{c|}{ }   & \multicolumn{2}{c|}{$\begin{array}{c}
\mbox{\bf Floquet} \\ 
     \bullet \mbox{Secular approximation} \\
     \bullet \mbox{Consistency: } \\
     \mbox{Full consistency} \\
\end{array}$ }  \\
\hline
\end{tabular}\caption{Summary of the approximations used to derive the Bloch and Floquet master equations, of their regimes of validity and of their thermodynamic consistency. Full consistency means satisfying the symmetries (\ref{eq:detailed-2}, \ref{eq:sym-qme-na}) and the strict energy conservation conditions (\ref{eq:strict-qme}, \ref{eq:strict-a}); this implies that the laws of thermodynamics are satisfied at the average and fluctuating levels. }
\label{table2}
\end{table*}

\subsubsection{Generalized Bloch equation}

In order to derive the generalized Bloch equation, we do the following approximation, inspired by the procedure employed in \cite{ptaszynski2019thermodynamics,mccauley2020accurate,soret2022},
\begin{align}
  &  \sum_k G_\pm(\omega_k)e^{i\lambda_B\omega_k}\delta_0\text{ sinc}\left(\frac{\omega_k-\omega_\alpha}{2}\delta_0\right)\text{ sinc}\left(\frac{\omega_k-\omega_{\alpha'}}{2}\delta_0\right) \nonumber \\
    &\approx 
    \left\{
    \begin{array}{l}
         \sqrt{(G_\pm(\omega_\alpha)G_\pm(\omega_{\alpha'})}e^{i\lambda_B\omega_\alpha}e^{i\lambda_B\omega_{\alpha'}} \mbox{if } |\omega_\alpha-\omega_{\alpha'}|<\gamma_{max}  \\
         0 \mbox{ otherwise}
    \end{array} \right. . \label{eq:approx-gob}
\end{align}
%
This procedure makes the superoperator \eqref{eq:gme-aux} symmetric, although the resulting superoperator takes different forms in the three driving regimes of \eqref{eq:regimes-1}. Since the operators \eqref{eq:s-ops} are factorized in the basis $\{|1n\rangle,|2n\rangle\}$, we may readily trace out $\mathcal{H}_{DL}$, and we obtain a master equation for the dressed qubit. In the weak driving regime ($2\Omega<\gamma_{max}$), the generalized Bloch equation is
\begin{align}
  &  \mathcal{L}_{\boldsymbol{\lambda}}^{aG}(\hat{\rho}_{DA}^{\boldsymbol{\lambda}}) =\label{eq:gob} \\
  & -i[\hat{H}_{DA},\hat{\rho}_{DA}^{\boldsymbol{\lambda}}] + \mathcal{D}_+^{aG\boldsymbol{\lambda}}(\hat{\rho}_{DA}^{\boldsymbol{\lambda}})+\mathcal{D}_-^{aG\boldsymbol{\lambda}}(\hat{\rho}_{DA}^{\boldsymbol{\lambda}}) \nonumber
\end{align}
with
\begin{widetext}
\begin{align}
   & \mathcal{D}_+^{aG\boldsymbol{\lambda}}(\hat{\rho}) = \hat{T}_+^{\boldsymbol{\lambda}}\hat{\rho} \hat{T}_+^{\boldsymbol{-\lambda}\dagger} - \frac{1}{2}\left(\hat{T}_+^{(\lambda_{DA},\lambda_{DL},0) \dagger }\hat{T}_+^{(\lambda_{DA},\lambda_{DL},0) }\hat{\rho} +\hat{\rho}\hat{T}_+^{(-\lambda_{DA},-\lambda_{DL},0) \dagger }\hat{T}_+^{(-\lambda_{DA},-\lambda_{DL},0) }\right)\nonumber \\
   \nonumber\\
   & \mathcal{D}_-^{aG\boldsymbol{\lambda}}(\hat{\rho}) = \hat{T}_-^{\boldsymbol{\lambda}\dagger}\hat{\rho} \hat{T}_-^{\boldsymbol{-\lambda}} - \frac{1}{2}\left(\hat{T}_-^{(\lambda_{DA},\lambda_{DL},0)  }\hat{T}_-^{(\lambda_{DA},\lambda_{DL},0)\dagger }\hat{\rho} +\hat{\rho}\hat{T}_-^{(-\lambda_{DA},-\lambda_{DL},0 )}\hat{T}_-^{(-\lambda_{DA},-\lambda_{DL},0 )\dagger}\right)\nonumber 
\end{align}
\end{widetext}
where the superscript $aG$ stands for autonomous generalized Bloch equation (the non-autonomous counterpart is derived in section~\ref{subsec:na-qme}), $\boldsymbol{\lambda} = (\lambda_{DA},\lambda_{DL},\lambda_B)$ and
\begin{align}
    \hat{T}_\pm^{\boldsymbol{\lambda}} = \sum_{\alpha=+,-,z} \sqrt{G}_\pm(\omega_\alpha)\hat{s}_\alpha^{\boldsymbol{\lambda}} \, .
\end{align}
In the strong driving regime ($\Omega>\gamma_{max}$), the generalized Bloch equation is equal to the Floquet master equation (in the rotating frame), which we derive in the next section; the expression is given in Eq.\eqref{eq:fme-2}. The expression of the generalized Bloch equation in the intermediate regime ($\Omega<\gamma_{max}<2\Omega$) is given in appendix~\ref{app:gob} in order to alleviate the text.

It is straightforward to check that the generalized Bloch master equation satisfies the condition \eqref{eq:detailed-2} in all three regimes (see e.g. Fig.\ref{fig:MGF-WDL-OBE-OBEMaps} for the case $2\Omega<\gamma_{max}$). Consequently, the second law of thermodynamics is satisfied on average. The strict energy conservation condition \eqref{eq:strict-a}, on the other hand, is only valid when $\Omega>\gamma_{max}$, since, in this case, \eqref{eq:approx-gob} amounts to performing the secular approximation \cite{breuer2002theory} (the product of sinc functions in \eqref{eq:approx-gob} is then non zero only in the case $\alpha=\alpha'$). 

We now derive the rates $\dot W_{DL}, \dot Q$ and $d_t E_{DA}$. Let's introduce 
\begin{align}
P_1(t)&= \langle 1|\hat{\rho}_{DA}(t)|1\rangle \nonumber\\ 
P_2(t)&= \langle 2|\hat{\rho}_{DA}(t)|2\rangle \label{eq:p12}\\ 
P_{21}(t)&= \langle 2|\hat{\rho}_{DA}(t)|1\rangle \nonumber
\end{align}
and 
\begin{equation}
    \begin{array}{ll}
\gamma_{0,\downarrow}&=\frac{g^2}{4\Omega^2}G_+(\omega_L)\\
\\
\gamma_{0,\uparrow}&=\frac{g^2}{4\Omega^2}G_-(\omega_L)\\
\\
\gamma_{1,\downarrow}&=\frac{(\Omega+\delta)^2}{4\Omega^2}G_+(\omega_L+\Omega)\\
\\
\gamma_{1,\uparrow}&=\frac{(\Omega+\delta)^2}{4\Omega^2}G_-(\omega_L+\Omega)\\
\\
\gamma_{2,\downarrow}&=\frac{(\Omega-\delta)^2}{4\Omega^2}G_-(\omega_L-\Omega)\\
\\
\gamma_{2,\uparrow}&=\frac{(\Omega-\delta)^2}{4\Omega^2}G_+(\omega_L-\Omega) \, .
    \end{array}
    \label{eq:coefs-fme}
\end{equation}
Taking the derivatives in $\lambda_{DL}, \lambda_B, \lambda_{DA}$ in the trace of $\mathcal{L}_{\boldsymbol{\lambda}}^{aG}$, we obtain respectively the rates $\dot W_{DL}, \dot Q$ and $ d_t\Delta E_A$. In the weak and intermediate driving regimes, the results are
\begin{align}
 \dot W_{DL} =& \omega_L(\gamma_{0\downarrow}-\gamma_{0\uparrow}) \label{eq:wdl-gob}\\
 &  +\omega_L\left[(\gamma_{2\uparrow}-\gamma_{1\uparrow})P_1(t)+(\gamma_{1\downarrow}-\gamma_{2\downarrow})P_2(t)\right] \nonumber \\
&-2\omega_L (\sqrt{\gamma_{0\uparrow}\gamma_{2\downarrow}}+\sqrt{\gamma_{0\uparrow}\gamma_{1\uparrow}}) \mbox{Re}[P_{21}(t)] \nonumber \\
&-2\omega_L (\sqrt{\gamma_{0\downarrow}\gamma_{2\uparrow}}+\sqrt{\gamma_{0\downarrow}\gamma_{1\downarrow}})\mbox{Re}[P_{21}(t)] \nonumber\\
\dot Q =& \omega_L(\gamma_{0\uparrow}-\gamma_{0\downarrow}) \label{eq:q-gob}\\
&+\left[ \omega_L(\gamma_{1\uparrow}-\gamma_{2\uparrow})+\Omega(\gamma_{2\uparrow}+\gamma_{1\uparrow}) \right] P_1(t) \nonumber \\
&+\left[ \omega_L(\gamma_{2\downarrow}-\gamma_{1\downarrow})-\Omega(\gamma_{2\downarrow}+\gamma_{1\downarrow}) \right] P_2(t) \nonumber \\
&+2(\omega_L-\Omega/2)(\sqrt{\gamma_{0\downarrow}\gamma_{2\uparrow}}+\sqrt{\gamma_{0\downarrow}\gamma_{2\downarrow}})\mbox{Re}[P_{21}(t)] \nonumber\\
&+2(\omega_L+\Omega/2)(\sqrt{\gamma_{0\downarrow}\gamma_{1\uparrow}}+\sqrt{\gamma_{0\downarrow}\gamma_{1\downarrow}})\mbox{Re}[P_{21}(t)] \nonumber \\
d_t E_{DA} =& \Omega(\gamma_{2\uparrow}+\gamma_{1\uparrow})P_1(t) -\Omega(\gamma_{2\downarrow}+\gamma_{1\downarrow}) P_2(t) \label{eq:eda-gob} \\
&-\Omega/2(\sqrt{\gamma_{0\downarrow}\gamma_{2\uparrow}}+\sqrt{\gamma_{0\downarrow}\gamma_{2\downarrow}})\mbox{Re}[P_{21}(t)] \nonumber \\
&+\Omega/2(\sqrt{\gamma_{0\downarrow}\gamma_{1\uparrow}}+\sqrt{\gamma_{0\downarrow}\gamma_{1\downarrow}})\mbox{Re}[P_{21}(t)] \, .\nonumber
\end{align}
It is straightforward to check that the first law is satisfied at the level of the rates. The rates in the strong driving regime are given in the next section, on the Floquet equation.

\subsubsection{Strong qubit-laser coupling: Floquet ME}\label{subsec:floquet-av}

We consider here the strong driving regime, defined in \eqref{eq:regimes-2}. 
In this case, the product of sinc functions \eqref{eq:approx-gob} is non zero only in the case $\alpha=\alpha'$, which is equivalent to the secular approximation \cite{breuer2002theory}. Performing the secular approximation on \eqref{eq:gob}, we obtain
%
%
(see appendix~\ref{app:tilted-fme} for the full expression)
\begin{align}
 \frac{d\hat{\rho}^{\boldsymbol{\lambda}}_{DA}}{dt} &= \mathcal{L}_{\boldsymbol{\lambda}}^{aF}(\hat{\rho}^{\boldsymbol{\lambda}}_{DA}) \label{eq:fme-2}\\
 &=-i[\hat{H}_{DA},\hat{\rho}^{\boldsymbol{\lambda}}_{DA}]+\mathcal{D}^{aF}_{\boldsymbol{\lambda}}(\hat{\rho}^{\boldsymbol{\lambda}}_{DA})\, . \nonumber
\end{align}
We call this equation ``autonomous Floquet'' master equation, denoted by the superscript ``aF'', since it is equivalent to the Floquet master equation -- traditionally used in the non-autonomous picture -- which we will derive in section~\ref{sec:cons-obe-fme}. 
The dissipator in \eqref{eq:fme-2} has three dissipation channels, corresponding to the three frequencies of the Mollow triplet $\omega_L, \omega_L\pm\Omega$; we give here its expression the counting fields are set to zero $\boldsymbol{\lambda=0}$, 
\begin{align}
\mathcal{D}^{aF} \equiv & \mathcal{D}^{aF}_+ +\mathcal{D}^{aF}_-\label{eq:diss-af} \\
\mathcal{D}^{aF}_+ =& \gamma_{0,\downarrow}\mathcal{D}_{\hat{\Sigma}_z}+\gamma_{1,\downarrow}\mathcal{D}_{\hat{\Sigma}_-}+\gamma_{2,\uparrow}\mathcal{D}_{\hat{\Sigma}_+} \, ,\nonumber\\
\mathcal{D}^{aF}_- =& \gamma_{0,\uparrow}\mathcal{D}_{\hat{\Sigma}^\dagger_z}+\gamma_{1,\uparrow}\mathcal{D}_{\hat{\Sigma}_+}+\gamma_{2,\downarrow}\mathcal{D}_{\hat{\Sigma}_-} \, .\nonumber
\end{align}
Since the autonomous Floquet equation is the restriction of the generalized Bloch equation to the strong driving regime, $\mathcal{L}^{aF}_{\boldsymbol{\lambda}}$ satisfies the condition \eqref{eq:detailed-2} and the strict energy conservation condition \eqref{eq:strict-qme} (this can also be seen directly from the explicit expression given in appendix~\ref{app:tilted-fme}), and is therefore fully thermodynamically consistent. 
%
%

Moreover, the dissipator in \eqref{eq:fme-2} satisfies the symmetry 
\begin{equation}
    \mathcal{D}^{aF\dagger}_{0,\lambda_{DL},\lambda_B+i\beta_B}(...) = \mathcal{D}^{aF}_{0,\lambda_{DL},\lambda_B}(...) \, .
    \label{eq:sym-diss}
\end{equation}
Since the steady state moment generating function, $\mathcal{G}^{ss}(\lambda_{DL},\lambda_B)\equiv\lim_{t\to +\infty}\frac{1}{t}\mathcal{G}(0,\lambda_{DL},\lambda_B,t)$, is given by the dominant eigenvalue of $\mathcal{D}_{0,\lambda_{DL},\lambda_B}$ \cite{cuetara2015rapid}, the identity \eqref{eq:sym-diss} implies the steady-state work fluctuation theorem
\begin{equation}
    \frac{p(W_{DL})}{p(-W_{DL})}\asymp e^{-\beta_B W_{DL}} \, .
    \label{eq:ft-ss}
\end{equation}

We now derive the explicit expressions of the work $W_{DL}$, dressed qubit energy $\Delta E_{DA}$ and heat, by taking the derivatives $\lambda_{DL}, \lambda_{DA}$ and $\lambda_B$ in the trace of $\mathcal{L}_{\boldsymbol{\lambda}}^{aF}(\hat{\rho}_{DA}^{\boldsymbol{\lambda}})$. We obtain
\begin{align}
\dot{W}_{DL} =& \omega_L(\gamma_{0,\downarrow}-\gamma_{0,\uparrow}) \label{eq:w-fme} \\
&+\omega_L[(\gamma_{1,\downarrow}-\gamma_{2,\downarrow})P_2(t) -(\gamma_{1,\uparrow}-\gamma_{2,\uparrow})P_1(t) ] ,
\nonumber
\end{align}
\begin{equation}
d_t E_{DA}= \Omega[(\gamma_{1,\uparrow}+\gamma_{2,\uparrow})P_1(t) - (\gamma_{1,\downarrow}+\gamma_{2,\downarrow})P_2(t)], \label{eq:dea-fme}
\end{equation}
and
\begin{align}
\dot Q =& \omega_L(\gamma_{0\uparrow}-\gamma_{0\downarrow}) \label{eq:q-fme}\\
&+\left[ \omega_L(\gamma_{1\uparrow}-\gamma_{2\uparrow})+\Omega(\gamma_{2\uparrow}+\gamma_{1\uparrow}) \right] P_1(t) \nonumber \\
&+\left[ \omega_L(\gamma_{2\downarrow}-\gamma_{1\downarrow})-\Omega(\gamma_{2\downarrow}+\gamma_{1\downarrow}) \right] P_2(t) \nonumber.
\end{align}
Notice that these expressions are equal to the rates \eqref{eq:wdl-gob}, \eqref{eq:q-gob} and \eqref{eq:eda-gob} without the coherent terms $P_{21}(t)$, which is a consequence of the secular approximation.

We also point out that the rate of the heat \eqref{eq:q-fme} is consistent with the results obtained in \cite{langemeyer2014,elouard2015}, and the rate of the work \eqref{eq:w-fme} is consistent with the expression derived in \cite{elouard2020thermodynamics} in the case of a non-autonomous description of the laser. However, it was then assumed that \eqref{eq:w-fme} should correspond to the work performed by the laser (and not the dressed laser), which seemed in contradiction with quantum thermodynamics, according to which that work is expected to be of the form \eqref{eq:wl}. Our approach unveils that the work \eqref{eq:w-fme} exerted on the dressed qubit is in fact produced by the dressed laser, which partially explains the difference from the anticipated form \eqref{eq:w-fme}. The expression \eqref{eq:w-fme} also suggests that work is mediated by transitions and thus originates from a non conservative force. This was also pointed out in \cite{elouard2020thermodynamics,cuetara2015rapid}, in the non-autonomous picture. Non conservative forces typically arise when eliminating an underlying degree of freedom (here the laser), that over sufficiently long time-scales, delivers energy without being affected by it \cite{rao2018conservation}. This is manifest in the form of the rates \eqref{eq:coefs-fme} which satisfy a local detailed balance condition \cite{esposito2012coarsegraining,rao2018conservation}, which relates the log ration of the transition rates \eqref{eq:coefs-fme} to the change in entropy in the bath resulting from that transition  
\begin{equation}
  \log  \frac{\gamma_{j,\downarrow}}{\gamma_{j,\uparrow}}=\beta_B(\omega_L+l_j\Omega) \, ,
  \label{eq:db}
\end{equation}
with $j=0,1,2$ and $l_0=0, l_1=1, l_2=-1$. Interestingly, the rates \eqref{eq:coefs-fme} are the same as those appearing in the dissipator of the master equation for $\hat{\rho}_X$, denoted $\mathcal{L}^{aF,X}$ above \eqref{eq:fme-2}. At the level of the master equation for $\hat{\rho}_X$, the r.h.s. of \eqref{eq:db} is equal to (up to the factor $\beta_B$) the energy variation of the system $X$ during a transition induced by the dissipator; 
this implies that the dynamics is detailed balanced, and will relax to equilibrium. 
However, at the level of the master equation for the dressed qubit \eqref{eq:fme-2}, the r.h.s. of \eqref{eq:db} does not only contain the difference of energies of the system (i.e., $\pm\Omega$), but also the laser's energy. The fact that the term $\omega_L$ remains even after we have traced out the degrees of freedom of the laser results directly from the assumption 2. Indeed, this assumption allows to neglect the variations of the number of photons in the laser during a transition between a state of $|j(n)\rangle\rightarrow|i(n\pm1)\rangle$ ($i,j\in\{1,2\}$), and justifies the mapping \eqref{eq:phida}, which in turn  leads to the simple product structure of the operators \eqref{eq:s-ops}. This product structure then allows to trace out the dressed laser $DL$ without changing the rates in the dissipator of the master equation. 
During this procedure, the variation of the number of photons is treated as an underlying degree of freedom which is traced out and does not appear explicitly in the dynamics, but leaves a fingerprint in the thermodynamics through the term $\omega_L$ in \eqref{eq:db}, which is at the origin of the non conservative force \eqref{eq:w-fme}. As a consequence, the steady state solution of $\mathcal{L}^{aF}$ is a non equilibrium steady state, as can be checked by noticing that the entropy production rate is strictly positive in the steady state (see appendix~\ref{app:ss-fme}). 

\subsubsection{Weak and intermediate coupling: Bloch ME}\label{subsec:obe-av}

We now consider the weak driving regime defined in \eqref{eq:regimes-2}. 
Provided that $G_\pm(\nu)$ are smooth on the intervals $[\pm\omega_L-\Omega,\pm\omega_L+\Omega]$, we may replace
\begin{equation}
    G_\pm(\omega_L), G_\pm(\omega_L\pm\Omega)\approx G_\pm(\omega_A)\equiv\overline{G}_\pm
    \label{eq:approx-obe}
\end{equation}
in \eqref{eq:gob}, which yields the tilted master equation $d_t\hat{\rho}_{DA}^{\boldsymbol{\lambda}}(t)\equiv\mathcal{L}_{\boldsymbol{\lambda}}^{aB}(\hat{\rho}_{DA}(t) )$ (see Appendix~\ref{app:tilted-obe} for the explicit expression). When the counting fields are set to zero, it is equal to
\begin{align}
&\frac{d\hat{\rho}_{DA}(t)}{dt} =\mathcal{L}^{aB}(\hat{\rho}_{DA}(t)) \label{eq:obe-2}\\
&= -i[\hat{H}_{DA},\hat{\rho}_{DA}(t)]+ \overline{G}_+\mathcal{D}_{\hat{\sigma}_-}(\hat{\rho}_{DA}(t))+\overline{G}_-\mathcal{D}_{\hat{\sigma}_+}(\hat{\rho}_{DA}(t)) \, , \nonumber
\end{align}
where we recall that $\hat{\sigma}_+=|b\rangle\langle a|=\hat{\sigma}_-^\dagger$.
%
%
The regime of validity of \eqref{eq:obe-2} can be extended to the intermediate driving regime in \eqref{eq:regimes-2}, since, in this regime, the condition $\delta_0>\Omega^{-1}$ is still satisfied. 
The notation $aB$ stands for autonomous Bloch equation, since we will see, in section~\ref{sec:cons-obe-fme}, that \eqref{eq:obe-2} is equivalent to the optical Bloch master equation \cite{grynberg1998, elouard2020thermodynamics}, usually derived in the non-autonomous picture.

It is straightforward to check that $\mathcal{L}^{aB}_{\boldsymbol{\lambda}}$ (see appendix~\ref{app:tilted-obe}) satisfies the condition \eqref{eq:detailed-2}. This implies the second law, at the level of the rates \cite{soret2022}, 
\begin{equation}
    d_t S_{DA}-\beta_B\dot Q\geq 0 \, .
\end{equation}
Setting $\lambda_{DA}=\lambda_{DL}=\lambda_B=\lambda$, we find that $\mathcal{L}^{aB}_{\boldsymbol{\lambda}}$ is not $\lambda$ independent, which means that the strict energy conservation \eqref{eq:strict-qme} is not satisfied, hence that the first law of thermodynamics does not hold at the fluctuating level. 

We obtain $\dot Q,\dot W_{DL}$ and $d_t E_{DA}$ by taking the derivatives in $\lambda_{B}, \lambda_{DL}$ and $\lambda_{DA}$ of $\text{Tr}[\mathcal{L}^{aB}_{\boldsymbol{\lambda}}(\hat{\rho}_{DA})]$,
\begin{align}
\dot{Q} =& -\omega_A(\overline{G}_+P_{b}(t)-\overline{G}_-P_{a}(t))  \label{eq:heat-obe} \\
&-  \frac{g}{2}(\overline{G}_++\overline{G}_-)Re(P_{ab}(t))\nonumber\\
\dot{W}_{DL} =& \omega_L(\overline{G}_+P_{b}(t)-\overline{G}_-P_{a}(t)) \label{eq:wdl-obe}\\
d_t E_{DA} =& -\delta(\overline{G}_+P_{b}(t)-\overline{G}_-P_{a}(t))  \label{eq:eda-obe} \\
&- \frac{g}{2}(\overline{G}_++\overline{G}_-)Re(P_{ab}(t)) \, ,
 \nonumber
\end{align}
with 
\begin{align}
P_{b}(t)&\equiv  \langle b|\hat{\rho}_{DA}(t)|b\rangle    \nonumber   \\
&=\frac{1}{2}+  \frac{\delta}{2\Omega}[P_2(t)-P_1(t)] - \frac{g}{\Omega}\text{Re}[P_{21}(t)] \nonumber \\
P_{a}(t)&\equiv \langle a|\hat{\rho}_{DA}(t)|a\rangle \nonumber\\
&= \frac{1}{2}-  \frac{\delta}{2\Omega}[P_2(t)-P_1(t)] \frac{g}{\Omega}+\text{Re}[P_{21}(t)] \nonumber\\
P_{ba}(t)&\equiv  \langle b|\hat{\rho}_{DA}(t)|a\rangle \nonumber  \\
&=\frac{g}{2\Omega}(P_2(t)-P_1(t)) + \frac{\delta}{\Omega}\text{Re}[P_{21}(t)] + i \text{Im}[P_{21}(t)] \nonumber\\
P_{ab}(t)&\equiv \langle a|\hat{\rho}_{DA}(t)|b\rangle   \, .
    \label{eq:pop-obe}
\end{align}
One can check, using the identity \eqref{eq:id-s}, that the above expressions are consistent with \eqref{eq:wdl-gob}, \eqref{eq:q-gob} and \eqref{eq:eda-gob}. 
We also check that the first law is satisfied,
\begin{equation}
    d_t E_{DA} =\dot{Q}+\dot{W}_{DL} \, . \label{eq:1st-law-obe-1}
\end{equation}

\subsection{Non-autonomous qubit }\label{subsec:na-qme}
 
As showed in section~\ref{subsubsec:floquet-rot}, and summarized in Fig.\ref{fig:recap},  the evolution of the dressed qubit (in the autonomous description) is equivalent to the evolution of the qubit in the rotating frame (in the non-autonomous description). This remains true at the level of master equations. Specifically, the non-autonomous counterparts of the generalized Bloch equation \eqref{eq:gob} and of the autonomous Bloch \eqref{eq:obe-2} and Floquet \eqref{eq:fme-2} master equations can be obtained simply by using the correspondence \eqref{eq:floquet-states}. This leads to the following relation between the autonomous master equations and there non-autonomous versions,
\begin{align}
& \tilde{\mathcal{L}}^{G,F,B}(\hat{\tilde{\rho}}_A(t))=e^{-i\omega_L\hat{\sigma}_zt/2}
\label{eq:qme-rot} \\
 &\left\{\mathcal{L}^{aG,aF,aB}[\hat{\rho}_{DA}(t)]-i\frac{\omega_L}{2}[\hat{\sigma}_z,\hat{\rho}_{DA}(t)]\right\}e^{i\omega_L \hat{\sigma}_z t/2}\nonumber \, ,
\end{align}
where we use ``G'' for generalized Bloch, ``F'' for Floquet and ``B'' for optical Bloch. See Fig.~\ref{fig:recap} for a summary of the correspondences. 

We now examine the thermodynamics in the qubit picture, similarly as in section~\ref{sec:uni} where we focused on the dressed qubit picture. We therefore derive tilted master equations with counting fields on the laser and bath, $\boldsymbol{\lambda} = (\lambda_L,\lambda_B)$. We leave out the qubit part, since measuring $\hat{H}_A+\hat{V}(t)$ is difficult in practice and deriving a tilted master equation with a counting field on $\hat{H}_A+\hat{V}(t)$ is technically cumbersome. The non-autonomous master equations with counting fields $\boldsymbol{\lambda} = (\lambda_L,\lambda_B)$ can be obtained using the substitutions $\hat{s}_{z,\pm}^{\boldsymbol{\lambda}}\rightarrow \hat{s}_{z,\pm}^{\boldsymbol{\lambda}}(t)$, where
\begin{align}
  &  \hat{s}_z^{\boldsymbol{\lambda}}(t)\equiv \frac{g}{2\Omega}e^{-i(\lambda_L-\lambda_B)\omega_L/2}\hat{\Sigma}_z\left(t+\frac{\lambda_L}{2}\right) \label{eq:slambda-na} \\
  &   \hat{s}_+^{\boldsymbol{\lambda}}(t) \equiv - \frac{\Omega-\delta}{2\Omega}e^{-i\lambda_L\omega_L/2}e^{i\lambda_B(\omega_L-\Omega)/2}\hat{\Sigma}_+\left(t+\frac{\lambda_L}{2}\right) \nonumber \\
  &   \hat{s}_-^{\boldsymbol{\lambda}}(t) \equiv \frac{\Omega+\delta}{2\Omega}e^{-i\lambda_L\omega_L/2}e^{i\lambda_B(\omega_L+\Omega)/2}\hat{\Sigma}_-\left(t+\frac{\lambda_L}{2}\right) \nonumber
\end{align}
with
\begin{align}
    \hat{\Sigma}_z(t) &\equiv |u_2(t)\rangle\langle u_2(t)|- |u_1(t)\rangle\langle u_1(t)| \label{eq:floquet-pauli}\\
    \hat{\Sigma}_+(t) &\equiv |u_2(t)\rangle\langle u_1(t)|\nonumber\\
    \hat{\Sigma}_-(t) &\equiv |u_1(t)\rangle\langle u_2(t)| \, .\nonumber
\end{align}
To see this, let's treat $\hat{H}_L,\hat{H}_B$ separately. The tilted master equation dressed with the counting field $\lambda_B$ can be obtained directly from the autonomous equations \eqref{eq:gob}, \eqref{eq:fme-2} and \eqref{eq:obe-2}, by setting $\boldsymbol{\lambda}=(0,0,\lambda_B)$ in those equations and using the identity \eqref{eq:qme-rot}. For $\hat{H}_L$, we use the identity $\hat{H}_L=\hat{H}_{DL}-\frac{\omega_L}{2}\hat{\sigma}_z$. Since $[\hat{H}_{DL},\frac{\omega_L}{2}\hat{\sigma}_z]=0$, we may measure them separately, using counting fields $\lambda_{DL},\lambda_{\sigma}$; then, setting $\lambda_L=\lambda_{DL}=-\lambda_{\sigma}$ yields the tilted master equation dressed with the counting field $\lambda_L$ for $\hat{H}_L$. The convenience of this approach is that, since $\frac{\omega_L}{2}\hat{\sigma}_z$ only acts on the Hilbert space of the qubit, the terms $e^{i\lambda_{\sigma}\omega_L\hat{\sigma}/2}$ can be factorized out of the Kraus operators when tracing out the Hilbert spaces $\mathcal{H}_L,\mathcal{H}_B$, similarly as the situation in \eqref{eq:kraus-da} (but for $\mathcal{H}_{DL},\mathcal{H}_B$). The exponential term $e^{-i\omega_L\lambda_L}$ comes from counting $\hat{H}_{DL}$, while the shifts in the operators $\hat{\Sigma}_z^F(t),\hat{\Sigma}^F_\pm(t)$ are due to counting $\frac{\omega_L}{2}\hat{\sigma}_z$.

\subsubsection{Generalized Bloch equation}

Using \eqref{eq:gob}, \eqref{eq:qme-rot} and \eqref{eq:slambda-na}, we obtain the generalized Bloch equation with counting fields $\boldsymbol{\lambda} = (\lambda_L,\lambda_B)$,
\begin{align}
 &   \mathcal{L}_{\boldsymbol{\lambda}}^{G}(\hat{\tilde{\rho}}_A(t)) = \label{eq:gob-tilted-na}\\
 &   -i[\hat{H}_A(t+\lambda_L/2)\hat{\tilde{\rho}}_A(t)-\hat{\tilde{\rho}}_A(t)\hat{H}_A(t-\lambda_L/2)]  \nonumber\\
 &+\mathcal{D}_+^{G\boldsymbol{\lambda}}(\hat{\tilde{\rho}}_A(t))+\mathcal{D}_-^{G\boldsymbol{\lambda}}(\hat{\tilde{\rho}}_A(t))
\end{align}
where we note 
\begin{equation}
    \hat{H}_A(t)\equiv \hat{H}_A+\hat{V}(t) \, .
\end{equation}
In the weak driving regime,
\begin{widetext}
\begin{align}
   & \mathcal{D}_+^{G\boldsymbol{\lambda}}(\hat{\rho}) = \hat{T}_+^{\boldsymbol{\lambda}}(t)\hat{\rho} \hat{T}_+^{\boldsymbol{-\lambda}\dagger}(t) - \frac{1}{2}\left(\hat{T}_+^{(\lambda_{L},0) \dagger }(t)\hat{T}_+^{(\lambda_{L},0) }(t)\hat{\rho} +\hat{\rho}\hat{T}_+^{(-\lambda_{L},0) \dagger }(t)\hat{T}_+^{(-\lambda_{L},0) }(t)\right)\nonumber \\
   \nonumber\\
   & \mathcal{D}_-^{G\boldsymbol{\lambda}}(\hat{\rho}) = \hat{T}_-^{\boldsymbol{\lambda}\dagger}(t)\hat{\rho} \hat{T}_-^{\boldsymbol{-\lambda}}(t) - \frac{1}{2}\left(\hat{T}_-^{(\lambda_{L},0)  }(t)\hat{T}_-^{(\lambda_{L},0)\dagger }(t)\hat{\rho} +\hat{\rho}\hat{T}_-^{(-\lambda_{L},0 )}\hat{T}_-^{(-\lambda_{L},0 )\dagger}(t)\right)\nonumber 
\end{align}
\end{widetext}
with
\begin{align}
    \hat{T}_\pm^{\boldsymbol{\lambda}}(t) = \sum_{\alpha=+,-,z} \sqrt{G}_\pm(\omega_\alpha)\hat{s}_\alpha^{\boldsymbol{\lambda}}(t) \, .
\end{align}
The expression of the dissipators in the strong driving regime corresponds to the Floquet equation, given in the next section in \eqref{eq:diss-f} and Appendix~\ref{app:tilted-fme-2}. The expression in the intermediate driving regime is obtained from the autonomous generalized Bloch equation in that regime, given in Appendix~\ref{app:gob}, by using \eqref{eq:qme-rot} and \eqref{eq:slambda-na}.

The heat is the same as in the autonomous case, and the rate of the work $\dot W_L$ is given in the appendix~\ref{app:wl-gob}.

\subsubsection{Floquet master equation}

The strong coupling limit in the non-autonomous description leads to the Floquet quantum master equation \cite{langemeyer2014,elouard2020thermodynamics}. We obtain it from \eqref{eq:fme-2}, using the correspondence \eqref{eq:qme-rot},
\begin{equation}
    \mathcal{L}^F(\hat{\tilde{\rho}}_A(t)) = -i[\hat{H}_A+\hat{V}(t),\hat{\tilde{\rho}}_A(t)]+\mathcal{D}^F(\hat{\tilde{\rho}}_A(t))
\end{equation}
where the dissipator is obtained by replacing the operators $\hat{\Sigma}_z,\hat{\Sigma}_\pm$ in \eqref{eq:diss-af} by $\hat{\Sigma}_z(t),\hat{\Sigma}_\pm(t)$, defined in \eqref{eq:floquet-pauli}.

The tilted Floquet master equation with counting fields $\boldsymbol{\lambda}=(\lambda_L,\lambda_B)$ is obtained by performing the secular approximation in \eqref{eq:gob-tilted-na},
\begin{align}
   & \mathcal{L}^F_{\boldsymbol{\lambda}}(\hat{\tilde{\rho}}_A(t))= \label{eq:fme-3}\\
   &-i[\hat{H}_A(t+\lambda_L/2)\hat{\tilde{\rho}}_A(t)-\hat{\tilde{\rho}}_A(t)\hat{H}_A(t-\lambda_L/2)] \nonumber \\
    &+\mathcal{D}^F_{\boldsymbol{\lambda}}(\hat{\tilde{\rho}}_A(t)) \nonumber
\end{align}
The dissipator $\mathcal{D}^F_{\boldsymbol{\lambda}}$ is decomposed as 
\begin{align}
\mathcal{D}^{F}_{\boldsymbol{\lambda}}\equiv & \mathcal{D}^{F}_{\boldsymbol{\lambda},+} +\mathcal{D}^{F}_{\boldsymbol{\lambda}-} \label{eq:diss-f} \\
\mathcal{D}^{F}_{\boldsymbol{\lambda}+} =& \gamma_{0,\downarrow}\mathcal{D}_{0,+}^{F\boldsymbol{\lambda}}+\gamma_{1,\downarrow}\mathcal{D}_{1,+}^{F\boldsymbol{\lambda}}+\gamma_{2,\uparrow}\mathcal{D}_{2,+}^{F\boldsymbol{\lambda}} \, ,\nonumber\\
\mathcal{D}^{F}_{F\boldsymbol{\lambda}-}=& \gamma_{0,\uparrow}\mathcal{D}_{0,-}^{F\boldsymbol{\lambda}}+\gamma_{1,\uparrow}\mathcal{D}_{1,-}^{F\boldsymbol{\lambda}}+\gamma_{2,\downarrow}\mathcal{D}_{2,-}^{F\boldsymbol{\lambda}} \, .\nonumber
\end{align}
The full dissipator is given in appendix~\ref{app:tilted-fme-2}. It is straightforward to check that $\mathcal{L}^F_{\boldsymbol{\lambda}}$ satisfies the symmetries \eqref{eq:sym-qme-na} and \eqref{eq:strict-a}. The dissipator also satisfies the symmetry
\begin{equation}
    \mathcal{D}^{F\dagger}_{0,\lambda_{L},\lambda_B+i\beta_B}(...) = \mathcal{D}^{F}_{0,\lambda_{L},\lambda_B}(...) \, ,
    \label{eq:sym-diss-2}
\end{equation}
which, as explained under \eqref{eq:ft-ss}, implies the following steady-state work fluctuation theorem, this time for $W_L$,
\begin{equation}
    \frac{p(W_{L})}{p(-W_{L})}\asymp e^{\beta_B W_{L}} \, .
    \label{eq:ft-ss-2}
\end{equation}
We may now obtain the rate of the work $\dot W_L$,
\begin{align}
    \dot W_L^F \equiv &-\frac{1}{i}\partial_{\lambda_L} \text{Tr}[\mathcal{L}^F_{\boldsymbol{\lambda}}(\hat{\tilde{\rho}}_A(t))]|_{\boldsymbol{\lambda=0}} \label{eq:wl-fme}\\
=& \dot W_{DL}+\text{Tr}[\frac{\omega_L}{2}\hat{\sigma}_z\mathcal{L}^F(\hat{\tilde{\rho}}_A(t))] \nonumber \\
=&\text{Tr}[d_t\hat{V}(t)\hat{\tilde{\rho}}_A(t))] +\omega_L(\gamma_{0\downarrow}-\gamma_{0\uparrow})\nonumber\\
&+\omega_L\frac{g}{\Omega}\text{Re}[P_{21(t)}]\left(\gamma_{0\downarrow}+\gamma_{0\uparrow}+\frac{\gamma_{1\downarrow}+\gamma_{1\uparrow}+\gamma_{2\downarrow}+\gamma_{2\uparrow}}{2}\right)\nonumber \\
&+\omega_L P_1(t)\left[\gamma_{2\uparrow} - \gamma_{1\uparrow} +\frac{\delta}{\Omega}(\gamma_{2\uparrow} + \gamma_{1\uparrow})\right]\nonumber \\
&-\omega_L P_2(t)\left[\gamma_{2\downarrow} - \gamma_{1\downarrow} +\frac{\delta}{\Omega}(\gamma_{2\downarrow} + \gamma_{1\downarrow})\right] \, .\nonumber
\end{align}
where $P_{21}(t)\equiv \langle 2|\hat{\rho}_{DA}(t)|1\rangle$ and where $P_{1,2}(t)$ were defined in \eqref{eq:p12}. Notice that $\text{Tr}[d_t\hat{V}(t)\hat{\tilde{\rho}}_A(t))]$ may be re-written as
\begin{align}
    \text{Tr}[d_t\hat{V}(t)\hat{\tilde{\rho}}_A(t))]& = -g\omega_L \text{Im}(\langle 2|\hat{\rho}_{DA}|1\rangle)\nonumber \\
    &= -g\omega_L \text{Im}(\langle b|\hat{\rho}_{DA}|a\rangle) \, ,
\end{align}
similarly to the unitary case \eqref{eq:wl-a}.

\subsubsection{Bloch master equation}\label{subsec:bloch-semiclass}

In the non-autonomous picture, the weak/intermediate driving regimes correspond to the regimes of validity of the optical Bloch master equation \cite{grynberg1998,elouard2020thermodynamics}. Using \eqref{eq:qme-rot} with \eqref{eq:obe-2}, we obtain the optical Bloch master equation \cite{grynberg1998}
\begin{align}
    \mathcal{L}^{B}(\hat{\tilde{\rho}}_A(t))=&-i[\hat{H}_A+\hat{V}(t),\hat{\tilde{\rho}}_A(t)]\\
    &+\overline{G}_+\mathcal{D}_{\hat{\sigma}_-}(\hat{\tilde{\rho}}_A(t))+\overline{G}_-\mathcal{D}_{\hat{\sigma}_+}(\hat{\tilde{\rho}}_A(t)) \, . \nonumber
\end{align}
From the discussion in section~\ref{subsec:obe-av}, we deduce that the Bloch master equation satisfies the symmetry \eqref{eq:sym-qme-na} as well as the first law \eqref{eq:1st-law-a-da}.

The optical Bloch master equation dressed with the counting fields $\boldsymbol{\lambda}=(\lambda_L,\lambda_B)$ is obtained by performing the approximation $G_\pm(\nu)\rightarrow \overline{G}_\pm$ in \eqref{eq:gob-tilted-na},
\begin{align}
\frac{d\hat{\rho}_A^{\boldsymbol{\lambda}}}{dt} &\equiv \mathcal{L}^{B}_{\boldsymbol{\lambda}}(\hat{\tilde{\rho}}_A) \label{eq:tilted-obe-na-2}\\
&=-i[\tilde{H}_A+\hat{V}_{\lambda_L}(t+\lambda_L/2),\hat{\tilde{\rho}}_A]\\
&+\overline{G}_+\mathcal{D}^{\lambda_B}_{\hat{\Sigma}}(\hat{\tilde{\rho}}_A)+\overline{G}_-\mathcal{D}^{\lambda_B}_{\hat{\Sigma}^\dagger}(\hat{\tilde{\rho}}_A)\nonumber \, ,
\end{align}
where $\mathcal{D}^{\lambda_B}$ is in fact equal to by setting to the dissipator of $\mathcal{L}^{aB}$ with $\boldsymbol{\lambda}=(0,0,\lambda_B)$. We can then obtain the rate of the work, 
$\dot W_L = \frac{1}{i}\partial_{\lambda_{L}} \text{Tr}[\hat{\rho}_A^{\boldsymbol{\lambda}}] $, 
and we find 
\begin{equation}
 \dot W_L = \text{Tr}[d_t\hat{V}(t)\hat{\tilde{\rho}}_A(t)] = -g\omega_L \text{Im}(\langle b|\hat{\rho}_{DA}|a\rangle)  \, .
 \label{eq:wl-obe}
\end{equation}
The heat is the same as in \eqref{eq:heat-obe}; we may decompose it as
\begin{align}
     \dot{Q}
    &= \omega_A((\overline{n}+1)P_{a}(t)-\overline{n}P_{b}(t)) + \gamma gRe(P_{ab}(t))\nonumber\\
    &=  \text{Tr}[\mathcal{L}^{B}(\hat{\tilde{\rho}}_A)\hat{H}_A]+\text{Tr}[\mathcal{L}^{B}(\hat{\tilde{\rho}}_A)\hat{V}(t)] \, ,\label{eq:q-obe}
\end{align}
which then gives the first law 
\begin{align}
d_t \tilde{E}_A &= \text{Tr}[(\hat{H}_A+\hat{V}(t))\mathcal{L}^{B}(\hat{\rho}_A(t))] + \text{Tr}[d_t \hat{V}(t) \hat{\rho}_A(t)]\nonumber\\
&= \dot Q +\dot W_L \, .\label{eq:1st-law-obe-2}
\end{align}
%



\section{Quantum maps vs Redfield equation}\label{sec:red}

In the section~\ref{sec:cons-obe-fme}, we derived the Bloch and Floquet equation using the formalism of quantum maps and the Kraus operators \eqref{eq:kraus-i}. We saw that this procedure preserves the fluctuation theorems. However, the Bloch equation is usually derived using the Redfield equation \cite{redfield1965,breuer2002theory,elouard2020thermodynamics}. We show in this short section that, although both procedures result in the same equation in the absence of counting fields, the tilted master equations differ, and that the Bloch equation derived from the Redfield method breaks the symmetry \eqref{eq:detailed-2}.

\begin{figure}
    \centering
    \includegraphics[width=0.7\linewidth]{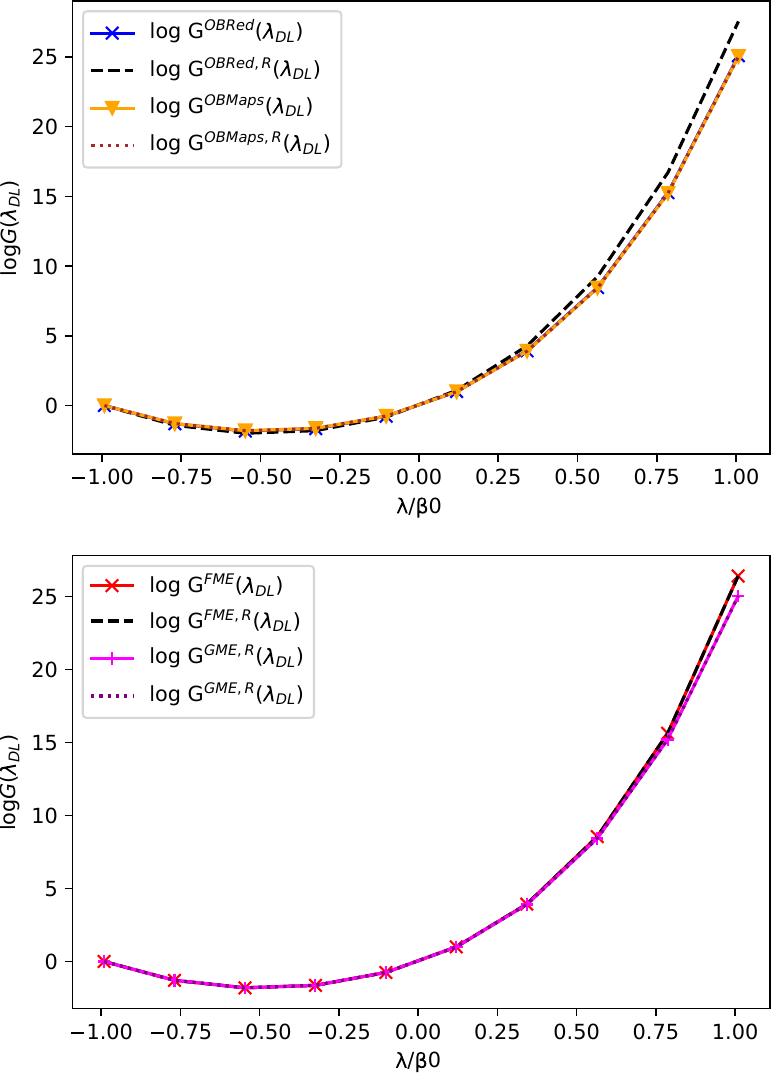}
    \caption{Work $W_{DL}$ moment generating functions for the master equations discussed. Top: the work fluctuation theorem holds for the Bloch equation derived using quantum maps, but breaks down if the Bloch equation is derived with the Redfield equation. Bottom: the work fluctuation theorem is satisfied by the Floquet master equation and the generalized master equation. Parameters: $\alpha=4$, $\beta=10/D$ with the spectral width $D=20$, $\gamma_0=0.4\sqrt{D}$. }
    \label{fig:MGF-WDL-OBE-OBEMaps}
\end{figure}

To see this, we repeat the derivation of the Bloch equation via the Redfield equation, which can be found in \cite{breuer2002theory,elouard2020thermodynamics}, but adding here the counting fields. The first step is to write the Liouville equation in the interaction picture, with counting fields, which is done by taking the time derivative of $\hat{\rho}_{\boldsymbol{\lambda}}(t)$ in \eqref{eq:mgf}, and going to the interaction picture w.r.t. $\hat{H}_X+\hat{H}_B$,  
\begin{equation}
    \frac{d\hat{\rho}^{\boldsymbol{\lambda}I}(t)}{dt} = -i\left(\hat{V}_{AB}^{\boldsymbol{\lambda}}(t)\hat{\rho}^{\boldsymbol{\lambda}I}(t)-\hat{\rho}^{\boldsymbol{\lambda}I}(t)\hat{V}_{AB}^{-\boldsymbol{\lambda}}(t)\right) \, .
    \label{eq:red-aux}
\end{equation}
The system $X$ evolves over a time scale $\sim \gamma_{max}^{-1}$. Assuming that the relaxation time $\tau_B$ of the heat bath $B$ satisfies $\tau_B\ll \gamma_{max}^{-1}$, we can coarse-grain the dynamics over a time-scale $\delta_0$ such that $\tau_B\ll\delta_0\ll \gamma_{max}^{-1}$. Integrating \eqref{eq:red-aux} over $\delta_0$ and re-injecting the solution in \eqref{eq:red-aux} yields, assuming that $[\hat{V}_{AB}(0),\hat{\rho}_{AL}^I(0)]=0$, an equation similar to the Redfield master equation,
\begin{equation}
\begin{array}{l}
\frac{d\hat{\rho}_{X}^{\boldsymbol{\lambda}}}{dt}=\\
\\
-Tr_B\left[ \frac{1}{\delta_0}\int\limits_t^{t+\delta_0} dt'\int\limits_t^{t'}du\right.\\
\\
\hat{V}_{AB}^{\boldsymbol{\lambda}/2}(t') \hat{V}_{AB}^{\boldsymbol{\lambda}/2}(u)\hat{\rho}_X^{\boldsymbol{\lambda}}(u)\otimes\hat{\rho}_B\\
\\
+\hat{\rho}_X^{\boldsymbol{\lambda}}(u)\otimes\hat{\rho}_B\hat{V}_{AB}^{-\boldsymbol{\lambda}/2}(u)\hat{V}_{AB}^{-\boldsymbol{\lambda}/2}(t') \\
\\
-\hat{V}_{AB}^{\boldsymbol{\lambda}/2}(t')\hat{\rho}_X^{\boldsymbol{\lambda}}(u)\otimes\hat{\rho}_B \hat{V}_{AB}^{-\boldsymbol{\lambda}/2}(u)\\
\\
\left.-\hat{V}_{AB}^{\boldsymbol{\lambda}/2}(u)\hat{\rho}_X^{\boldsymbol{\lambda}}(u)\otimes\hat{\rho}_B \hat{V}_{AB}^{-\boldsymbol{\lambda}/2}(t') \right] \, ,
\end{array}
\end{equation}
In the weak coupling limit, we can perform the Born-Markov approximation \cite{breuer2002theory} and replace $\hat{\rho}(u)\equiv \hat{\rho}_X(t)\otimes\hat{\rho}_B$. We then do the change of variable $\tau=t'-u$ in the second integral; since $\tau_B\ll\delta_0$, we replace the upper bound of the second integral by $+\infty$, and we obtain finally, 
\begin{equation}
\begin{array}{l}
\frac{d\hat{\rho}_{X}^{\boldsymbol{\lambda}}}{dt}=\\
\\
-Tr_B\left[ \frac{1}{\delta_0}\int\limits_t^{t+\delta_0} dt'\int\limits_0^{+\infty}d\tau\right.\\
\\
\hat{V}_{AB}^{\boldsymbol{\lambda}/2}(t') \hat{V}_{AB}^{\boldsymbol{\lambda}/2}(t'-\tau)\hat{\rho}_X^{\boldsymbol{\lambda}}(t)\otimes\hat{\rho}_B\\
\\
+\hat{\rho}_X^{\boldsymbol{\lambda}}(t)\otimes\hat{\rho}_B\hat{V}_{AB}^{-\boldsymbol{\lambda}/2}(t'-\tau)\hat{V}_{AB}^{-\boldsymbol{\lambda}/2}(t') \\
\\
-\hat{V}_{AB}^{\boldsymbol{\lambda}/2}(t')\hat{\rho}_X^{\boldsymbol{\lambda}}(t)\otimes\hat{\rho}_B \hat{V}_{AB}^{-\boldsymbol{\lambda}/2}(t'-\tau)\\
\\
\left.-\hat{V}_{AB}^{\boldsymbol{\lambda}/2}(t'-\tau)\hat{\rho}_X^{\boldsymbol{\lambda}}(t)\otimes\hat{\rho}_B \hat{V}_{AB}^{-\boldsymbol{\lambda}/2}(t') \right] \, .
\end{array}
\label{eq:gen-eq-cf}
\end{equation}
The key difference between \eqref{eq:gen-eq-cf} and the master equation \eqref{eq:gme-aux}, obtained from a perturbative expansion of \eqref{eq:l}, is the final approximation made for the Redfield: replacing the upper bound of the second integral by $+\infty$. We did not make this approximation in the section~\ref{sec:cons-qme}, see Eqs.\eqref{eq:gme-aux} to \eqref{eq:approx-sinc}. This approximation is known to break the positivity of the Redfield equation \cite{breuer2002theory}. We also showed in a previous work that it breaks a fluctuation theorem in the case of a quantum system connected to heat baths \cite{soret2022}. The same happens here, in the presence of the laser: to see this, suffices to finish the calculation of the Bloch equation, using the approximation $G_\pm(\nu)\approx\overline{G}_\pm$; details are provided in the appendix~\ref{app:redfield}. The expression of the Bloch equation with counting fields is then 
\begin{align}
    \frac{d\hat{\rho}_{DA}^{\boldsymbol{\lambda}}}{dt}&\equiv\mathcal{L}_{\boldsymbol{\lambda}}^{aB}\\
    &=-i[\hat{H}_{DA},\hat{\rho}_{DA}^{\boldsymbol{\lambda}}]+\overline{G}_+\mathcal{D}_{\hat{s}}^{\boldsymbol{\lambda}}(\hat{\rho}_{DA}^{\boldsymbol{\lambda}})+\overline{G}_-\mathcal{D}_{\hat{s}^\dagger}^{\boldsymbol{\lambda}}(\hat{\rho}_{DA}^{\boldsymbol{\lambda}}) \, , \nonumber
\end{align}
with
\begin{align}
\mathcal{D}_{\hat{s}}^{\boldsymbol{\lambda}}(\hat{\rho}) =&-\frac{1}{2}(\hat{s}_{\lambda_{DA},0,0}^\dagger \hat{s}_{\lambda_{DA},0,0}\hat{\rho}+\hat{\rho}\hat{s}_{-\lambda_{DA},0,0}^\dagger \hat{s}_{-\lambda_{DA},0,0}) \nonumber  \\
&+\frac{1}{2}e^{i\omega_L\lambda_B}\hat{s}_{\lambda_{DA},\lambda_{DL},0}\hat{\rho}\hat{s}_{\boldsymbol{-\lambda}}^\dagger \label{eq:diss-obe}\\
&+\frac{1}{2}e^{i\omega_L\lambda_B}\hat{s}_{\boldsymbol{\lambda}}\hat{\rho}\hat{s}_{-\lambda_{DA},-\lambda_{DL},0}^\dagger \nonumber \\
\mathcal{D}_{\hat{s}^\dagger}^{\boldsymbol{\lambda}}(\hat{\rho}) =&-\frac{1}{2}(\hat{s}_{\lambda_{DA},0,0} \hat{s}_{\lambda_{DA},0,0}^\dagger\hat{\rho}+\hat{\rho}\hat{s}_{-\lambda_{DA},0,0} \hat{s}^\dagger_{-\lambda_{DA},0,0}) \nonumber \\
&+\frac{1}{2}e^{-i\omega_L\lambda_B}\hat{s}_{\lambda_{DA},\lambda_{DL},0}^\dagger\hat{\rho}\hat{s}_{\boldsymbol{-\lambda}}\\
&+\frac{1}{2}e^{-i\omega_L\lambda_B}\hat{s}_{\boldsymbol{\lambda}}^\dagger\hat{\rho}\hat{s}_{-\lambda_{DA},-\lambda_{DL},0} \, . \nonumber
\end{align}
The explicit expressions in \eqref{eq:diss-obe} allow to see directly that the symmetry \eqref{eq:detailed-2} is not satisfied. See also Fig.\ref{fig:MGF-WDL-OBE-OBEMaps} for a numerical check.

The positivity can however be restored by applying the secular approximation \cite{breuer2002theory}. The secular approximation also restores the fluctuation theorems \cite{soret2022}. Note that the Floquet master equation with counting fields is identical with both methods, as a consequence of the secular approximation.

\section{Steady-state solutions}\label{sec:steady-state}

Here, we briefly discuss and compare the steady-state solutions for thermodynamics quantities predicted by the Floquet and Bloch master equations. 

First, we point out that the rates $\dot W_L$ and $\dot W_{DL}$ become equal in the steady-state: both the Floquet and Bloch master equations, we have 
\begin{equation}
    \dot W^{ss}_L = \dot W^{ss}_{DL} \, .
\end{equation}
This result is obtained by replacing the steady-state solutions of the Floquet and Bloch master equations, given in appendix~\ref{app:ss-fme} and \ref{app:ss-obe}, in \eqref{eq:w-fme} and \eqref{eq:wl-fme} (Floquet) and \eqref{eq:wdl-obe} and \eqref{eq:wl-obe} (Bloch). This result was expected, since, at the microscopic level, $W_L$ and $W_{DL}$ only differ by the expectation value of $\frac{\omega_L}{2}\hat{\sigma}_z$, which vanishes in the steady-state since it is a state variable.

\medskip

We make a second observation, that, in the common regime of validity of the Floquet and Bloch master equations, characterized by $\omega_L,\omega_A\gg\Omega\gg \gamma_{max}$, with the assumption that the spectral density $\Gamma(\nu)$ is smooth, the steady-state expectation values of heat, $\dot Q^{ss}$, and work, $\dot W_L^{ss}$, can equivalently be obtained from either equation. More precisely, using the steady-state solutions given in appendix~\ref{app:ss-fme} and \ref{app:ss-obe} in \eqref{eq:w-fme}, \eqref{eq:q-fme}, \eqref{eq:wdl-obe} and \eqref{eq:heat-obe}, we find
\begin{align}
    \frac{\dot Q^{ss, F} -\dot Q^{ss, B} }{\overline{\Gamma}\omega_L} &= -\frac{1/2}{\frac{\delta^2+\Omega^2}{g^2}(1+2\frac{\delta^2+\Omega^2}{\gamma^2(2\overline{n}+1)^2})} \\
    \frac{\dot W_L^{ss, F} -\dot W_L^{ss, B} }{\overline{\Gamma}\omega_L} &= \frac{\delta^2/g^2+\overline{\Gamma}^2(2\overline{n}+1)^2/4g^2}{1+2\frac{\delta^2}{g^2}+\overline{\Gamma}^2\frac{(2\overline{n}+1)^2}{2g^2}}
\end{align}
where the superscripts $F$ and $B$ stand respectively for Floquet and Bloch, where $\overline{\Gamma}\equiv \overline{G}_+/(\overline{n}+1)$ with $\overline{n}\equiv n_B(\omega_L)$, and where we approximated $\gamma_{0\downarrow},\gamma_{1\downarrow},\gamma_{2\uparrow}\sim \overline{G}_+/4$ and $\gamma_{0\uparrow},\gamma_{1\uparrow},\gamma_{2\downarrow}\sim \overline{G}_-/4$. Since $\omega_L,\omega_A\gg\Omega\gg \gamma_{max}$ and $\gamma_{max} = \overline{G}_+$, and since $\Omega\sim g$, we find that 
\begin{align}
    \frac{\dot Q^{ss, F} -\dot Q^{ss, B} }{\overline{\Gamma}\omega_L} &= \mathcal{O}\left(\frac{\gamma_{max}^2}{g^2}\right) \label{eq:q-ss} \\
    \frac{\dot W_L^{ss, F} -\dot W_L^{ss, B} }{\overline{\Gamma}\omega_L} &= \mathcal{O}\left(\frac{\gamma_{max}^2}{g^2}\right)  \, . \label{eq:w-ss}
\end{align}
Let's now show that variations of the order $\gamma_{max}^2/g^2$ are too small to be captured by the Floquet and Bloch master equations in the common regime of validity. Given that the master equations were obtained from a perturbative expansion, to second order, of $\hat{W}^{\boldsymbol{\lambda}I}_{\mu,\nu}(t+\delta_0,t)$ defined in \eqref{eq:kraus-i}, the accuracy of the master equations is of the order $\delta_0^2\gamma_{max}^2$. In the common regime of validity, $\delta_0$ needs to satisfy $\Omega \gg \delta_0^{-1}\gg \gamma_{max}$. Choosing for example $\delta_0^{-1}=\sqrt{\gamma_{max}\Omega}$, it follows that 
\begin{equation}
    \frac{\dot Q^{ss, F} -\dot Q^{ss, OB} }{\overline{\Gamma}\omega_L}= \frac{\dot W_L^{ss, F} -\dot W_L^{ss, OB} }{\overline{\Gamma}\omega_L}  = o(\delta_0^2\gamma_{max}^2) \, ,
    \label{eq:equivs}
\end{equation}
hence $\dot Q^{ss, F} =\dot Q^{ss, OB}$ and $\dot W_L^{ss, F} =\dot W_L^{ss, OB}$ up to negligible corrections.

\begin{figure*}
    \centering
    \includegraphics[width=1\linewidth]{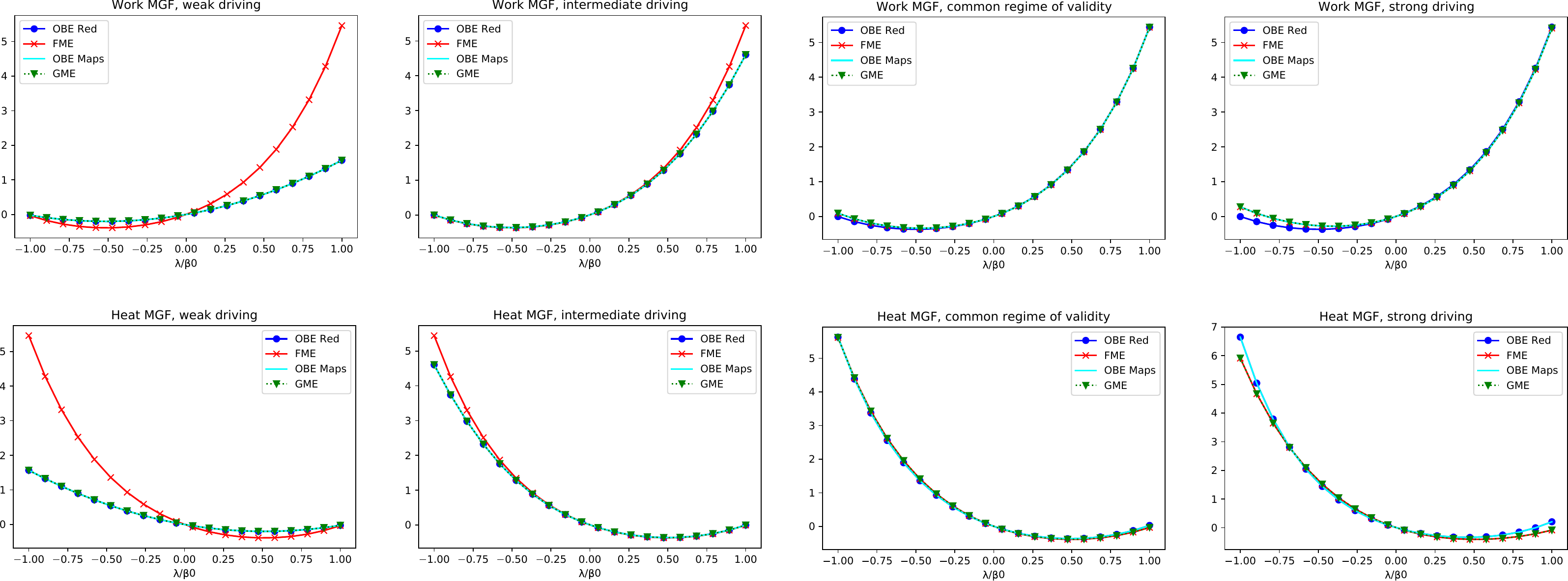}
    \caption{Steady-state moment generating functions for the work $W_{DL}$ (top) and heat $Q$ (bottom), for increasing values of the drive (from left to right). The parameters are $D=1000$, $\beta=100/D$ and $\gamma_0=0.1\sqrt{D}$, $\delta=10^{-8}D$ and $\omega_A=0.02 D$. The values of the laser-qubit coupling are such that: $g/\gamma_{max} = 0.8$ for the weak driving,  $g/\gamma_{max} = 8$ for the intermediate regime, $g/\gamma_{max} = 800$ for the common regime of validity and $g/\gamma_{max} = 2000$ for the strong driving regime.}
    \label{fig:all-mgf}
\end{figure*}

We highlight that the equivalences \eqref{eq:equivs} assume that the spectral density $\Gamma(\nu)$ is ``smooth enough'' on the interval $[\omega_A-\Omega,\omega_A+\Omega]$. If this was not the case, we expect the Floquet and Bloch master equation to predict different rates for $\dot W_L$. 

We conclude with a few plots of the heat and work moment generating functions, from times $t=0$ to the steady-state, for different values of the driving strength, see Fig.~\ref{fig:all-mgf}. The initial density matrix is $|b\rangle\langle b|$. We observe that, in the weak and intermediate driving regimes, the generalized Bloch equation coincides with the Bloch equations (derived whether from the Redfield equation or from the Kraus operators), but not with the Floquet master equation. In the common regime of validity, all the master equations give the same result, except at large $\lambda_{DL}$ where the Redfield Bloch equation slightly diverges. In the strong drive regime, the generalized Bloch equation matches instead with the Floquet master equation.

\section{Summary}\label{sec:summary}

In the sections \ref{sec:cons-qme} and \ref{sec:cons-obe-fme}, we developped a toolbox for deriving quantum master equations for coherently driven systems. In this section, we briefly sum up the results of most practical use.

\begin{itemize}
\item The generalized Bloch equation is valid at all driving regimes, thermodynamically consistent [satisfies the symmetries (\ref{eq:detailed-2}, \ref{eq:sym-qme-na}) and the laws of thermodynamics on average], and fully consistent in the strong drive regime. The Floquet master equation is valid at strong drives ($\Omega\gg\gamma_{max}$) and is fully consistent. The Bloch equation, derived from the maps, is valid at weak ($\Omega<\gamma_{max}$) and intermediate ($\Omega\sim\gamma_{max}$) drives and in the common regime of validity ($\omega_A,\omega_L\gg\Omega\gg\gamma_{max}$), and   satisfies the symmetries (\ref{eq:detailed-2}, \ref{eq:sym-qme-na}) and the laws of thermodynamics on average, but not the strict energy conservation. See Table.~\ref{table2} for a summary.
    \item At the unitary level and at the level of quantum master equations for the qubit, the evolution of the dressed qubit (in the autonomous description) is equivalent to the evolution in the rotating frame (in the non-autonomous description). See Fig.~\ref{fig:recap} for a summary of the unitary operations connecting the autonomous and non-autonomous pictures.
    \item The work source for the dressed qubit is the dressed laser; the work source for the qubit is the laser. The laws of thermodynamics in both approaches are summarized in Fig.~\ref{fig:summary}. 
    \item In their common regimes of validity, the Bloch and Floquet master equations predict similar steady-state thermodynamics, on average (\ref{eq:q-ss}, \ref{eq:w-ss}), and at the level of moment generating functions, see Fig. \ref{fig:all-mgf}.
\end{itemize}

\section{Conclusion}\label{sec:concl}

In this work, we analyzed the thermodynamics of a qubit interacting with a coherent, macroscopic, electromagnetic field, in the weak, intermediate and strong \eqref{eq:regimes-2} driving regimes. We point out that our method can be readily extended to $d-$level qubits and collective sets on qubits.  A summary of the result is presented in the Fig.~\ref{fig:summary} and Table~\ref{table2}. 
We derived two new symmetries \eqref{eq:ft-2} and \eqref{eq:ft-3}, which serve as criteria of thermodynamic consistency for quantum master equations, and translate into the symmetries \eqref{eq:detailed-2} and \eqref{eq:sym-qme-na} at the level of the master equations. We derived a new master equation, the generalized Bloch equation, valid at all drive regimes and satisfying \eqref{eq:detailed-2} and \eqref{eq:sym-qme-na}. The generalized Bloch equation also satisfies the strict energy conservation condition at strong drives, making it fully consistent in this limit. The Floquet master equation corresponds to the restriction of the generalized Bloch equation in the strong drive regime, while the Bloch master equation is obtained by performing an additional approximation in the weak/intermediate driving regimes, which preserve the symmetries \eqref{eq:detailed-2} and \eqref{eq:sym-qme-na}. We also pointed out the importance of using quantum maps rather than the Redfield equation when deriving master equations, since the Redfield equation breaks the symmetries \eqref{eq:detailed-2} and \eqref{eq:sym-qme-na}.

The present work could be useful for assessing the energy cost of qubit manipulation using coherent light sources, in the spirit of \cite{huard2022}. 
Furthermore, our findings are relevant in the context 
designing and optimizing autonomous heat engines \cite{perarnau2015,polini2018, mazzoncini2023} using far from equilibrium states of radiation as work sources, in the spirit of \cite{cavina2023tba}. 
Our findings could also be relevant for studying energy fluctuations in hybrid opto-mechanical systems \cite{arcizet2011,elouard2015} where the development of the precision of such systems \cite{poizat2021} might allow one to measure work fluctuations directly. 
More generally, our framework could be easily adapted to models in low temperature solid state physics, to study the interaction between phonons and defects of the material (often modelled as two-level systems), which reproduces Jaynes-Cummings like physics \cite{enss2005low}.
In the context of work measurement schemes in cold atoms systems, our findings complements the method developed by \cite{roncaglia2014,dechiara2015} using coherent light as a probe to reconstruct the work statistics from homodyne detection: in these works, the laser is solely seen as a probe, and the work transferred by the laser is not taken into account. Applying our results to these schemes could yield a complete thermodynamic description of work measurement in cold atoms setups. 
Finally, our results should motivate further investigation of the thermodynamics of non equilibrium steady states generated when coherent light drives a system out of equilibrium. 
Such states have been studied in mesoscopic physics, in setups where coherent light propagates through random scattering media \cite{Spivak87,soret2020fif,soret2021qmur}, and have been shown to yield fluctuation induced forces, but a thorough thermodynamic description of these features is still lacking.  

\begin{acknowledgments}
This work was supported by the Luxembourg National Research Fund, Project ThermoQO C21/MS/15713841. We also acknowledge fruitful discussions with Cyril Elouard.
\end{acknowledgments}

\appendix
\widetext 

\newpage

\newpage

\section{Laser as a work source}\label{app:work-source}



Assume that $\hat{\rho}_L(0)$ is either a coherent state or a Poisson state, and that the final state is not far from $\hat{\rho}_L(0)$; we write it in the form 
 \begin{equation}
 \hat{\rho}_L(t) = \hat{\rho}_L(0)+\epsilon\hat{r}
 \label{eq:near-coh-1}
 \end{equation}
with $\text{Tr}[\hat{r}]=0$ and $\epsilon$ small. Consider now the relative entropy 
\begin{equation}
    D[\hat{\rho}_L(t)||\hat{\rho}_L(0)]=\text{Tr}[\hat{\rho}_L(t)\log\hat{\rho}_L(t)]-\text{Tr}[\hat{\rho}_L(t)\log\hat{\rho}_L(0)] \, . \label{eq:rel-ent-coh-1}
\end{equation}
Substituting \eqref{eq:near-coh-1} in \eqref{eq:rel-ent-coh-1} and expanding the logarithm to second order in $\epsilon$, we find that 
\begin{equation}
    D[\hat{\rho}_L(t)||\hat{\rho}_L(0)] = \mathcal{O}(\epsilon^2)\, .
    \label{eq:rel-ent-0}
\end{equation}
On the other hand, we have the identity
\begin{align}
D[\hat{\rho}_L(t)||\hat{\rho}_L(0)] &= -\Delta S_L + \text{Tr}[(\hat{\rho}_L(0)-\hat{\rho}_L(t))\log \hat{\rho}_L(0)] \nonumber \\
&=  -\Delta S_L -\epsilon \text{Tr}[\hat{r}\log \hat{\rho}_L(0)] .
\label{eq:negl0}
\end{align} 
We now show that $\text{Tr}[\hat{r}\log \hat{\rho}_L(0)] $ is negligible both for a coherent state and a Poisson state. We begin with the case of a Poisson state. As discussed in section~\ref{sec:laser}, a Poisson state is equivalent to a Gibbs state at infinite temperature $\beta_L^{-1}\to +\infty$, which implies that $\text{Tr}[\hat{r}\log \hat{\rho}_L(0)]\propto -\beta_L$, hence $\text{Tr}[\hat{r}\log \hat{\rho}_L(0)]\sim 0$. 

In the case of a coherent state, the logarithm of $\hat{\rho}_L(0)$ is in fact ill defined; to fix this issue, let us introduce
\begin{equation}
    \hat{\rho}_\eta\equiv \hat{D}(\alpha)\left[|0\rangle_L\langle 0|_L +  e^{-\eta \hat{A}}\right ]\hat{D}^\dagger(\alpha) \, ,
\end{equation}
where 
\begin{equation}
    \hat{A}\equiv e^{-\eta} \sum\limits_{N\geq 1}|N\rangle_L\langle N|_L \, ,
\end{equation}
such that $\hat{\rho}_L(0)=\lim_{\eta\to +\infty}\hat{\rho}_\eta$ (recall that $\hat{D}(\alpha) |0\rangle_L\langle 0|_L \hat{D}^\dagger(\alpha) = |\alpha\rangle_L\langle\alpha|_L$). Then,
\begin{equation}
    \text{Tr}[\hat{r}\log\hat{\rho}_\eta] = \eta e^{-\eta} \langle 0| \hat{D}^\dagger(\alpha)\hat{r}\hat{D}(\alpha)|0\rangle =\mathcal{O}(\eta e^{-\eta})\, ,
\end{equation}
where we used $\text{Tr}[\hat{r}]=0$ and the fact that $|\langle 0| \hat{D}^\dagger(\alpha)\hat{r}\hat{D}(\alpha)|0\rangle|$ is bounded. Hence, taking the limit $\eta\to +\infty$ and using \eqref{eq:rel-ent-0} and \eqref{eq:negl0}, we obtain finally that
\begin{equation}
    \Delta S_L = \mathcal{O}(\epsilon^2) \, .
\end{equation}

To complete the proof, it is sufficient to notice that 
\begin{equation}
    \Delta E_L =  \text{Tr}[\hat{H}_L(\hat{\rho}_L(t)-\hat{\rho}_L(0))] = \mathcal{O}(\epsilon) \, .
\end{equation}

\section{Correspondance between the Floquet and the dressed qubit bases}\label{app:floquet-basis}

\subsection{Proof of \eqref{eq:floquet-states}}

In order to prove the relation \eqref{eq:floquet-states}, it is sufficient to show that the states $e^{-i\omega_L\hat{\sigma}_z t/2}|j\rangle$, $j=1,2$, are solutions of the eigenvalue problem \eqref{eq:eig-val-prob}. This straightforward using the facts that 
\begin{equation}
e^{i\omega_L\hat{\sigma}_z t/2}(\hat{H}_A+\hat{V}(t) )e^{-i\omega_L\hat{\sigma}_z t/2} = \hat{H}_{DA}+\frac{\omega_L}{2}\hat{\sigma}_z
\end{equation}
and
\begin{equation}
-i\partial_t e^{-i\omega_L\hat{\sigma}_z t/2}|j\rangle = -\frac{\omega_L}{2}\hat{\sigma}_z e^{-i\omega_L\hat{\sigma}_z t/2}|j\rangle \, .
\end{equation}
Indeed, replacing $|u_j(t)\rangle=e^{-i\omega_L\hat{\sigma}_z t/2}|j\rangle$ in \eqref{eq:eig-val-prob}, and applying $e^{i\omega_L\hat{\sigma}_z t/2}$ on both sides of the equality, we obtain
\begin{align}
\left(\hat{H}_{DA}+\frac{\omega_L}{2}\hat{\sigma}_z-\frac{\omega_L}{2}\hat{\sigma}_z\right)|j\rangle = \epsilon_j |j\rangle \, ,
\end{align}
which is true when choosing $\epsilon_1=-\frac{\Omega}{2}$, $\epsilon_2=\frac{\Omega}{2}$. Note that this feature can be generalized for any system with a $\mbox{SU}(2)$ symmetry, described by Pauli operators $\hat{\sigma}_x,\hat{\sigma}_y,\hat{\sigma}_z$.

\subsection{Proof of \eqref{eq:equiv-rot-da}}

We now show that the relation \eqref{eq:floquet-states} implies that the evolution of the system in the dressed basis, in the autonomous picture, is equivalent to the evolution in the rotating frame, in the non-autonomous picture. 

We start with the simple case where the bath is not taken into account. In the autonomous picture, the total Hamiltonian is then given 
$\hat{H}_X=\hat{H}_{DA}+\hat{H}_{DL}$, and we assume that the initial density matrix is factorized
\begin{equation}
    \hat{\rho}(0)=\hat{\rho}_{DA}(0)\otimes\hat{\rho}_{DL}(0) \, , 
\end{equation}
hence
\begin{equation}
    \hat{\rho}(t) = e^{-it\hat{H}_X}\hat{\rho}_{DA}(0)\otimes\hat{\rho}_{DL}(0) e^{it\hat{H}_X}\, .
    \label{eq:app-floquet-aux2}
\end{equation}
Is then straightforward to obtain that
\begin{equation}
\begin{array}{ll}
    \hat{\rho}_{DA}(t) &= \text{Tr}_{DL}[\hat{\rho}(t)]\\
    \\
    &= \sum_{j,j'=1,2} e^{-it(\epsilon_j-\epsilon_{j'})} \hat{\rho}^{jj'}_{DA}(0)|j\rangle\langle j'| \, ,
\end{array}   
\end{equation}
where $\hat{\rho}_{DA}^{jj'}(0)\equiv \langle j |\hat{\rho}_{DA}(0)|j'\rangle$. On the other hand, in the non-autonomous picture, we have
\begin{equation}
    \hat{\tilde{\rho}}_A(t) = \mathcal{T}_{\leftarrow} [e^{-i\int_0^t ds\hat{H}_A+\hat{V}(s)}]\hat{\tilde{\rho}}_A(0)\mathcal{T}_{\leftarrow} [e^{i\int_0^t ds\hat{H}_A+\hat{V}(s)}] \, .
\end{equation}
Using Floquet theory \cite{langemeyer2014}, we can write the propagator in the Floquet basis
\begin{equation}
    \mathcal{T}_{\leftarrow} [e^{-i\int_0^t ds\hat{H}_A+\hat{V}(s)}] = \sum_{j=1,2}e^{-it\epsilon_j}|u_j(t)\rangle\langle u_{j}(0)| \, .
    \label{eq:app-floquet-aux}
\end{equation}
Using \eqref{eq:floquet-states}, we may replace $|u_j(t)\rangle = e^{-i\omega_j\hat{\sigma}_z t/2}|j\rangle$ in \eqref{eq:app-floquet-aux}, and, using the definition \eqref{eq:rho-tilde-rot}, we obtain
\begin{equation}
    \hat{\tilde{\rho}}^{\text{rot}}(t) = \sum_{j,j'=1,2} e^{-it(\epsilon_j-\epsilon_{j'})} \hat{\tilde{\rho}}^{\text{rot}jj'}_{A}(0)|j\rangle\langle j'| \, .
\end{equation}
Comparing with \eqref{eq:app-floquet-aux2}, it is then sufficient to assume that $\hat{\rho}_{DA}(0)=\hat{\tilde{\rho}}^{\text{rot}}(0)$ to conclude that the two density matrices coincide at all times.

\medskip

Let's now turn to the general proof of \eqref{eq:equiv-rot-da}, when the coupling with the bath is taken into account. It is convenient to go in the interaction picture. In the autonomous case, the density matrix of the total system in the interaction picture w.r.t. $\hat{H}_X+\hat{H}_B$ is given by
\begin{equation}
    \hat{\rho}^I(t) = \hat{U}_0^\dagger(t,0)\hat{\rho}(t)\hat{U}_0(t,0) =  \hat{U}^I(t,0)\hat{\rho}(0) \hat{U}^{I\dagger}(t,0)\
\end{equation}
with $\hat{U}_0(t,0) = e^{-it(\hat{H}_X+\hat{H}_B)}$ and
\begin{equation}
    \hat{U}^I(t,0)\equiv \mathcal{T}[e^{-i\int_0^t ds \hat{V}_{AB}(s)}] \, ,
    \label{eq:ui}
\end{equation}
where $\hat{V}_{AB}(t)$ is the Hamiltonian $\hat{V}_{AB}$ in the interaction picture, given by 
\begin{equation}
\hat{V}_{AB}(t)\equiv e^{i\hat{H}_0t}\hat{V}_{AB}e^{-i\hat{H}_0t}   \, .  
\end{equation}
To compute $\hat{V}_{AB}(t)$, we express the operators $\hat{\sigma}_-\otimes\mathds{I}_L$, $\hat{\sigma}_+\otimes\mathds{I}_L$ in the eigenbasis of $\hat{H}_X$ \eqref{eq:eig-bas}. In this basis, the operator $\hat{\sigma}_-\otimes\mathds{I}_L$ writes
\begin{equation}
\begin{array}{ll}
    \hat{\sigma}_-\otimes\mathds{I}_L&=|a\rangle\langle b|\otimes\sum\limits_{N_{L}} |N_{L}\rangle\langle N_{L}|\\
    \\
    &=\sum\limits_{n} \left(\sqrt{\frac{\Omega-\delta}{2\Omega}}|1(n-1)\rangle+\sqrt{\frac{\Omega+\delta}{2\Omega}}|2(n-1)\rangle\right)\left(\sqrt{\frac{\Omega+\delta}{2\Omega}}\langle 1(n)|-\sqrt{\frac{\Omega-\delta}{2\Omega}}\langle 2(n)|\right) \\
    \\
    &= \hat{S}_z + \hat{S}_+ + \hat{S}_- \, 
\end{array}
\label{eq:sigma-base}
\end{equation}
where
\begin{align}
\hat{S}_z&=\frac{g}{2\Omega}(|2\rangle\langle 2|-|1\rangle\langle 1|) \otimes\sum\limits_{n\geq 0} |n-1\rangle\langle n|\nonumber \\
&\equiv \hat{s}_z\otimes \sum\limits_{n\geq 0} |n-1\rangle\langle n|\nonumber \\
\hat{S}_+&\equiv-\frac{\Omega-\delta}{2\Omega}|2\rangle\langle1|\otimes\sum\limits_{n\geq 0} |n-1\rangle\langle n|\nonumber\\
&\equiv \hat{s}_+\otimes\sum\limits_{n\geq 0} |n-1\rangle\langle n|\nonumber\\
\hat{S}_-&\equiv\frac{\Omega+\delta}{2\Omega}|1\rangle\langle 2|\otimes\sum\limits_{n\geq 0} |n-1\rangle\langle n|\nonumber\\
&\equiv \hat{s}_- \otimes\sum\limits_{n\geq 0} |n-1\rangle\langle n|\, ,\label{eq:s-ops-app}
\end{align}
with 
\begin{align}
    \hat{s}_z&\equiv\frac{g}{2\Omega}(|2\rangle\langle 2|-|1\rangle\langle 1|) \equiv \frac{g}{2\Omega}\hat{\Sigma}_z\nonumber \\
\hat{s}_+&\equiv-\frac{\Omega-\delta}{2\Omega}|2\rangle\langle1|\equiv -\frac{\Omega-\delta}{2\Omega}\hat{\Sigma}_+\nonumber\\
\hat{s}_-&\equiv\frac{\Omega+\delta}{2\Omega}|1\rangle\langle 2|\equiv 
\frac{\Omega+\delta}{2\Omega}\hat{\Sigma}_-  ,
\end{align}
and where $\hat{\Sigma}_z=|2\rangle\langle 2|-|1\rangle\langle 1|,\hat{\Sigma}_+=|2\rangle\langle 1|=\hat{\Sigma}_-^\dagger$, as defined in the main text in \eqref{eq:s-ops} and \eqref{eq:s-ops-red}. The term $\hat{\sigma}_+\otimes\mathds{I}_L$ is obtained by taking the hermitian conjugate of \eqref{eq:sigma-base}. We therefore obtain
\begin{equation}
    \hat{V}_{AB}(t)=\left(\hat{S}_z(t)+\hat{S}_-(t)+\hat{S}_+(t)\right) \hat{B}^\dagger(t) + h.c.
    \label{eq:vab-eig-x}
\end{equation}
with  
\begin{align}
\hat{S}_z(t)&=e^{-i\omega_L t}\hat{S}_z\nonumber \\
\hat{S}_+(t)&=e^{-i(\omega_L -\Omega)t}\hat{S}_+\nonumber\\
\hat{S}_-(t)&=e^{-i(\omega_L+\Omega)t}\hat{S}_-\, .
\end{align}

Let's assume that the dressed laser is initially in a Poisson state \eqref{eq:poisson}. The generalization to a coherent state is straightforward using the fact that the distribution $e^{-|\alpha|^2/2}\frac{\alpha^{N}}{\sqrt{N!}}$ is peaked around $N= |\alpha|^2$ in the macroscopic limit $|\alpha|\gg 1$. Since the density matrix is initially factorized \eqref{eq:rho-ini-dadlb}, the evolution of the density matrix of the dressed qubit, obtained after tracing out the degrees of freedom of the dressed laser $DL$ and of the bath, is given by the quantum map \cite{breuer2002theory},
\begin{align}
    \hat{\rho}_{DA}^I(t)&\equiv \text{Tr}_{DL,B}(\hat{\rho}^I(t)) \label{eq:kraus-aux}\\
    & = \sum_{\boldsymbol{\kappa,\kappa'}} \hat{W}^I_{\boldsymbol{\kappa,\kappa'}}(t,0) \hat{\rho}_{DA}(0) 
\hat{W}^I_{\boldsymbol{\kappa,\kappa'}}(t,0) \nonumber
\end{align}
where the sum runs over the pairs $\boldsymbol{\kappa}=(\mu,n), \boldsymbol{\kappa'}=(\nu,n')$, with
\begin{equation}
    \hat{W}_{\boldsymbol{\kappa,\kappa'}}^I = \sqrt{\eta_{\nu}\xi_{n}} \langle n, \mu | \hat{U}^I(t,0)  |n',\nu \rangle \, ,
\end{equation}
where $\xi_{n}=e^{-|\alpha|^2} |\alpha|^{2n}/n!$, and with $\{|\nu\rangle\}$ the eigenstates of $\hat{H}_B$ of eigenvalues $\nu$ and $ \eta_\nu=e^{-\beta_B\omega_\nu}/Z_B$. At this stage, we do not want to carry out the trace over the bath (this is the object of the sections~\ref{sec:cons-qme} and \ref{sec:cons-obe-fme}). However, we now use the fact that the expectation value of the product of operators $\hat{B}^\dagger,\hat{B}$ over the Gibbs state $\hat{\rho}_B$ is non-zero only when there is the same number of repetition of $\hat{B}^\dagger$ and $\hat{B}$. This allows us, after writing the exponential in \eqref{eq:ui} in a series, to deduce that the only relevant terms are such that the partial trace over $DL$ is equal to one. Indeed, the relevant terms are those with an equal number of $\hat{B}^\dagger$ and $\hat{B}$, which, given the form of $\hat{V}_{AB}(t)$ and \eqref{eq:s-ops-app}, implies that the operator acting on $DL$ is the identity (since a product of the same number of $\sum_{n}|n-1\rangle\langle n|$ and $\sum_{n}|n\rangle\langle n-1|$ is the identity). Therefore, we obtain
\begin{equation}
    \hat{\rho}_{DA}^I(t) = \text{Tr}_B\left[\hat{U}^{I,\text{red}}(t,0)\hat{\rho}_{DA}(0)\otimes\hat{\rho}_B\hat{U}^{I,\text{red}\dagger}(t,0)  \right]
    \label{eq:app-da-aux}
\end{equation}
where the reduced propagator $\hat{U}^{I,\text{red}}(t,0)$ is
\begin{equation}
    \hat{U}^{I,\text{red}}(t,0) = \mathcal{T}\left[ 
e^{-i\int_0^t ds \hat{V}_{AB}^{I,\text{red}}(s)} \right]
\end{equation}
with
\begin{equation}
    \hat{V}_{AB}^{I,\text{red}}(t) = (e^{-i\omega_L t}\hat{s}_z+e^{-i(\omega_L -\Omega)t}\hat{s}_++e^{-i(\omega_L+\Omega)t}\hat{s}_-)\hat{B}^\dagger (t) + \text{h.c.} \, .
    \label{eq:vabired}
\end{equation}

Let's now turn to the non-autonomous picture. The goal is to show that the evolution of the qubit in the rotating frame is equivalent to that of the dressed qubit (in the autonomous picture), given by \eqref{eq:app-da-aux}. We therefore go to the interaction picture, here w.r.t. $\hat{H}_{DA}+\hat{H}_B$ (since the degrees of the freedom of the laser have already been traced out),
\begin{equation}
    \hat{\tilde{\rho}}^{\text{rot},I}\equiv \hat{\tilde{U}}^I(t,0) \hat{\tilde{\rho}}^{\text{rot}}(0)\hat{\tilde{U}}^{I\dagger}(t,0) 
\end{equation}
where
\begin{equation}
    \hat{\tilde{U}}^I(t,0) \equiv \mathcal{T}\left[ 
e^{-i\int_0^t ds \hat{V}_{AB}'^I(s)} \right]
\end{equation}
with
\begin{equation}
    \hat{V}_{AB}'^I(t) = e^{it(\hat{H}_{DA}+\hat{H}_B)}\hat{V}_{AB}'(t)e^{-it(\hat{H}_{DA}+\hat{H}_B)}
\end{equation}
where $\hat{V}_{AB}'(t)$ is given in \eqref{eq:vabprimet}. Using \eqref{eq:sigma-base}, and comparing with \eqref{eq:vabired}, it is straightforward to check that 
\begin{equation}
    \hat{V}_{AB}^{I,\text{red}}(t) = \hat{V}_{AB}'^{I}(t) \, , 
\end{equation}
which concludes the proof.

\section{First law in the rotating frame}\label{app:proof60}

We provide here details on the derivation of \eqref{eq:1st-law-rot}.
From the definition of $\tilde{E}^{\text{rot}}_A(t)$, the conservation of energy yields 
\begin{equation}
    d_t \tilde{E}^{\text{rot}}_A(t) = \dot Q+ \text{Tr}[d_t\hat{V}_{AB}'(t)\hat{\tilde{\rho}}^{\text{rot}}(t)] \, .
    \label{eq:1st-law-rot-aux}
\end{equation}
Then, using the definition $\hat{\tilde{\rho}}^{\text{rot}}(t)=e^{i\omega_L \hat{\sigma}_z t/2}\hat{D}^\dagger(\alpha e^{-i\omega_L t})\hat{\rho}(t) \hat{D}^\dagger(\alpha e^{-i\omega_L t})e^{-i\omega_L \hat{\sigma}_z t/2}$, we obtain
\begin{equation}
    \begin{array}{ll}
\text{Tr}[d_t\hat{V}_{AB}'(t)\hat{\tilde{\rho}}^{\text{rot}}(t)] &= \text{Tr}[i\omega_L[\hat{\sigma}_z,\hat{V}_{AB}']\hat{\tilde{\rho}}(t)]
= \text{Tr}[i\omega_L[\hat{\sigma}_z,\hat{V}_{AB}+\hat{V}_{AL}+g(\hat{\sigma}_+\alpha(t)+\hat{\sigma}_-\alpha(t)^*)]\hat{\rho}(t)]\\
\\
&= i\omega_L\text{Tr}[\hat{\sigma}_z[\hat{V}_{AB}+\hat{V}_{AL},\hat{\rho}(t)]]+\text{Tr}[d_t \hat{V}(t)\hat{\rho}(t)]\\
\\
&= -\omega_L \text{Tr}[\hat{\sigma}_zd_t\hat{\rho}(t)]-i\omega_L \underbrace{Tr([\hat{\sigma}_z,\hat{H}_A+\hat{H}_L+\hat{H}_B]\hat{\rho}(t))}_{=0} +\text{Tr}[d_t \hat{V}(t)\hat{\rho}(t)] =\dot W_{DL}\, ,
    \end{array}
\end{equation}
where in the last equality we used \eqref{eq:wdl-split} and \eqref{eq:wl-vt}. 
Replacing in \eqref{eq:1st-law-rot-aux}, we obtain the first law \eqref{eq:1st-law-rot}.

\section{Proof of the symmetry \eqref{eq:ft-2}}\label{app:ft}

We begin by justifying \eqref{eq:cond-rhod}. In the macroscopic limit $|\alpha|\gg 1$, the Poisson distribution effectively becomes equivalent to a Gaussian distribution,
\begin{equation}
    e^{-|\alpha|^2}\frac{|\alpha|^{2N}}{N!}\sim \frac{1}{\sqrt{2\pi}\sigma}e^{-\frac{(N-\mu)^2}{2\sigma^2}} \, ,
\end{equation}
where $\sigma\equiv |\alpha|$ and $\mu\equiv |\alpha|$. In turn, a Gaussian state can be understood as a Gibbs state, by re-writing
\begin{equation}
    \frac{1}{\sqrt{2\pi}\sigma}e^{-\frac{(N-\mu)^2}{2\sigma^2}}  = \frac{1}{Z(\beta_{DL})}e^{-\beta_{DL} E(N)}
\end{equation}
where $\beta_{DL} \equiv 1/\sigma^2 = 1/|\alpha|^2$, $E(N)\equiv (N-\mu)^2/2$, and $Z(\beta_{DL})\equiv \sum_N {-\beta_{DL} E(N)}$, which leads to \eqref{eq:cond-rhod}.

\medskip

Let's now prove \eqref{eq:ft-2}. Using \eqref{eq:init-cond} and \eqref{eq:cond-rhod}, we can write the generating function of the forward \eqref{eq:mgf} and time-reversed \eqref{eq:mgfrev} processes explicitly,
\begin{align}
&G(\boldsymbol{\lambda},t)  = \frac{1}{Z} \text{Tr}[ \hat{U}(t,0) e^{-i\lambda_{DA} \hat{H}_{DA} - i\lambda_{DL} \hat{H}_{DL}-i\lambda_{B}\hat{H}_B} e^{- \beta_{DA} \hat{H}_{DA} -\beta_{DL}\hat{H}'_{DL}-\beta_B\hat{H}_B} \hat{U}^{\dagger}(t,0) e^{i\lambda_{DA} \hat{H}_{DA} + i\lambda_{DL} \hat{H}_{DL}+i\lambda_{B}\hat{H}_B} ], \label{eq:genapp1} \\
&G^R(\boldsymbol{-\lambda}+i\boldsymbol{\beta},t)  \label{eq:genapp2}  \\
&=\frac{1}{Z} \text{Tr}[ \hat{U}^{\dagger}(t,0) e^{i\lambda_{DA} \hat{H}_{DA} +i\lambda_{DL} \hat{H}_{DL}+i\lambda_{B}\hat{H}_B} e^{\beta_{DL}\hat{H}_{DL}}e^{-\beta_{DL}\hat{H}'_{DL}} \hat{U}(t,0) e^{-i\lambda_{DA} \hat{H}_{DA} -i\lambda_{DL} \hat{H}_{DL}-i\lambda_{B}\hat{H}_B} e^{- \beta_{DA} \hat{H}_{DA} -\beta_{DL}\hat{H}_{DL}-\beta_B\hat{H}_B} ]. \nonumber
\end{align}
where $Z=Z_{DA} Z(\beta_{DL}) Z_B$. Since, in the limit $\beta_{DL}\rightarrow 0$,
\begin{equation}
    e^{\beta_{DL}\hat{H}_{DL}}e^{-\beta_{DL}\hat{H}'_{DL}}\rightarrow \mathds{I}
\end{equation}
we may replace $e^{-\beta_{DL}\hat{H}_{DL}}\sim e^{-\beta_{DL}\hat{H}'_{DL}}$. Using finally the cyclic property of the trace, it is then straightforward to check that
\begin{equation}
    G^R(-\boldsymbol{\lambda}+i\boldsymbol{\nu},t)=G(\boldsymbol{\lambda},t) \, ,
\end{equation}
where $\boldsymbol{\nu}=(\beta_{DA},\beta_ {DL},\beta_B)$, which proves the fluctuation theorem \eqref{eq:ft-2}. 

Note also that, by linearity, the theorem \eqref{eq:ft-2} can readily be extended to the case where the system is weakly coupled to many heat baths like $B$, as long as the baths do not interact with each other.

\section{Work fluctuation theorem in the non-autonomous picture}\label{app:ft-na}

We prove here the relation $\hat{\tilde{\rho}}^R(t) = \hat{\tilde{\rho}}(-t)$. 
Let's consider first, for simplicity, the case where the bath is not taken into account, such that the total Hamiltonian is $\hat{H}_A+\hat{V}(t)$. Using \eqref{eq:app-floquet-aux}, the density matrices of the forward and backward processes write 
\begin{equation}
    \begin{array}{ll}
        \hat{\tilde{\rho}}_A(t) &= \sum\limits_{j,j'=1,2} e^{-it(\epsilon_j-\epsilon_{j'})} |u_j(t)\rangle\langle u_j(0)|  \hat{\tilde{\rho}}_A(0) |u_{j'}(0)\rangle\langle u_{j'}(t)|\\
        \\
       \hat{\tilde{\rho}}^R_A(t) &= \sum\limits_{j,j'=1,2} e^{it(\epsilon_j-\epsilon_{j'})} |u_j(0)\rangle\langle u_j(t)|  \hat{\tilde{\rho}}^R_A(0) |u_{j'}(t)\rangle\langle u_{j'}(0)| \, .
    \end{array}
\end{equation}
Using now \eqref{eq:floquet-states}, the fact that $e^{i\hat{\omega_L\sigma}_zt/2}\hat{V}(t)e^{-i\omega_L\hat{\sigma}_zt/2}=\hat{V}(0)$, we notice that, given the initial conditions \eqref{eq:ini-cond-na}, we have
\begin{equation}
    \begin{array}{ll}
   \langle u_j(0)|    \hat{\tilde{\rho}}^R_A(t) |u_{j'}(0)\rangle  &=  e^{it(\epsilon_j-\epsilon_{j'})} \langle u_j(t)|  \hat{\tilde{\rho}}^R_A(0) |u_{j'}(t)\rangle \\
   \\
         & =e^{it(\epsilon_j-\epsilon_{j'})} \langle j|  \hat{\tilde{\rho}}_A(0) |j'\rangle \\
         \\
         &= \langle u_j(-t)|\hat{\tilde{\rho}}_A(-t)|u_{j'}(-t)\rangle  \, 
    \end{array}
\end{equation}
from which we deduce that
\begin{equation}
     \hat{\tilde{\rho}}^R_A(t) = \hat{\tilde{\rho}}_A(-t) \, .
\end{equation}
The generalization to the case where the bath is taken into account is obtained by repeating the reasoning, tracing out first the degrees of freedom of the bath and introducing Kraus operators as in \eqref{eq:kraus-aux}.

\section{Derivation of the master equation \eqref{eq:qme-x}}\label{app:qme-x}

We provide details on the derivation of the master equation \eqref{eq:qme-x}. 

The tilted operator $\hat{V}^{\boldsymbol{\lambda}}_{AB}(t)$ in \eqref{eq:vab} is simply obtained using the expression \eqref{eq:vab-eig-x} and the identities $e^{i(\lambda_{DA}\hat{H}_{DA}+\lambda_{DL}\hat{H}_{DL})}|1,n\rangle=e^{i(-\lambda_{DA}\Omega+\lambda_{DL}n\omega_L}|1,n\rangle$, $e^{i(\lambda_{DA}\hat{H}_{DA}+\lambda_{DL}\hat{H}_{DL})}|2,n\rangle=e^{i(\lambda_{DA}\Omega+\lambda_{DL}n\omega_L}|2,n\rangle$ and $e^{i\lambda_B\hat{H}_B}\hat{b}_k e^{-i\lambda_B\hat{H}_B}=e^{-i\lambda_B\omega_k}\hat{b}_k$. 

\medskip

Let us now introduce
 \begin{align}
&d^{\boldsymbol{\lambda}}_{mn,m'n'}(t)= \label{eq:dij}\\
&\sum\limits_{\mu,\nu}\eta_\nu \text{Tr}_S[\hat{\sigma}_{mn}^\dagger\hat{W}^{\boldsymbol{\lambda}I}_{\mu,\nu}(t+\delta,t)]\text{Tr}_S[\hat{\sigma}_{m'n'}\hat{W}_{\mu,\nu} ^{-\boldsymbol{\lambda}I\dagger}(t+\delta,t)] \nonumber\\
&=\sum\limits_{\mu,\nu}\eta_\nu \langle E_{n},\mu|\mathcal{T}_{\leftarrow}\{e^{-i\int_t^{t+\delta} ds \hat{V}_{AB}^{\boldsymbol{\lambda}}(s)}\}|E_{m},\nu\rangle \langle E_{m'},\nu|\mathcal{T}_{\rightarrow}\{e^{i\int_t^{t+\delta} ds \hat{V}_{AB}^{-\boldsymbol{\lambda}\dagger}(s)}\}|E_{n'},\mu\rangle,
\nonumber
\end{align}
which lead to \eqref{eq:aux-lfl}. 
A perturbative expansion to second order in $\hat{V}_{AB}^{\boldsymbol{\lambda}}$ then yields \eqref{eq:gme-aux}. 
Since $\{\hat{\sigma}_{mn}\}$ form an orthogonal basis, the only terms $\hat{\sigma}_{mn}$ which remain in \eqref{eq:gme-aux}, and  $\sum_{mn}\text{Tr}_X[\hat{\sigma}_{mn}]\hat{\sigma}_{mn}=\mathds{I}$, we can re-write \eqref{eq:gme-aux} as
\begin{equation}
\hspace{-0.5cm}   \begin{array}{l}
\mathcal{L}_{\boldsymbol{\lambda}}(\hat{\rho}^I_X(t))=\\
\\
\frac{1}{\delta_0}\sum\limits_{\alpha,\alpha'=z,+,-}\int_t^{t+\delta_0} ds\int_t^{t+\delta_0} ds' \text{Tr}_B[ \hat{B}_{\lambda_B}^\dagger(s)\hat{\rho}_B \hat{B}_{-\lambda_B}(s')]
e^{-i(\omega_{\alpha}s-\omega_{\alpha'}s')}\hat{S}^{\boldsymbol{\lambda}}_{\alpha}\hat{\rho}_X(t)\hat{S}_{\alpha'}^{-\boldsymbol{\lambda}\dagger}\\
    \\
    +\frac{1}{\delta_0}\sum\limits_{\alpha,\alpha'}\int_t^{t+\delta_0} ds\int_t^{t+\delta_0} ds' \text{Tr}_B[ \hat{B}_{\lambda_B}(s)\hat{\rho}_B \hat{B}_{-\lambda_B}^\dagger(s')]e^{i(\omega_{\alpha}s-\omega_{\alpha'}s')}\hat{S}^{\boldsymbol{\lambda}\dagger}_{\alpha}\hat{\rho}_X(t)\hat{S}^{-\boldsymbol{\lambda}}_{\alpha'}\\
    \\
    -\frac{1}{2}\frac{1}{\delta_0}\sum\limits_{\alpha,\alpha'} \int_t^{t+\delta} ds \int_t^{s} ds' \text{Tr}_B[\hat{B}_{\lambda_B}^\dagger(s)\hat{B}_{\lambda_B}(s')\hat{\rho}_B]e^{-i(\omega_{\alpha}s-\omega_{\alpha'}s')}\hat{S}^{\boldsymbol{\lambda}}_{\alpha}\hat{S}^{\boldsymbol{\lambda}\dagger}_{\alpha'}\hat{\rho}_X(t)\\
    \\
    -\frac{1}{2}\frac{1}{\delta_0}\sum\limits_{\alpha,\alpha'} \int_t^{t+\delta_0} ds \int_t^{s} ds' \text{Tr}_B[\hat{B}_{\lambda_B}(s)\hat{B}_{\lambda_B}^\dagger(s')\hat{\rho}_B]e^{i(\omega_{\alpha}s-\omega_{\alpha'}s')}\hat{S}^{\boldsymbol{\lambda}\dagger}_{\alpha}\hat{S}^{\boldsymbol{\lambda}}_{\alpha'}\hat{\rho}_X(t)\\
    \\
    -\frac{1}{2}\frac{1}{\delta_0}\sum\limits_{\alpha,\alpha'} \int_t^{t+\delta_0} ds \int_X^{t+\delta_0} ds' \text{Tr}_B[\hat{B}_{-\lambda_B}^\dagger(s)\hat{B}_{-\lambda_B}(s')\hat{\rho}_B]e^{-i(\omega_{\alpha}s-\omega_{\alpha'}s')}\hat{\rho}_X(t)\hat{S}^{-\boldsymbol{\lambda}}_{\alpha}\hat{S}^{-\boldsymbol{\lambda}\dagger}_{\alpha'}\\
    \\
    -\frac{1}{2}\frac{1}{\delta_0}\sum\limits_{\alpha,\alpha'} \int_t^{t+\delta_0} ds \int_s^{t+\delta_0} ds' \text{Tr}_B[\hat{B}_{-\lambda_B}(s)\hat{B}_{-\lambda_B}^\dagger(s')\hat{\rho}_B]e^{i(\omega_{\alpha}s-\omega_{\alpha'}s')}\hat{\rho}_X(t)\hat{S}^{-\boldsymbol{\lambda}\dagger}_{\alpha}\hat{S}^{-\boldsymbol{\lambda}}_{\alpha'}.
    \end{array}
    \label{eq:gme-aux-2}
\end{equation}
where the operators $\hat{S}^{\boldsymbol{\lambda}}_z, \hat{S}^{\boldsymbol{\lambda}}_+, \hat{S}^{\boldsymbol{\lambda}}_-$ are the operators \eqref{eq:s-ops} in the Heisenberg picture,
\begin{equation}
    \begin{array}{ll}
\hat{S}^{\boldsymbol{\lambda}}_z&\equiv e^{-i\omega_L\lambda_{DL}/2}\hat{S}_z\\
\\
\hat{S}^{\boldsymbol{\lambda}}_+&\equiv e^{-i(\omega_L\lambda_{DL}-\Omega\lambda_{DA})/2}\hat{S}_+\\
\\
\hat{S}^{\boldsymbol{\lambda}}_-&\equiv e^{-i(\omega_L\lambda_{DL}+\Omega\lambda_{DA})/2}\hat{S}_-\, .
    \end{array}
    \label{eq:slambda}
\end{equation}

Performing the double integrals then leads to \eqref{eq:qme-x}. 
We write explicitly the double integral in first line of the r.h.s. of \eqref{eq:gme-aux-2} (the other terms have similar forms),
\begin{align}
& \frac{1}{\delta_0}   \int_t^{t+\delta_0} ds\int_t^{t+\delta_0} ds' \text{Tr}_B[ \hat{B}_{\lambda_B}^\dagger(s)\hat{\rho}_B \hat{B}_{-\lambda_B}(s')]
e^{-i(\omega_{\alpha}s-\omega_{\alpha'}s')}\hat{S}^{\boldsymbol{\lambda}}_{\alpha}\hat{\rho}_X(t)\hat{S}_{\alpha'}^{-\boldsymbol{\lambda}\dagger} \nonumber \\
&=\sum_k |g_k|^2(n_B(\omega_k)+1)e^{i\lambda_B\omega_k}\delta_0\text{ sinc}\left(\frac{\omega_k-\omega_\alpha}{2}\delta_0\right)\text{ sinc}\left(\frac{\omega_k-\omega_{\alpha'}}{2}\delta_0\right) \hat{S}^{\boldsymbol{\lambda}}_{\alpha}\hat{\rho}_X(t)\hat{S}_{\alpha'}^{-\boldsymbol{\lambda}\dagger} e^{i(t+\delta_0/2)(\omega_{\alpha'}-\omega_\alpha)} \, .
\end{align}

\section{Expression of the generalized Bloch equation when $\Omega<\gamma_{max}<2\Omega$}\label{app:gob}

When $\Omega<\gamma_{max}<2\Omega$, the terms of \eqref{eq:gme-aux} involving the jump operators $\hat{S}_+$ and $\hat{S}_-$ are removed. We obtain
\begin{align}
    \mathcal{L}_{\boldsymbol{\lambda}}^{G}(\hat{\rho}_{DA}^{\boldsymbol{\lambda}}) = -i[\hat{H}_{DA},\hat{\rho}_{DA}^{\boldsymbol{\lambda}}] + \mathcal{D}_+^{G\boldsymbol{\lambda}}(\hat{\rho}_{DA}^{\boldsymbol{\lambda}})+\mathcal{D}_-^{G\boldsymbol{\lambda}}(\hat{\rho}_{DA}^{\boldsymbol{\lambda}}) \nonumber
\end{align}
with
\begin{align}
   & \mathcal{D}_+^{G\boldsymbol{\lambda}}(\hat{\rho}) = 
   \sum_{\alpha=+,-,z} G_+(\omega_\alpha)\hat{s}_\alpha^{\boldsymbol{\lambda}}\hat{\rho} \hat{s}_\alpha^{-\boldsymbol{\lambda}\dagger} +\sqrt{G_+(\omega_z)G_+(\omega_+)}(\hat{s}_+^{\boldsymbol{\lambda}}\hat{\rho}\hat{s}_z^{-\boldsymbol{\lambda}\dagger}+\hat{s}_z^{\boldsymbol{\lambda}}\hat{\rho}\hat{s}_+^{-\boldsymbol{\lambda}\dagger})+\sqrt{G_+(\omega_z)G_+(\omega_+)}(\hat{s}_+^{\boldsymbol{\lambda}}\hat{\rho}\hat{s}_z^{-\boldsymbol{\lambda}\dagger}+\hat{s}_z^{\boldsymbol{\lambda}}\hat{\rho}\hat{s}_+^{-\boldsymbol{\lambda}\dagger}) \nonumber\\
   &- \frac{1}{2}\left( \sum_{\alpha=+,-,z} G_+(\omega_\alpha)\hat{s}_\alpha^{\dagger}\hat{s}_\alpha +\sqrt{G_+(\omega_z)G_+(\omega_+)}(\hat{s}_z^{(\lambda_{DA},\lambda_{DL},0)\dagger}\hat{s}_+^{(\lambda_{DA},\lambda_{DL},0)}+\hat{s}_+^{(\lambda_{DA},\lambda_{DL},0)\dagger}\hat{s}_z^{(\lambda_{DA},\lambda_{DL},0)}) \right.\nonumber\\
  & \left. +\sqrt{G_+(\omega_z)G_+(\omega_-)}(\hat{s}_z^{(\lambda_{DA},\lambda_{DL},0)\dagger}\hat{s}_-^{(\lambda_{DA},\lambda_{DL},0)}+\hat{s}_-^{(\lambda_{DA},\lambda_{DL},0)\dagger}\hat{s}_z^{(\lambda_{DA},\lambda_{DL},0)}) \right)\hat{\rho}\nonumber \\
  &-\frac{1}{2}\hat{\rho}\left( \sum_{\alpha=+,-,z} G_+(\omega_\alpha)\hat{s}_\alpha^{\dagger}\hat{s}_\alpha+\sqrt{G_+(\omega_z)G_+(\omega_+)}(\hat{s}_z^{(-\lambda_{DA},-\lambda_{DL},0)\dagger}\hat{s}_+^{(-\lambda_{DA},-\lambda_{DL},0)}+\hat{s}_+^{(-\lambda_{DA},-\lambda_{DL},0)\dagger}\hat{s}_z^{(-\lambda_{DA},-\lambda_{DL},0)}) \right.\nonumber\\
  & \left. +\sqrt{G_+(\omega_z)G_+(\omega_-)}(\hat{s}_z^{(-\lambda_{DA},-\lambda_{DL},0)\dagger}\hat{s}_-^{(-\lambda_{DA},-\lambda_{DL},0)}+\hat{s}_-^{(-\lambda_{DA},-\lambda_{DL},0)\dagger}\hat{s}_z^{(-\lambda_{DA},-\lambda_{DL},0)}) \right)\nonumber \\
   \nonumber\\
    & \mathcal{D}_-^{G\boldsymbol{\lambda}}(\hat{\rho}) = 
   \sum_{\alpha=+,-,z} G_-(\omega_\alpha)\hat{s}_\alpha^{\boldsymbol{\lambda}\dagger}\hat{\rho} \hat{s}_\alpha^{-\boldsymbol{\lambda}} +\sqrt{G_-(\omega_z)G_-(\omega_+)}(\hat{s}_+^{\boldsymbol{\lambda}\dagger}\hat{\rho}\hat{s}_z^{-\boldsymbol{\lambda}}+\hat{s}_z^{\boldsymbol{\lambda}\dagger}\hat{\rho}\hat{s}_+^{-\boldsymbol{\lambda}})+\sqrt{G_-(\omega_z)G_-(\omega_+)}(\hat{s}_+^{\boldsymbol{\lambda}\dagger}\hat{\rho}\hat{s}_z^{-\boldsymbol{\lambda}}+\hat{s}_z^{\boldsymbol{\lambda}\dagger}\hat{\rho}\hat{s}_+^{-\boldsymbol{\lambda}}) \nonumber\\
   &- \frac{1}{2}\left( \sum_{\alpha=+,-,z} G_-(\omega_\alpha)\hat{s}_\alpha\hat{s}^\dagger_\alpha +\sqrt{G_-(\omega_z)G_-(\omega_+)}(\hat{s}_z^{(\lambda_{DA},\lambda_{DL},0)}\hat{s}_+^{(\lambda_{DA},\lambda_{DL},0)\dagger}+\hat{s}_+^{(\lambda_{DA},\lambda_{DL},0)}\hat{s}_z^{(\lambda_{DA},\lambda_{DL},0)\dagger}) \right.\nonumber\\
  & \left. +\sqrt{G_-(\omega_z)G_-(\omega_-)}(\hat{s}_z^{(\lambda_{DA},\lambda_{DL},0)}\hat{s}_-^{(\lambda_{DA},\lambda_{DL},0)\dagger}+\hat{s}_-^{(\lambda_{DA},\lambda_{DL},0)}\hat{s}_z^{(\lambda_{DA},\lambda_{DL},0)\dagger}) \right)\hat{\rho}\nonumber \\
  &-\frac{1}{2}\hat{\rho}\left( \sum_{\alpha=+,-,z} G_-(\omega_\alpha)\hat{s}_\alpha\hat{s}^\dagger_\alpha+\sqrt{G_-(\omega_z)G_-(\omega_+)}(\hat{s}_z^{(-\lambda_{DA},-\lambda_{DL},0)}\hat{s}_+^{(-\lambda_{DA},-\lambda_{DL},0)\dagger}+\hat{s}_+^{(-\lambda_{DA},-\lambda_{DL},0)}\hat{s}_z^{(-\lambda_{DA},-\lambda_{DL},0)\dagger}) \right.\nonumber\\
  & \left. +\sqrt{G_-(\omega_z)G_-(\omega_-)}(\hat{s}_z^{(-\lambda_{DA},-\lambda_{DL},0)}\hat{s}_-^{(-\lambda_{DA},-\lambda_{DL},0)\dagger}+\hat{s}_-^{(-\lambda_{DA},-\lambda_{DL},0)}\hat{s}_z^{(-\lambda_{DA},-\lambda_{DL},0)\dagger}) \right)\nonumber
\end{align}
where $\boldsymbol{\lambda} = (\lambda_{DA},\lambda_{DL},\lambda_B)$.

\section{Expression of $\mathcal{L}^{aF}$ with counting fields}\label{app:tilted-fme}

In the strong qubit-laser coupling limit defined in  \eqref{eq:regimes-2}, the product of sinc functions \eqref{eq:approx-sinc} is non zero only in the case $\alpha=\alpha'$, which is equivalent to the secular approximation in \eqref{eq:gob}. The resulting master equation is
\begin{equation}
    \begin{array}{ll}
\mathcal{L}_{\boldsymbol{\lambda}}^{aF,X}(\hat{\rho}^{\boldsymbol{\lambda}}_{X})=& -i[\hat{H}_{X}+\hat{H}_{LS},\hat{\rho}^{\boldsymbol{\lambda}}_{X}]\\
\\
&-\frac{1}{2}\{G_+(\omega_L)\hat{S}_z^\dagger\hat{S}_z+G_+(\omega_L-\Omega)\hat{s}_+^\dagger\hat{S}_++G_+(\omega_L+\Omega)\hat{s}_-^\dagger\hat{s}_-,\hat{\rho}^{\boldsymbol{\lambda}}_{X}\}\\
\\
&-\frac{1}{2}\{G_-(\omega_L)\hat{s}_z\hat{s}^\dagger_z+G_-(\omega_L-\Omega)\hat{s}_+\hat{s}^\dagger_++G_+(\omega_L+\Omega)\hat{s}_-\hat{s}^\dagger_-,\hat{\rho}^{\boldsymbol{\lambda}}_{X}\}\\
\\
&+G_+(\omega_L)e^{i\omega_L\lambda_B}\hat{s}^{\boldsymbol{\lambda}}_z\hat{\rho}^{\boldsymbol{\lambda}}_{X}\hat{s}_z^{-\boldsymbol{\lambda}\dagger}
+G_+(\omega_L-\Omega)e^{i(\omega_L-\Omega)\lambda_B}\hat{s}^{\boldsymbol{\lambda}}_+\hat{\rho}^{\boldsymbol{\lambda}}_{X}\hat{s}_+^{-\boldsymbol{\lambda}\dagger}\\
\\
&+G_+(\omega_L+\Omega)e^{i(\omega_L+\Omega)\lambda_B}\hat{s}^{\boldsymbol{\lambda}}_-\hat{\rho}^{\boldsymbol{\lambda}}_{X}\hat{s}_-^{-\boldsymbol{\lambda}\dagger}\\
\\
&+G_-(\omega_L)e^{-i\omega_L\lambda_B}\hat{s}^{\boldsymbol{\lambda}\dagger}_z\hat{\rho}^{\boldsymbol{\lambda}}_{X}\hat{s}^{-\boldsymbol{\lambda}}_z
+G_-(\omega_L-\Omega)e^{-i(\omega_L-\Omega)\lambda_B}\hat{s}_+^{\boldsymbol{\lambda}\dagger}\hat{\rho}^{\boldsymbol{\lambda}}_{X}\hat{s}^{-\boldsymbol{\lambda}}_+\\
\\
&+G_-(\omega_L+\Omega)e^{-i(\omega_L+\Omega)\lambda_B}\hat{s}^{\boldsymbol{\lambda}\dagger}_-\hat{\rho}^{\boldsymbol{\lambda}}_{X}\hat{s}_-^{-\boldsymbol{\lambda}} \, ,
    \end{array}
\end{equation}
where the Lamb shift contribution is 
\begin{align}
\hat{H}_{LS}=& I_+(\omega_L)\hat{S}_z^\dagger\hat{S}_z+I_+(\omega_L-\Omega)\hat{S}_+^\dagger\hat{S}_+ \nonumber\\
&+I_+(\omega_L+\Omega)\hat{S}_-^\dagger\hat{S}_-+I_-(\omega_L)\hat{S}_z\hat{S}_z^\dagger \nonumber\\
&+I_-(\omega_L-\Omega)\hat{S}_+\hat{S}_+^\dagger +I_-(\omega_L+\Omega)\hat{S}_-\hat{S}_-^\dagger \, ,\nonumber
\end{align}
with 
\begin{align}
    I_+(\omega)\equiv Im\left( \frac{1}{2}\int\limits_{0}^{+\infty}d\tau \text{Tr}[\hat{B}(\tau)\hat{B}^\dagger(0)\hat{\rho}_B] e^{i\nu\tau} \right)   \\
    I_-(\omega)\equiv Im\left( \frac{1}{2}\int\limits_{0}^{+\infty}d\tau \text{Tr}[\hat{B}^\dagger(\tau)\hat{B}(0)\hat{\rho}_B] e^{i\nu\tau} \right) \, .
\end{align}
As explained in the main text, the Lamb shift contribution may be neglected.

It is straightforward to trace out the degrees of freedom of $DL$, which gives the master equation \eqref{eq:fme-2},
\begin{equation}
    \begin{array}{ll}
\mathcal{L}_{\boldsymbol{\lambda}}^{aF}(\hat{\rho}^{\boldsymbol{\lambda}}_{DA})=& -i[\hat{H}_{DA},\hat{\rho}^{\boldsymbol{\lambda}}_{DA}]\\
\\
&-\frac{1}{2}\{\gamma_{0,\downarrow}\hat{\Sigma}_z^\dagger\hat{\Sigma}_z+\gamma_{2,\uparrow}\hat{\Sigma}_+^\dagger\hat{\Sigma}_++\gamma_{1,\downarrow}\hat{\Sigma}_-^\dagger\hat{\Sigma}_-,\hat{\rho}^{\boldsymbol{\lambda}}_{DA}\}\\
\\
&-\frac{1}{2}\{\gamma_{0,\uparrow}\hat{\Sigma}_z\hat{\Sigma}^\dagger_z+\gamma_{2,\downarrow}\hat{\Sigma}_+\hat{\Sigma}^\dagger_++\gamma_{1,\uparrow}\hat{\Sigma}_-\hat{\Sigma}^\dagger_-,\hat{\rho}^{\boldsymbol{\lambda}}_{DA}\}\\
\\
&+\gamma_{0,\downarrow}e^{i\omega_L(\lambda_B-\lambda_{DL})}\hat{\Sigma}_z\hat{\rho}^{\boldsymbol{\lambda}}_{DA}\hat{\Sigma}_z^\dagger
+\gamma_{2,\uparrow}e^{i\Omega\lambda_{DA}-i\omega_L\lambda_{DL}+i(\omega_L-\Omega)\lambda_B}\hat{\Sigma}_+\hat{\rho}^{\boldsymbol{\lambda}}_{DA}\hat{\Sigma}_+^\dagger\\
\\
&+\gamma_{1,\downarrow}e^{-i\Omega\lambda_{DA}-i\omega_L\lambda_{DL}+i(\omega_L+\Omega)\lambda_B}\hat{\Sigma}_-\hat{\rho}^{\boldsymbol{\lambda}}_{DA}\hat{\Sigma}_-^\dagger\\
\\
&+\gamma_{0,\uparrow}e^{-i\omega_L(\lambda_B-\lambda_{DL})}\hat{\Sigma}_z^\dagger\hat{\rho}^{\boldsymbol{\lambda}}_{DA}\hat{\Sigma}_z
+\gamma_{2,\downarrow}e^{-i\Omega\lambda_{DA}+Wi\omega_L\lambda_{DL}-i(\omega_L-\Omega)\lambda_B}\hat{\Sigma}_+^\dagger\hat{\rho}^{\boldsymbol{\lambda}}_{DA}\hat{\Sigma}_+\\
\\
&+\gamma_{1,\uparrow}e^{i\Omega\lambda_{DA}+i\omega_L\lambda_{DL}-i(\omega_L+\Omega)\lambda_B}\hat{\Sigma}_-^\dagger\hat{\rho}^{\boldsymbol{\lambda}}_{DA}\hat{\Sigma}_- \, .
    \end{array}
    \label{eq:fme-lambda}
\end{equation}
From \eqref{eq:fme-lambda}, it is clear that $\mathcal{L}_{\boldsymbol{\lambda}}^{aF}$ satisfies the strict energy conservation condition \eqref{eq:strict-qme}.

\section{Steady state of $\mathcal{L}^{aF}$}\label{app:ss-fme}

As explained in the main text, the fixed point of $\mathcal{L}^{aF}$ is defined by $\mathcal{L}^{aF}(\hat{\rho}_{DA}^{ss}) = 0 $, 
which implies $d_t E_{DA}=0$. Replacing in \eqref{eq:dea-fme} leads to 
\begin{equation}
    P_1^{ss} = \frac{\gamma_{1,\downarrow}+\gamma_{2,\downarrow}}{\gamma_{1,\uparrow}+\gamma_{2,\uparrow}}P_2^{ss},
\end{equation}
where $P_1^{ss},P_2^{ss}$ denote the values of $P_1(t),P_2(t)$ in the steady state. Combined with the normalisation condition, $P_1^{ss}+P_2^{ss}=1$ we obtain
\begin{equation}
\begin{array}{ll}
 P_1^{ss}&=\frac{\gamma_{1,\downarrow}+\gamma_{2,\downarrow}}{\gamma_{1,\downarrow}+\gamma_{2,\downarrow}+\gamma_{1,\uparrow}+\gamma_{2,\uparrow}},\\
 \\
  P_2^{ss}&=\frac{\gamma_{1,\uparrow}+\gamma_{2,\uparrow}}{\gamma_{1,\downarrow}+\gamma_{2,\downarrow}+\gamma_{1,\uparrow}+\gamma_{2,\uparrow}} \, .
\end{array}
\end{equation}
The resonances vanish in the steady state, as can be seen directly from the form of $\mathcal{L}^{aF}$.

Substituting now in \eqref{eq:q-fme} yields
\begin{equation}
    \dot Q^{ss} = -\omega_L(\gamma_{0,\downarrow}-\gamma_{0,\uparrow})-2\omega_L\frac{\gamma_{1,\downarrow}\gamma_{2,\uparrow}-\gamma_{1,\uparrow}\gamma_{2,\downarrow}}{\gamma_{1,\downarrow}+\gamma_{2,\downarrow}+\gamma_{1,\uparrow}+\gamma_{2,\uparrow}} .
\end{equation}
Using \eqref{eq:coefs-fme} and recalling that $G_+(\omega)>G_-(\omega)$ for all $\omega$, we deduce that $\dot Q^{ss}<0$. Replacing now in \eqref{eq:sigma-da-dl}, and using the fact that, in the steady state, $d_t S_{DA}=0$, we obtain that the entropy production rate of the dressed qubit, defined in \eqref{eq:sigma-da}, is strictly positive in the steady state,
\begin{equation}
    d_t S_{DA}^{ss} - \beta_B\dot Q^{ss} >0 \, .
\end{equation}

\section{Expression of $\mathcal{L}^{aB}$ with counting fields}\label{app:tilted-obe}

Assuming that $G_\pm$ are smooth enough in the range $[\omega_L-\Omega,\omega_L+\Omega]$, we replace $G_\pm(\omega_L), G_\pm(\omega_L\pm\Omega)\approx G_\pm(\omega_A)\equiv \overline{G}_\pm$ in \eqref{eq:gob}. Tracing out $DL$, which is straightforward using \eqref{eq:s-ops}, we obtain the tilted master equation $\mathcal{L}_{\boldsymbol{\lambda}}^{aB}$,
\begin{align}
    \mathcal{L}_{\boldsymbol{\lambda}}^{aB}(\hat{\rho}_{DA}^{\boldsymbol{\lambda}}) =& -i[\hat{H}_{DA},\hat{\rho}_{DA}^{\boldsymbol{\lambda}}] + \overline{G}_+\mathcal{D}_+^{\boldsymbol{\lambda}}(\hat{\rho}_{DA}^{\boldsymbol{\lambda}})+ \overline{G}_-\mathcal{D}_-^{\boldsymbol{\lambda}}(\hat{\rho}_{DA}^{\boldsymbol{\lambda}})
\end{align}
with
\begin{align}
    \mathcal{D}_+^{\boldsymbol{\lambda}}(\hat{\rho}) =& \left(\sum_{\alpha=+,-,z} e^{i\lambda_B\omega_\alpha/2}\hat{s}_\alpha^{\boldsymbol{\lambda}}\right)\hat{\rho} \left(\sum_{\alpha=+,-,z} e^{-i\lambda_B\omega_\alpha/2}\hat{s}_\alpha^{-\boldsymbol{\lambda}}\right)^\dagger \\
    &- \frac{1}{2}\left(\sum_{\alpha=+,-,z} \hat{s}_\alpha^{\boldsymbol{\lambda}}\right)^\dagger \left(\sum_{\alpha=+,-,z} \hat{s}_\alpha^{\boldsymbol{\lambda}}\right)\hat{\rho} - \frac{1}{2}\hat{\rho}\left(\sum_{\alpha=+,-,z} \hat{s}_\alpha^{-\boldsymbol{\lambda}}\right)^\dagger \left(\sum_{\alpha=+,-,z} \hat{s}_\alpha^{-\boldsymbol{\lambda}}\right) \nonumber \\
    \mathcal{D}_-^{\boldsymbol{\lambda}}(\hat{\rho}) =& \left(\sum_{\alpha=+,-,z} e^{i\lambda_B\omega_\alpha/2}\hat{s}_\alpha^{\boldsymbol{\lambda}}\right)^\dagger\hat{\rho} \left(\sum_{\alpha=+,-,z} e^{-i\lambda_B\omega_\alpha/2}\hat{s}_\alpha^{-\boldsymbol{\lambda}}\right) \\
    &- \frac{1}{2}\left(\sum_{\alpha=+,-,z} \hat{s}_\alpha^{\boldsymbol{\lambda}}\right) \left(\sum_{\alpha=+,-,z} \hat{s}_\alpha^{\boldsymbol{\lambda}}\right)^\dagger\hat{\rho} - \frac{1}{2}\hat{\rho}\left(\sum_{\alpha=+,-,z} \hat{s}_\alpha^{-\boldsymbol{\lambda}}\right)\left(\sum_{\alpha=+,-,z} \hat{s}_\alpha^{-\boldsymbol{\lambda}}\right)^\dagger \, . \nonumber 
\end{align}
where the operators $\hat{s}_{+,-,z}^{\boldsymbol{\lambda}}$ were defined in \eqref{eq:slambda-1}.

\section{$\dot W_L$ for the generalized Bloch equation}\label{app:wl-gob}

The rate $\dot W_L$ for the generalized Bloch equation is, in the weak and intermediate regimes,
\begin{equation}
    \dot W_L = \dot W_{DL} +  \text{Tr}\left[\frac{\omega_L}{2}\hat{\sigma}_z \mathcal{L}^{G}(\hat{\rho})\right]
\end{equation}
with
\begin{equation}
    \text{Tr}\left[\frac{\omega_L}{2}\hat{\sigma}_z \mathcal{L}^{aG}(\hat{\rho})\right] = -g\omega_L\text{Im}[P_{21}(t)] + \frac{\omega_L}{2}\text{Tr}[\hat{\sigma}_z\mathcal{D}^{G}_+(\hat{\rho})]+\frac{\omega_L}{2}\text{Tr}[\hat{\sigma}_z\mathcal{D}^{G}_-(\hat{\rho})]
\end{equation}
where
\begin{align}
    \text{Tr}[\hat{\sigma}_z\mathcal{D}^{G}_+(\hat{\rho})] =&-\frac{g^2}{4\Omega^2}\left(\sqrt{G_+(\omega_L)G_+(\omega_L-\Omega)}\frac{\Omega-\delta}{\Omega}+\sqrt{G_+(\omega_L)G_+(\omega_L+\Omega)}\frac{\Omega+\delta} {\Omega}\right) \nonumber \\
    &+P_1(t)\left( -\frac{g^2}{2\Omega^2} \sqrt{G_+(\omega_L)G_+(\omega_L-\Omega)}\frac{\Omega-\delta}{\Omega} +\frac{\delta}{\Omega} G_+(\omega_L-\Omega) \left(\frac{\Omega-\delta}{2\Omega}\right)^2\right) \nonumber \\
  &  + P_2(t)\left(-\frac{g^2}{2\Omega^2}\sqrt{G_+(\omega_L)G_+(\omega_L+\Omega)}\frac{\Omega+\delta}{\Omega} -\frac{\delta}{\Omega} G_+(\omega_L+\Omega) \left(\frac{\Omega+\delta}{2\Omega}\right)^2\right) \nonumber \\
    &+[P_{21}(t)+P_{12}(t)]\left(\frac{g}{2\Omega}\right)^3\left[4G_+(\omega_L) + 2\sqrt{G_+(\omega_L+\Omega)G_+(\omega_L-\Omega)}\right] \nonumber \\
    &+P_{21}(t) \left[\frac{g}{\Omega}G_+(\omega_L-\Omega)\left(\frac{\Omega-\delta}{2\Omega}\right)^2+\frac{\delta g}{2\Omega^2}\sqrt{G_+(\omega_L)G_+(\omega_L+\Omega)}\frac{\Omega+\delta}{\Omega} \right] \nonumber \\
    &+P_{12}(t)\left[\frac{g}{\Omega}G_+(\omega_L+\Omega)\left(\frac{\Omega+\delta}{2\Omega}\right)^2 -\frac{\delta g}{2\Omega^2}\sqrt{G_+(\omega_L)G_+(\omega_L-\Omega)}\frac{\Omega-\delta}{\Omega}\right] \nonumber
\end{align}
and
\begin{align}
    \text{Tr}[\hat{\sigma}_z\mathcal{D}^{G}_-(\hat{\rho})] =&-\frac{g^2}{4\Omega^2}\left(\sqrt{G_-(\omega_L)G_-(\omega_L-\Omega)}\frac{\Omega-\delta}{\Omega}+\sqrt{G_-(\omega_L)G_-(\omega_L+\Omega)}\frac{\Omega+\delta}{\Omega}\right) \nonumber \\
&+ P_1(t)\left(-\frac{g^2}{2\Omega^2}\sqrt{G_-(\omega_L)G_-(\omega_L+\Omega)}\frac{\Omega+\delta}{\Omega} -\frac{\delta}{\Omega} G_-(\omega_L+\Omega) \left(\frac{\Omega+\delta}{2\Omega}\right)^2\right) \nonumber \\
&+P_2(t)\left( -\frac{g^2}{2\Omega^2} \sqrt{G_-(\omega_L)G_-(\omega_L-\Omega)}\frac{\Omega-\delta}{\Omega} +\frac{\delta}{\Omega} G_-(\omega_L-\Omega) \left(\frac{\Omega-\delta}{2\Omega}\right)^2\right) \nonumber \\
    &-[P_{21}(t)+P_{12}(t)]\left(\frac{g}{2\Omega}\right)^3\left[4G_-(\omega_L) + 2\sqrt{G_-(\omega_L+\Omega)G_-(\omega_L-\Omega)}\right] \nonumber \\
     &-P_{21}(t)\left[\frac{g}{\Omega}G_-(\omega_L+\Omega)\left(\frac{\Omega+\delta}{2\Omega}\right)^2 -\frac{\delta g}{2\Omega^2}\sqrt{G_-(\omega_L)G_+(\omega_L-\Omega)}\frac{\Omega-\delta}{\Omega}\right] \nonumber \\
    &-P_{12}(t) \left[\frac{g}{\Omega}G_-(\omega_L-\Omega)\left(\frac{\Omega-\delta}{2\Omega}\right)^2+\frac{\delta g}{2\Omega^2}\sqrt{G_-(\omega_L)G_-(\omega_L+\Omega)}\frac{\Omega+\delta}{\Omega} \right]
    \, .
\end{align}

In the strong driving regimes, it is equal to the rate obtained with the Floquet master equation, in \eqref{eq:wl-fme}.

\section{Derivation of the Floquet master equation dressed with counting fields}\label{app:tilted-fme-2}

We give here the full expression of the equation \eqref{eq:fme-3}. Following the method explained in the main text, we obtain
\begin{align}
&\mathcal{D}^{F\boldsymbol{\lambda}}_{0,+}(\hat{\tilde{\rho}}_A) = e^{-i\omega_L\lambda_L}e^{i\omega_L\lambda_B}\hat{\Sigma}_z^F(t+\lambda_L/2)\hat{\tilde{\rho}}_A\hat{\Sigma}_z^F(t-\lambda_L/2) \nonumber\\
&-\frac{1}{2}\hat{\Sigma}_z^F(t+\lambda_L/2)\hat{\Sigma}_z^F(t+\lambda_L/2)\hat{\tilde{\rho}}_A\\
&-\frac{1}{2}\hat{\tilde{\rho}}_A\hat{\Sigma}_z^F(t-\lambda_L/2)\hat{\Sigma}_z^F(t-\lambda_L/2) \, ,\nonumber
\end{align}

\begin{align}
&\mathcal{D}^{F\boldsymbol{\lambda}}_{0,-}(\hat{\tilde{\rho}}_A) = e^{i\omega_L\lambda_L}e^{-i\omega_L\lambda_B}\hat{\Sigma}_z^F(t+\lambda_L/2)\hat{\tilde{\rho}}_A\hat{\Sigma}_z^F(t-\lambda_L/2) \nonumber\\
&-\frac{1}{2}\hat{\Sigma}_z^F(t+\lambda_L/2)\hat{\Sigma}_z^F(t+\lambda_L/2)\hat{\tilde{\rho}}_A\\
&-\frac{1}{2}\hat{\tilde{\rho}}_A\hat{\Sigma}_z^F(t-\lambda_L/2)\hat{\Sigma}_z^F(t-\lambda_L/2) \, ,\nonumber
\end{align}

\begin{align}
&\mathcal{D}^{F\boldsymbol{\lambda}}_{1,+}(\hat{\tilde{\rho}}_A) = e^{-i\omega_L\lambda_L}e^{i(\omega_L+\Omega)\lambda_B}\hat{\Sigma}_-^F(t+\lambda_L/2)\hat{\tilde{\rho}}_A\hat{\Sigma}_+^F(t-\lambda_L/2) \nonumber\\
&-\frac{1}{2}\hat{\Sigma}_+^F(t+\lambda_L/2)\hat{\Sigma}_-^F(t+\lambda_L/2)\hat{\tilde{\rho}}_A\\
&-\frac{1}{2}\hat{\tilde{\rho}}_A\hat{\Sigma}_+^F(t-\lambda_L/2)\hat{\Sigma}_-^F(t-\lambda_L/2) \, ,\nonumber
\end{align}

\begin{align}
&\mathcal{D}^{F\boldsymbol{\lambda}}_{1,-}(\hat{\tilde{\rho}}_A) = e^{i\omega_L\lambda_L}e^{-i(\omega_L+\Omega)\lambda_B}\hat{\Sigma}_+^F(t+\lambda_L/2)\hat{\tilde{\rho}}_A\hat{\Sigma}_-^F(t-\lambda_L/2) \nonumber\\
&-\frac{1}{2}\hat{\Sigma}_+^F(t+\lambda_L/2)\hat{\Sigma}_-^F(t+\lambda_L/2)\hat{\tilde{\rho}}_A\\
&-\frac{1}{2}\hat{\tilde{\rho}}_A\hat{\Sigma}_+^F(t-\lambda_L/2)\hat{\Sigma}_-^F(t-\lambda_L/2) \, ,\nonumber
\end{align}

\begin{align}
&\mathcal{D}^{F\boldsymbol{\lambda}}_{2,+}(\hat{\tilde{\rho}}_A) = e^{-i\omega_L\lambda_L}e^{i(\omega_L-\Omega)\lambda_B}\hat{\Sigma}_+^F(t+\lambda_L/2)\hat{\tilde{\rho}}_A\hat{\Sigma}_-^F(t-\lambda_L/2) \nonumber\\
&-\frac{1}{2}\hat{\Sigma}_-^F(t+\lambda_L/2)\hat{\Sigma}_+^F(t+\lambda_L/2)\hat{\tilde{\rho}}_A\\
&-\frac{1}{2}\hat{\tilde{\rho}}_A\hat{\Sigma}_-^F(t-\lambda_L/2)\hat{\Sigma}_+^F(t-\lambda_L/2) \, ,\nonumber
\end{align}

\begin{align}
&\mathcal{D}^{F\boldsymbol{\lambda}}_{2,-}(\hat{\tilde{\rho}}_A) = e^{i\omega_L\lambda_L}e^{-i(\omega_L-\Omega)\lambda_B}\hat{\Sigma}_-^F(t+\lambda_L/2)\hat{\tilde{\rho}}_A\hat{\Sigma}_+^F(t-\lambda_L/2) \nonumber\\
&-\frac{1}{2}\hat{\Sigma}_+^F(t+\lambda_L/2)\hat{\Sigma}_-^F(t+\lambda_L/2)\hat{\tilde{\rho}}_A\\
&-\frac{1}{2}\hat{\tilde{\rho}}_A\hat{\Sigma}_+^F(t-\lambda_L/2)\hat{\Sigma}_-^F(t-\lambda_L/2) \, .\nonumber
\end{align}

\section{Redfield equation with counting fields}\label{app:redfield}

We give here the first and last terms of the tilted master equation \eqref{eq:gen-eq-cf},

\begin{equation}
    \begin{array}{l}
-Tr_B\left[ \frac{1}{\delta_0}\int\limits_t^{t+\delta_0} dt'\int\limits_0^{+\infty}d\tau \hat{V}_{AB}^{\boldsymbol{\lambda}/2}(t') \hat{V}_{AB}^{\boldsymbol{\lambda}/2}(t'-\tau)\hat{\rho}(t)\otimes\hat{\rho}_B\right] \\
\\
=-\frac{1}{\delta_0}\int\limits_t^{t+\delta_0} dt'\left[G_+(\omega_L)\hat{S}_z\hat{S}_z^\dagger+G_+(\omega_L-\Omega)\hat{S}_-\hat{S}_-^\dagger+G_+(\omega_L+\Omega)\hat{S}_+\hat{S}_+^\dagger \right.\\
\\
+G_+(\omega_L-\Omega)e^{i\Omega t'}e^{i\Omega \lambda_{DA}/2}\hat{S}_z\hat{S}_-^\dagger+G_+(\omega_L+\Omega)e^{-i\Omega t'}e^{-i\Omega\lambda_{DA}/2}\hat{S}_z\hat{S}_+^\dagger\\
\\
+G_+(\omega_L)e^{-i\Omega\lambda_{DA}/2}e^{-i\Omega t'}\hat{S}_-\hat{S}_z^\dagger+G_+(\omega_L)e^{i\Omega\lambda_{DA}/2}e^{i\Omega t'}\hat{S}_+\hat{S}_z^\dagger\\
\\
+G_-(\omega_L)\hat{S}_z^\dagger\hat{S}_z+G_+(\omega_L-\Omega)\hat{S}_-^\dagger \hat{S}_-+G_+(\omega_L+\Omega)\hat{S}_+^\dagger \hat{S}_+\\
\\
+G_-(\omega_L-\Omega)e^{-i\Omega t'}e^{-i\Omega \lambda_{DA}/2}\hat{S}_z^\dagger\hat{S}_-+G_-(\omega_L+\Omega)e^{i\Omega t'}e^{i\Omega\lambda_{DA}/2}\hat{S}_z^\dagger\hat{S}_+\\
\\
\left.+G_-(\omega_L)e^{i\Omega\lambda_{DA}/2}e^{i\Omega t'}\hat{S}_-^\dagger\hat{S}_z+G_-(\omega_L)e^{-i\Omega\lambda_{DA}/2}e^{-i\Omega t'}\hat{S}_+^\dagger\hat{S}_z\right]\hat{\rho}(t)
    \end{array}
\end{equation}
and
\begin{equation}
\begin{array}{l}
Tr_B\left[ \frac{1}{\delta_0}\int\limits_t^{t+\delta_0} dt'\int\limits_0^{+\infty}d\tau \hat{V}_{AB}^{\boldsymbol{\lambda}/2}(t'-\tau)\hat{\rho}(t)\otimes\hat{\rho}_B \hat{V}_{AB}^{-\boldsymbol{\lambda}/2}(t')\right]\\
\\
=\frac{1}{\delta_0}\int\limits_t^{t+\delta_0} dt' \left[e^{i\omega_L\lambda_{DL}}\left( G_-(\omega_L)e^{-i\omega_L\lambda_B}\hat{S}_z\hat{\rho}(t)\hat{S}_z^\dagger\right. \right.\\
\\
+G_-(\omega_L-\Omega)e^{-i(\omega_L-\Omega)\lambda_B}e^{-i\Omega\lambda_{DA}}\hat{S}_-\hat{\rho}(t)\hat{S}_-^\dagger+G_-(\omega_L+\Omega)e^{-i(\omega_L+\Omega)\lambda_B}e^{i\Omega\lambda_{DA}}\hat{S}_+\hat{\rho}(t)\hat{S}_+^\dagger \\
\\
+G_-(\omega_L)e^{-i\omega_L\lambda_B}e^{-i\Omega\lambda_{DA}/2}e^{i\Omega t}\hat{S}_z\hat{\rho}(t)\hat{S}_-^\dagger+G_-(\omega_L)e^{-i\omega_L\lambda_B}e^{i\Omega\lambda_{DA}/2}e^{-i\Omega t}\hat{S}_z\hat{\rho}(t)\hat{S}_+^\dagger\\
\\
+G_-(\omega_L-\Omega)e^{-i\lambda_B(\omega_L-\Omega)}e^{-i\Omega\lambda_{DA}/2}e^{-i\Omega t'}\hat{S}_-\hat{\rho}(t)\hat{S}_z^\dagger+G_-(\omega_L+\Omega)e^{-i\lambda_B(\omega_L+\Omega)}e^{i\Omega\lambda_{DA}/2}e^{i\Omega t'}\hat{S}_+\hat{\rho}(t)\hat{S}_z^\dagger\\
\\
\left.+G_-(\omega_L-\Omega)e^{-i\lambda_B(\omega_L-\Omega)}e^{-2i\Omega t'}\hat{S}_-\hat{\rho}(t)\hat{S}_+^\dagger+G_-(\omega_L+\Omega)e^{-i\lambda_B(\omega_L+\Omega)}e^{2i\Omega t'}\hat{S}_+\hat{\rho}(t)\hat{S}_-^\dagger\right)\\
\\
+e^{-i\omega_L\lambda_{DL}}\left( G_+(\omega_L)e^{i\omega_L\lambda_B}\hat{S}_z^\dagger\hat{\rho}(t)\hat{S}_z\right.\\
\\
+G_+(\omega_L-\Omega)e^{i(\omega_L-\Omega)\lambda_B}e^{i\Omega\lambda_{DA}}\hat{S}_-^\dagger\hat{\rho}(t)\hat{S}_-+G_+(\omega_L+\Omega)e^{i(\omega_L+\Omega)\lambda_B}e^{-i\Omega\lambda_{DA}}\hat{S}_+^\dagger\hat{\rho}(t)\hat{S}_+ \\
\\
+G_+(\omega_L)e^{i\omega_L\lambda_B}e^{i\Omega\lambda_{DA}/2}e^{-i\Omega t}\hat{S}_z^\dagger\hat{\rho}(t)\hat{S}_-+G_+(\omega_L)e^{i\omega_L\lambda_B}e^{-i\Omega\lambda_{DA}/2}e^{i\Omega t}\hat{S}_z^\dagger\hat{\rho}(t)\hat{S}_+\\
\\
+G_+(\omega_L-\Omega)e^{i\lambda_B(\omega_L-\Omega)}e^{i\Omega\lambda_{DA}/2}e^{i\Omega t'}\hat{S}_-^\dagger\hat{\rho}(t)\hat{S}_z+G_+(\omega_L+\Omega)e^{i\lambda_B(\omega_L+\Omega)}e^{-i\Omega\lambda_{DA}/2}e^{-i\Omega t'}\hat{S}_+^\dagger\hat{\rho}(t)\hat{S}_z\\
\\
\left.\left.+G_+(\omega_L-\Omega)e^{i\lambda_B(\omega_L-\Omega)}e^{2i\Omega t'}\hat{S}_-^\dagger\hat{\rho}(t)\hat{S}_++G_+(\omega_L+\Omega)e^{i\lambda_B(\omega_L+\Omega)}e^{-2i\Omega t'}\hat{S}_+^\dagger\hat{\rho}(t)\hat{S}_-\right)\right] \, .
\end{array}
\end{equation}

\begin{equation}
\begin{array}{ll}
    \frac{d\hat{\rho}_{DA}^{\boldsymbol{\lambda}}}{dt}&\equiv\mathcal{L}_{\boldsymbol{\lambda}}^{aB}\\
    \\
    &=-i[\hat{H}_{DA},\hat{\rho}_{DA}^{\boldsymbol{\lambda}}]+\overline{G}_+\mathcal{D}_{\hat{s}}^{\boldsymbol{\lambda}}(\hat{\rho}_{DA}^{\boldsymbol{\lambda}})+\overline{G}_-\mathcal{D}_{\hat{s}^\dagger}^{\boldsymbol{\lambda}}(\hat{\rho}_{DA}^{\boldsymbol{\lambda}}) \, ,
\end{array}
\end{equation}
with
\begin{equation}
    \begin{array}{ll}
\mathcal{D}_{\hat{s}}^{\boldsymbol{\lambda}}(\hat{\rho}) =&-\frac{1}{2}(\hat{s}_{\lambda_{DA},0}^\dagger \hat{s}_{\lambda_{DA},0}\hat{\rho}+\hat{\rho}\hat{s}_{-\lambda_{DA},0}^\dagger \hat{s}_{-\lambda_{DA},0})\\
\\
&+\frac{1}{2}e^{i\omega_L(\lambda_B-\lambda_{DL})}(\hat{s}_{\lambda_{DA},0}\hat{\rho}\hat{s}_{-\lambda_{DA},-\lambda_B}^\dagger+\hat{s}_{\lambda_{DA},\lambda_B}\hat{\rho}\hat{s}_{-\lambda_{DA},0}^\dagger)\\
\\
\mathcal{D}_{\hat{s}^\dagger}^{\boldsymbol{\lambda}}(\hat{\rho}) =&-\frac{1}{2}(\hat{s}_{\lambda_{DA},0} \hat{s}_{\lambda_{DA},0}^\dagger\hat{\rho}+\hat{\rho}\hat{s}_{-\lambda_{DA},0} \hat{s}^\dagger_{-\lambda_{DA},0})\\
\\
&+\frac{1}{2}e^{-i\omega_L(\lambda_B-\lambda_{DL})}(\hat{s}_{\lambda_{DA},0}^\dagger\hat{\rho}\hat{s}_{-\lambda_{DA},-\lambda_B}+\hat{s}_{\lambda_{DA},\lambda_B}^\dagger\hat{\rho}\hat{s}_{-\lambda_{DA},0}) \, ,
    \end{array}
\end{equation}
where
\begin{equation}
\hat{s}_{\lambda_{DA},\lambda_B}=\hat{s}_z+e^{i\Omega(\lambda_{DA}/2-\lambda_B)}\hat{s}_+ +e^{-i\Omega(\lambda_{DA}/2-\lambda_B)}\hat{s}_- \, .
\end{equation}
The explicit expressions \eqref{eq:diss-obe} allow to see directly that the symmetry \eqref{eq:detailed-2} is not satisfied. It is also clear that the strict energy conservation condition \eqref{eq:strict-qme} is not satisfied.

\section{Steady-state of $\mathcal{L}^{aB}$}\label{app:ss-obe}

The fixed point of $\mathcal{L}^{aB}$, defined by $\mathcal{L}^{aB}(\hat{\rho}_{DA}^{ss}) = 0 $, is

\begin{equation}
\begin{array}{ll}
P_b^{ss} &= \frac{1}{\overline{G}_+ + \overline{G}_-}\left( \overline{G}_- + \frac{1}{2}\frac{\overline{G}_+ - \overline{G}_-}{1+2\frac{\delta^2}{g^2} + \frac{(\overline{G}_+ + \overline{G}_-)^2}{2g^2}} \right) \\
\\
P_{ba}^{ss} &= - \frac{ \frac{\delta (\overline{G}_+ - \overline{G}_-) }{g(\overline{G}_+ + \overline{G}_-)} + i\frac{(\overline{G}_+ - \overline{G}_-)}{2g} }{1+2\frac{\delta^2}{g^2} + \frac{(\overline{G}_+ + \overline{G}_-)^2}{2g^2}} \, .
\end{array}
\end{equation}

\bibliography{bib}

\end{document}